\documentstyle[preprint,rmp,aps,floats,epsfig]{revtex}

\tightenlines
\def\etal{{\it et al.}}

\begin{document}

\title{Suppression of Bremsstrahlung and Pair Production
due to Environmental Factors}

\author{Spencer Klein}

\address{Lawrence  Berkeley National Laboratory, Berkeley, CA 94720}

\break
\maketitle

\begin{abstract}

The environment in which bremsstrahlung and pair creation occurs can
strongly affect cross sections for these processes.  Because
ultra-relativistic electromagnetic interactions involve very small
longitudinal momentum transfers, the reactions occur gradually, spread
over long distances.  During this time, even relatively weak factors
can accumulate enough to disrupt the interaction.

This review will discuss a variety of factors which can suppress
bremsstrahlung and pair production, as well as related effects
involving beamstrahlung and QCD processes.  After surveying different
theoretical approaches, experimental measurements will be covered.
Recent accurate measurements by the SLAC E-146 collaboration will be
highlighted, along with several recent theoretical works relating
to the experiment.

\end{abstract}
\vskip .1 in
\centerline{(To appear in {\it Reviews of Modern Physics})}
\pacs{PACS  Numbers: }
\narrowtext

\tableofcontents
%

\def\xo{$X_0$}
\section{Introduction}
\label{intro}

Bremsstrahlung and pair creation are two of the most common high
energy electromagnetic processes, and the interaction cross sections
are well known (Bethe and Heitler, 1934).  However, it is much less
well known that these cross sections can change dramatically depending
on the environment in which the interaction occurs.  The SLAC E-146
collaboration observed photon intensities as low as 1/4 of that
predicted by Bethe and Heitler, and much larger suppression is
possible.

The cross sections change because the kinematics of the processes
dictates the interactions are spread over a significant distance, in
contrast to the conventional picture where reactions occur at a single
point.  The momentum transfer between a highly relativistic
interacting particle and the target nucleus can be small, especially
along the direction of particle motion.  When this longitudinal
momentum transfer is small, the uncertainty principle dictates that
the interaction is spread out over a distance, known as the formation
length for particle production, or, more generally, as the coherence
length.

If the medium in the neighborhood of the interaction has enough
influence on the interacting particle during its passage through the
formation zone, then the pair production or bremsstrahlung can be
suppressed.  When the formation length is long, weak but cumulative
factors can be important.  Some of the factors that may suppress
electromagnetic radiation are multiple scattering, photon interactions
with the medium (coherent forward Compton scattering), and magnetic
fields.  In crystals, where the atoms are arranged in ordered rows, a
large variety of effects can suppress or enhance radiation.

The formation length has a number of interesting physical
interpretations. It is the wavelength of the exchanged virtual photon.
It is also the distance required for the final state particles to
separate enough (by a Compton wavelength) that they act as separate
particles.  It is also the distance over which the amplitudes from
several interactions can add coherently to the total cross section.
All of these pictures can be helpful in understanding different
aspects of suppression mechanisms.

The formation length was first considered by Ter-Mikaelian in 1952
(Feinberg, 1994).  It occurs in both classical and quantum mechanical
calculations.  Classically, all electromagnetic processes have a phase
factor
\begin{equation}
\Phi=\exp{\{i[\omega t -{\bf k}\cdot{\bf r}(t)]\}}
\label{cphase}
\end{equation}
where $\omega$ is the photon frequency, $t$ is the time, ${\bf k}$ is
the photon wave vector, and ${\bf r}(t)$ is the electron position.
The formation length is the distance over which this phase factor
maintains coherence; the electron and photon have slightly different
velocities, which causes them to gradually separate.  Coherence is
maintained over a distance for which $\omega t -{\bf k}\cdot{\bf
r}(t)\sim 1$.  Around an isolated atom $\omega=|{\bf k}|c$, so,
neglecting the (small) angle between ${\bf k}$ and ${\bf r}$, for a
straight trajectory
\begin{equation}
l_{f0} = {2 E^2 \over \omega m^2c^3}
\label{lf0c}
\end{equation}
where $E$ is the electron energy, $m$ is the electron mass, and $c$
the speed of light.  The $f0$ subscript denotes the unsuppressed (free
space) formation length.  Electromagnetic interactions are spread over
this distance.  If $E$ is large and $\omega$ small, $l_{f0}$ can be
very long.  For example, for a 25 GeV electron radiating a 10 MeV
bremsstrahlung photon, $l_{f0} = 100\mu$m.  This distance was the
basis for the suppression calculations by Landau and Pomeranchuk,
discussed in Sec. \ref{semiclassical}.

The formation length also occurs in quantum mechanical calculations.
Electron bremsstrahlung, where an electron interacts with a nucleus
and emits a real photon is a good example.  As the electron
accelerates, part of its surrounding virtual photon field shakes
loose.  Neglecting the photon emission angle and electron scattering,
the momentum transfer in the longitudinal ($+z$) direction is
\begin{equation}
q_\parallel = p_e - p_e' - p_\gamma =\ \sqrt{(E/c)^2\ -\ (mc)^2}
-\sqrt{((E-k)/c)^2\ -\ (mc)^2} - k/c
\label{eqlong}
\end{equation}
where $p_e$ and $p_e'$ are the electron momenta before and after the
interaction respectively and the photon momentum $p_\gamma=k/c$, with
$k=\hbar\omega=\hbar |{\bf k}|/c$ the photon energy. For $\gamma=
E/mc^2 \gg 1$ this simplifies to
\begin{equation}
q_\parallel\sim {m^2 c^3 k \over 2E (E - k) }.
\label{eqlongs}
\end{equation}
Here, we have used the small $y$ approximation $\sqrt{1-y}\sim 1-y/2$.
For high energy electrons emitting low energy photons, $q_\parallel$
can become very small.  For the above example, $q_\parallel= 0.002\
$eV/c.  Because $q_\parallel$ is so small, the uncertainty principle
requires that the radiation must take place over a long distance:
\begin{equation}
l_{f0} = {\hbar \over q_\parallel}= {2 \hbar E (E-k) \over m^2c^3 k }.
\label{lfzero}
\end{equation}
For $k\ll E$, this matches Eq. (\ref{lf0c}); the momentum transfer
approach is needed only for $k\sim E$.

This formation length appears in most electromagnetic processes
including pair production, transition radiation, \v{C}erenkov
radiation and synchrotron radiation.  Usually, $q_\parallel$ acquires
additional terms to account for specific interactions with the target
medium.  This review will discuss a number of these additions, and
show how they affect bremsstrahlung and pair production.  Formation
lengths also apply to processes involving other forces; this review
will consider a few non-electromagnetic interactions, showing how
kinematics can dictate that diverse reactions share many fundamental
traits.

The formation length is the distance over which the interaction
amplitudes can add coherently.  So, if an external force doubles
$q_\parallel$, thereby halving $l_f$, then the radiation intensity is
also halved, from $\sim l_f^2$ to $\sim 2\cdot(l_f/2)^2$ as the
trajectory splits into two independent emitting regions.  If an
electron traversing a distance $D$ radiates, the trajectory acts as
$D/l_f$ independent emitters, each with a strength proportional to
$|l_f|^2$, so the total radiation is proportional to $l_f$. As
external factors increase $q_\parallel$ and reduce $l_f$, the
radiation drops proportionally.

A similar coherence distance $\hbar/q_\perp$ limits the perpendicular
distance over which coherent addition is possible.  However, because
$q_\perp\gg q_\parallel$, this distance is much smaller, and of lesser
interest here.

This article begins by considering separate classical and quantum
mechanical (semi-classical) calculations of suppression due to
multiple scattering.  Suppression due to photon interactions with the
medium, pair creation and external magnetic fields will then be
similarly considered.

Other calculations have considered the multiple scattering in more
detail, as diffusion of the electron.  Migdal's 1956 calculation, in
Sec. \ref{smigdal}, was the first; it has become something of a
standard.  A number of recent calculations of suppression are
described in Secs. \ref{szakharov} to \ref{sbkatkov}.

Experimental work is surveyed in Sec. \ref{sexperiment}, beginning
with the first cosmic ray studies of Landau-Pomeranchuk-Migdal (LPM)
suppression shortly after Migdal's paper appeared.  These suffered
from very limited statistics, and hence had limited significance.  In
the 1970's and 1980's, a few accelerator experiments provided some
data, but still with limited accuracy.  In 1993, SLAC experiment E-146
made detailed measurements of suppression due to multiple scattering
and photon-medium interactions.  E-146 confirmed Migdal's formula to
good accuracy, and at least partly inspired the more recent
calculations in Secs.  \ref{szakharov} to \ref{sbkatkov}; because of
this timing, some reference will be made to E-146 before the
experiment is described in detail.

Some more specialized topics will also be considered.  Section\
\ref{splasmas} considers suppression in plasmas, where particles with
similar energies interact with each other. Section \ref{sshowers}
shows how suppression mechanisms can affect electromagnetic showers,
with a focus on cosmic ray air showers. Section\ \ref{sqcd} surveys
the application of LPM formalism to hadrons scattering inside nuclei,
where color charge replaces electric charge, and gluons replace
photons. Finally, a number of open questions will be presented.

Earlier reviews (Feinberg and Pomeranchuk, 1956) (Akhiezer and
Shul'ga, 1987) and monographs (Ter-Mikaelian, 1972) (Akhiezer and
Shul'ga, 1996) have discussed these topics.  A recent article
(Feinberg, 1994) discusses the subject's early development.

Unless otherwise indicated, energies will be given in electron volts
(eV) and momentum in eV/c; $\hbar c$ is $1.97\times10^{-7}$eV m and
the fine structure constant $\alpha=e^2/\hbar c \approx 1/137$, where
$e$ is the electric charge.  For both bremsstrahlung and pair
conversion, $E$ will represent electron/positron energy and $k$ photon
energy.

\section{Classical and Semi-classical Formulations}
\label{semiclassical}

Landau and Pomeranchuk (1953a) used classical electromagnetism to
demonstrate bremsstrahlung suppression due to multiple scattering; the
suppression comes from the interference between photons emitted by
different elements of electron pathlength.  For those with quantum
tastes, subsection \ref{sbqcd} gives a simple semi-classical
derivation based on the uncertainty principle.

\subsection{Classical Bremsstrahlung - Landau and Pomeranchuk}

The classical intensity for radiation from an accelerated charge in
the influence of a nucleus with charge $Z$ is
\begin{equation}
{d^2I \over d\omega d\Omega} = {Z^2 e^2\omega^2 \over 4\pi^2c^3}
\bigg|\int {\bf n}\times d{\bf r} \exp{\{i[(\omega t- {\bf k}\cdot
{\bf r}(t)]\} }\bigg|^2
\label{classicalb}
\end{equation}
where $d{\bf r}={\bf v}(t)dt$, {\bf n} is the photon direction and
${\bf v}(t)$ the electron velocity.  The angular integration $d\Omega$
is over all possible photon and outgoing electron directions.  The
classical bremsstrahlung spectrum can be found by assuming that {\bf
v} changes abruptly during the collision.  Then, the integral splits
into two pieces and the emission is easily found:
\begin{equation}
{d^2I \over d\omega d\Omega} = {Z^2e^2  \over 4\pi^2 c}
\bigg| { {\bf k} \times {\bf v_1} \over {\bf k}\cdot {\bf v_1} -|{\bf k}|c}
- { {\bf k} \times {\bf v_2} \over {\bf k}\cdot {\bf v_2} -|{\bf k}|c}
\bigg|^2.
\label{classicalb2}
\end{equation}
where ${\bf v}_1$ and ${\bf v}_2$ are the electron velocity before and
after the interaction respectively.  With small angle approximations,
${\bf k}\cdot {\bf v} =\omega(1-1/2\gamma^2)(1-\theta_\gamma^2/2)$,
where $\theta_\gamma$ is the angle between the photon and incident
electron direction and $|{\bf k}\times {\bf v}| = |{\bf k}||{\bf v}|
\theta_\gamma$, the bremsstrahlung angular distribution may be derived
(for $\gamma\gg 1, k\ll E$):
\begin{equation}
{dI^2 \over d\omega d\Omega} = {Z^2e^2\gamma^4 |{\bf \Delta v}|^2 \over
\pi^2 c^3} { (1+\gamma^4\theta_\gamma^4) \over
(1+\gamma^2\theta_\gamma^2)^4}
\label{eqnb}
\end{equation}
where ${\bf \Delta v}={\bf v}_1-{\bf v}_2$.  The overall emission
intensity can be found by integrating over $d\Omega$, and expressed in
terms of the change in electron momentum $q=\gamma m|{\bf \Delta v}|$.
Then,
\begin{equation}
{dI \over d\omega} = {2Z^2 e^2 q^2 \over 3\pi m^2c^3}.
\label{eqnd}
\end{equation}
The complete bremsstrahlung cross section can be found by multiplying
this by the elastic scattering cross section, as a function of $q$ and
converting from classical field intensity to cross section 
(Jackson, 1975, pg. 709):
\begin{equation}
{d\sigma \over dk} = { 16 Z^2 \alpha r_e^2  \over 3k}  
\int_{q_{min}}^{q_{max}} {dq \over q}
\label{eqbremsigma}
\end{equation}
where $r_e=e^2/mc^2=2.8\times10^{-15}$ m is the classical electron
radius. The integral evaluates to $\ln{(q_{max}/q_{min})}$, often
called the form factor.  This form factor is roughly equivalent to
$\ln{(\theta_{max}/\theta_{min})}$, where $\theta$ is the electron
scattering angle; this representation is convenient, because the
distributions of angles depend on the suppression.  The use of
$\alpha$, and implicitly $\hbar$ is needed to convert from field
intensity to number of photons.

The minimum and maximum momentum transfers correspond to maximum and
minimum effective impact parameters respectively.  For a bare nucleus,
the minimum impact parameter is the electron Compton wavelength
$\lambda_e= \hbar/mc = r_e/\alpha = 3.8\times 10^{-13}$ m, so
$q_{max}=2mc$. When the momentum transfer $q_{max}$ is larger than
$2mc$, the electron scatters by an angle larger than $1/\gamma$.
Then, coherence between the incoming and outgoing electron pathlength
(implied in Eq. (\ref{classicalb2})) is lost and the cross section
drops.  The probability of such a large scatter is very small.  This
loss of coherence foreshadows how multiple scattering affects the
electron trajectory and hence the radiation. The minimum momentum
transfer occurs when $q_\perp=0$ and $q_{min}=q_\parallel$.

However, nuclei are generally surrounded by electrons which screen the
nuclear charge.  For ultrarelativistic electrons, screening is
important over almost the entire range of $k$.  Screening shields the
radiating electron from the nucleus at large distances, so the maximum
impact parameter is the Thomas-Fermi atomic radius $a=0.8
Z^{-1/3}a_0$, for $Z\gg 1$ (Tsai, 1974), where the Bohr radius
$a_0=\hbar/\alpha mc=r_e/\alpha^2=5.3\times10^{-9}$ m. Then,
$q_{min}=1.25 \hbar/\alpha mcZ^{1/3}$ and
$\ln{(q_{max}/q_{min})}=\ln{(2.5/ \alpha Z^{-1/3})}$.  A more detailed
calculation finds $\ln{(184Z^{-1/3})}$ for the form factor (Bethe and
Heitler, 1934).

This classical calculation is valid only for $y=k/E\ll 1$.  A Born
approximation quantum mechanical calculation gives a similar result,
but covers the entire range of $y=k/E$ (Bethe and Heitler, 1934):
\begin{equation}
{d\sigma_{BH} \over dk} =
{4\alpha r_e^2 \over 3k}
\bigg(y^2 + 2[1+(1-y)^2] \bigg) Z^2 \ln{(184Z^{-1/3})} =
{1\over 3n X_0 k}
\bigg(y^2 + 2[1+(1-y)^2] \bigg)
\label{eqbh}
\end{equation}
where the radiation length 
\begin{equation}
X_0=[4n\alpha r_e^2 Z^2 \ln{(184Z^{-1/3})}]^{-1},
\end{equation}
with $n$ the number of atoms per unit volume.
More sophisticated calculations, discussed in Section IV, include
additional terms in the cross section and radiation length.  With a
more detailed screening calculation, a small constant may be
subtracted from the form factor.  However, Eq.\ (\ref{eqbh}) is a good
semi-classical benchmark.

If the interaction occurs in a dense medium, however, this treatment
may be inadequate.  The interaction is actually spread over the
distance $l_{f0}$; if the electron multiple scatters during this time,
it can affect the emission.  The multiple scattering changes the
electron path, so that
\begin{equation}
{\bf v}(t) = {\bf v}_z(t) + {\bf v}_\perp(t) = 
{\bf v}(0)(1-\theta_{MS}(t)^2/2) + 
|{\bf v}(0)|\theta_{MS}(t){\bf\Theta}
\label{voft}
\end{equation}
where $\bf\Theta$ is a unit vector perpendicular to the initial
direction of motion and $\theta_{MS}(t)$ is the electron multiple
scattering in the time interval 0 to $t$.  The particle trajectory is
then ${\bf r}(t) = \int {\bf v}(t) dt$, so that
\begin{equation}
{\bf k}\cdot {\bf r}(t)
= |{\bf v}(0)|\  |{\bf k}| (1-{\theta_\gamma^2\over 2})
\int_0^t (1- {\theta_{MS}^2 \over 2})dt + 
|{\bf k}| \theta_\gamma\cdot \int_0^t \theta_{MS}(t) dt
\label{kdotr}
\end{equation}
where $\theta_\gamma$ is the angle between the photon and the initial
electron direction.  Landau and Pomeranchuk took
$\theta_{MS}^2 = \langle \theta_{MS}^2 \rangle$.  Over a distance
$d={\bf v}_z t$, the rms multiple scattering is (Rossi, 1952)
\begin{equation}
\langle\theta_{MS}^2\rangle = ({E_s \over E })^2 {d \over X_0}
\label{eqthetams}
\end{equation}
where $X_0$ is the radiation length and $E_s= mc^2\sqrt{4\pi/\alpha} =
21.2 MeV$.  

The effect of multiple scattering can be found by inserting the
trajectory of Eq. (\ref{voft}) into Eq. (\ref{classicalb}).  This
calculation simplifies with a clever choice of coordinate system.  If
the origin is centered on the formation zone, with ${\bf v}_\perp=0$
at $z=0$, then the multiple scattering distributes evenly before and
after the `interaction'.  Then, ${\bf v}_1$ and ${\bf v}_2$ are
equally affected by multiple scattering, with
\begin{equation}
{\bf k}\cdot {\bf v}_1  = 
{\bf k}\cdot {\bf v}_2  = |{\bf k}| (1-1/2\gamma^2)
(1-\theta_\gamma^2/2)(1-\theta_{MS/2}^2/2)
\end{equation}
and
\begin{equation}
|{\bf k}\times {\bf v}_1 |= 
|{\bf k}\times {\bf v}_2 |= 
|{\bf k}|(1-1/2\gamma^2)\sqrt{\theta_\gamma^2 +
\theta_{MS/2}^2}, 
\label{thetacc}
\end{equation}
where $\theta_{MS/2}$ is the multiple scattering in half of the
formation length.  The angular addition in Eq. (\ref{thetacc}) is in
quadrature because the angles are randomly oriented in the plane
perpendicular to the electron direction.  If the multiple scattering
occurs on the same time scale as the interaction, then
Eq. (\ref{eqnb}) becomes
\begin{equation}
{dI^2 \over d\omega d\Omega} = {Z^2e^2\gamma^4 |{\bf \Delta v}|^2 \over
\pi^2 c^3} { [1+\gamma^4(\theta_\gamma^2 + \theta_{MS/2}^2)^2] \over
[1+\gamma^2(\theta_\gamma^2+\theta_{MS/2}^2)]^4}
\label{eqnbc}
\end{equation}
For $\theta_{MS/2}^2 > 1/\gamma^2 + \theta_\gamma^2$, the radiation
will be reduced and the angular distribution changed.  If the multiple
scattering in the formation zone is large enough, emission is
suppressed.  For $\theta_\gamma\ll 1/\gamma$, this happens when
\begin{equation}
\omega < {E_s^2 E^2 \over m^4c^7 X_0 }.
\end{equation}
Photons with lower energies will be suppressed. Equation (\ref{eqnbc})
also shows that the angular distributions will be affected.  Because
multiple scattering can reduce the coherence length, a more detailed
calculation is required to find the degree of suppression. Landau and
Pomeranchuk (1953b) started with the expression
\begin{equation}
{dI \over d\omega} =
{Z^2e^2\omega^2 \over 4\pi^2c^3}
\int_{-\infty}^\infty ({\bf n}\times d{\bf r}_1)
\cdot
\int_{-\infty}^\infty ({\bf n}\times d{\bf r}_2)
\int d{\bf n} 
\exp{\{i\omega[(t_1-t_2)-{\bf n}\cdot({\bf r_1}-{\bf r_2})/c]\} },
\end{equation}
which simplifies to
\begin{equation}
{dI\over d\omega} =  {Z^2e^2\omega  \over \pi c^2}
\int_{-\infty}^{+\infty} dt_1
\int_{-\infty}^{+\infty} dt_2
{\exp{[i\omega(t_1-t_2)]} \over |{\bf r}_{12}|} (J_1+J_2)
\label{eqlp1}
\end{equation}
where
\begin{equation}
J_1  = {\bf v}_1\cdot{\bf v}_2 - 
{({\bf v}_1\cdot{\bf r}_{12})({\bf v}_2\cdot{\bf r}_{12}) 
\over r_{12}^2 }\bigg(\sin{(g)} + {3g\cos(g)-sin(g)\over g^2}\bigg)
\label{eqlp2}
\end{equation}
and
\begin{equation}
J_2=(-2{\bf v}_1\cdot {\bf v}_2)
\bigg({g\cos(g)-\sin(g)\over g^2}\bigg).
\label{eqlp3}
\end{equation}
Here $g=\omega |{\bf r}_{12}|/c$ and ${\bf r_{12}}= {\bf r}(t_1)- {\bf
r}(t_2)$.  The $J_1$ term was evaluated by Landau and Pomeranchuk
(1953b) while the $J_2$ term, neglected by Landau and Pomeranchuk was
evaluated by Blankenbecler and Drell (1996), who found $J_1=J_2$.
Landau and Pomeranchuk still found the 'right' result, compensating by
using a perpendicular momentum transfer due to multiple scattering
twice the correct one.

These integrals can be evaluated by using the electron trajectory
given in Eqs. (\ref{voft}) and (\ref{kdotr}).  Landau and Pomeranchuk
gave a formula for the total (summed over time) emission.
\begin{eqnarray}
{dI \over d\omega} = {Z^2e^2 \omega \over \pi c^2} 
\int_{-\infty}^\infty dT
\int_{-\infty}^\infty {dt \over t^3} e^{i\omega t}
\bigg[ {\bigg(\int_0^t \theta_{MS}(t) dt\bigg)^2} -
{t\theta_{MS}\int_0^t \theta_{MS}(t)dt} 
\bigg]\times
\nonumber\\
\sin{\bigg[\omega
\bigg({vt\over c} - {1\over 2} \int_0^t\theta_{MS}^2(t)dt + {1 \over 2t}
\big(\int_0^t\theta_{MS} dt\big)^2 \bigg)\bigg]}
\label{longlp}
\end{eqnarray}
where the integral over $T=t_1-t_2$, will be factored out to get the
emission per unit time.  The remaining integral is evaluated by
replacing the angular quantities by their average values.  For
example,
\begin{equation}
\langle\int_0^t \theta_{MS}(t) dt\rangle = {E_s^2 ct|t| \over 2E^2X_0}
\end{equation} 
Landau and Pomeranchuk expressed their results in emission per unit
time.  In terms of the more prevalent total emission,
\begin{equation}
{dI \over d\omega} = {4 Z^2 e^2 \over 3c} \int_0^\infty dX \sin\big(X+ {E^2
E_s^2X^2 \over 3m^4c^7 \omega X_0}\big)
\end{equation}
For $\omega\ll E_s^2c/E^2 X_0$, the emission is
\begin{equation}
{dI \over d\omega} =
\sqrt{2\pi\over 3} {Z^2 e^2 m^2 c^3 \over E_s E} \sqrt{\omega X_0 \over c}
\end{equation}

In the strong suppression limit, the field intensity
$dI/d\omega\sim\sqrt{\omega}$, compared to an isolated interaction,
where $dI/d\omega$ is independent of $\omega$; the corresponding cross
sections scale as $1/\sqrt{\omega}$ and $1/\omega$ respectively. The
following subsection will describe a simple semi-classical derivation
of the same result, and also discuss some of its implications.

\subsection{Bremsstrahlung - Quantum approach}
\label{sbqcd}

Bremsstrahlung suppression due to multiple scattering can also be
found by starting with Eq. (\ref{lfzero}). Multiple scattering can
suppress bremsstrahlung if it contributes significantly to
$q_\parallel$ (Feinberg and Pomeranchuk, 1956).  Multiple scattering
affects $q_\parallel$ by reducing the electron longitudinal velocity.
For a rough calculation, the multiple scattering angle can be taken as
the average scattering angle.  As before, the multiple scattering is
divided into two regions: before and after the interaction.  The longitudinal
momentum transfer from the nucleus is
\begin{equation}
q_\parallel = \sqrt{(E\cos{\theta_{MS/2}}/c)^2-(mc)^2}
-\sqrt{[(E-k)\cos{\theta_{MS/2}}/c]^2-(mc)^2} -k/c
\label{eqlonglpm}
\end{equation}
where $\theta_{MS/2}$ is the multiple scattering in half the formation
length, $(E_s/E)\sqrt{l_f/2X_0}$. The post-interaction scattering is
based on the outgoing electron energy $E-k$. The inclusion of electron
energy loss is a significant advantage of the quantum formulation.
With some small angle approximations,
\begin{equation}
q_\parallel = {km^2c^3 \over 2E(E-k)} + {k\theta_{MS/2}^2 
\over 2c}
\end{equation}
Multiple scattering is significant if the second term is larger than
the first.  This happens if $\theta_{MS/2} > 1/\gamma$, or for
\begin{equation}
k < k_{LPM} = { E(E-k)\over E_{LPM}}
\label{eqy}
\end{equation}
where $E_{LPM}$ is a material dependent constant, given by
\begin {equation}
E_{LPM} = {m^4c^7 X_0 \over \hbar E_s^2 } = 
{m^2 c^3 X_0 \alpha \over 4\pi\hbar}
\approx 7.7 \ {\rm TeV/cm}\cdot X_0.
\label{eqelpm}
\end{equation}
Table I gives $E_{LPM}$ in a variety of materials.  This definition of
$E_{LPM}$ was used by Landau and Pomeranchuk (1953a), Galitsky and
Gurevich (1964), Klein (1993) and others, but is twice that used in
some papers (Anthony, 1995, 1997) (Baier, 1996), and $1/8$ of that in
others (Stanev, 1982).  The last choice is convenient for working with
Migdal's equations (Sec. \ref{smigdal}), but not for the
semi-classical derivation.

{\center{\epsfig{file=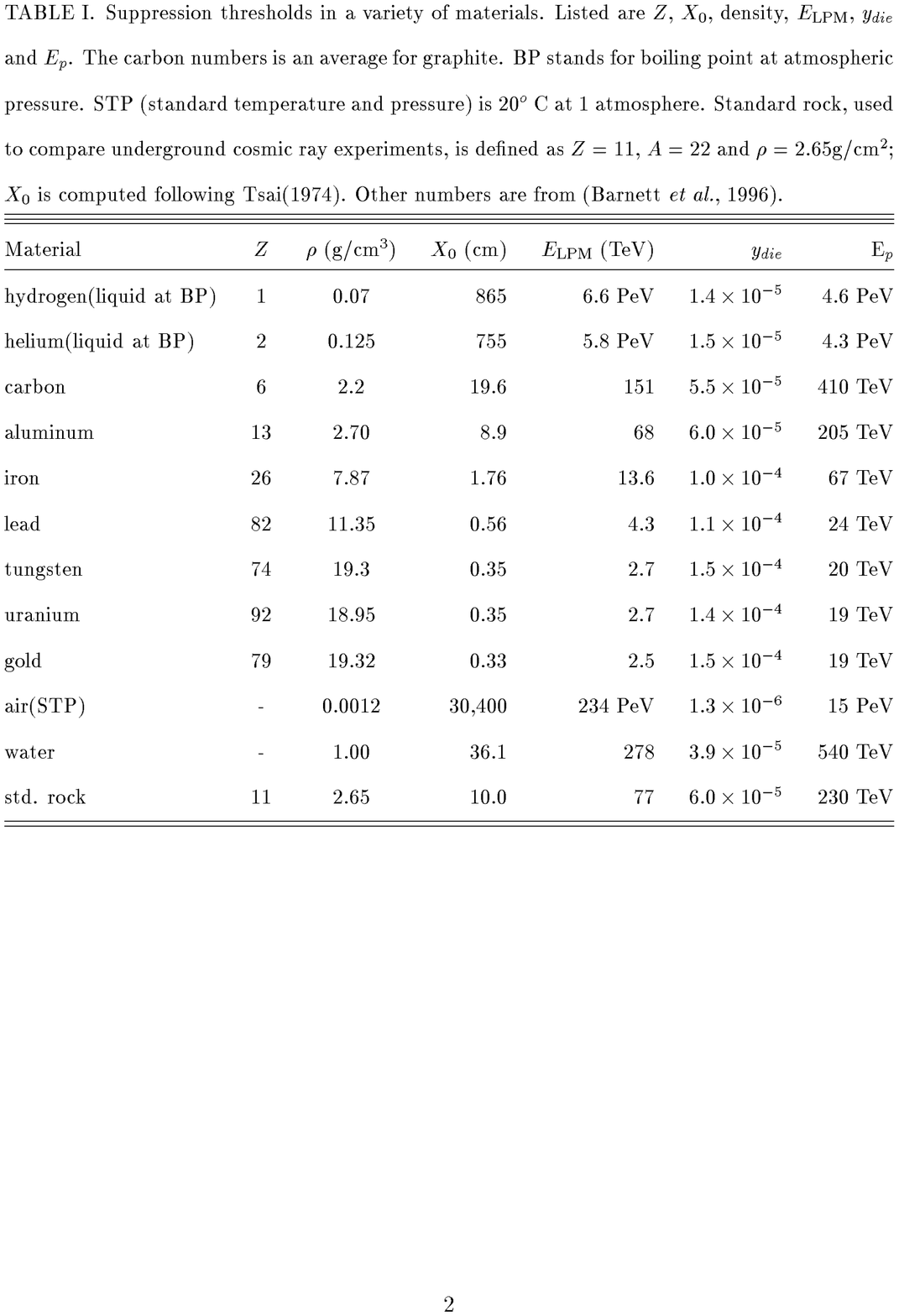,width=6.5in,%
bbllx=67,bblly=285,bburx=545,bbury=730,clip=}}}

For $k\ll E$, Eq. (\ref{eqy}) matches the classical result: photon
emission for $k<k_{LPM}=E^2/E_{LPM}$ is reduced.  For example, for 25
GeV electrons in lead, photon emission below 125 MeV is suppressed.
For higher energy electrons, a large portion of the spectrum can be
suppressed. For 4.3 TeV electrons in lead, fewer photons with $k<2.15$
TeV are radiated.  At the endpoint of the spectrum, $k\rightarrow E$,
$l_{f0}$ approaches zero, so the Bethe-Heitler cross section always
applies there.

Different calculations have found slightly different coefficients in
Eq. (\ref{eqy}).  Landau and Pomeranchuk match Eq. (\ref{eqy}), but
Feinberg \& Pomeranchuk (1956) and the introductions to Anthony (1995,
1997) used the simpler criteria $\langle\theta_{MS}\rangle >
1/\gamma$, and find a suppression twice that given here. Use of the
minimum uncertainty principle $\Delta p\Delta x> \hbar/2$ (Schiff,
1968, pg. 61) in Eq. (\ref{lfzero}) would shorten $l_{f0}$ and reduce
the suppression.

Because $\theta_{MS/2}$ both depends on $l_f$, and also partly
determines $l_f$, finding the suppression requires solving a quadratic
equation for $l_f$.
\begin{equation}
l_f = {\hbar\over q_\parallel}
=l_{f0} \bigg[1 + {E_s^2 l_f \over 2 m^2 c^4 X_0}\bigg]^{-1}.
\label{lflpm}
\end{equation}
If multiple scattering is small, this reduces to Eq.\ (\ref{lfzero}).
Where multiple scattering dominates 
\begin{equation}
l_f =l_{f0}\sqrt{kE_{LPM}\over E(E-k)}.
\label{lflpm2}
\end{equation}

The bremsstrahlung cross section scales linearly with the distance
over which coherence is maintained, or the formation length.  It is
convenient to define a suppression factor $S$, giving the emission
relative to Bethe Heitler:
\begin{equation}
S= {d\sigma/dk \over d\sigma_{BH}/dk} = {l_f
\over l_{f0}} =\sqrt{kE_{LPM}\over E(E-k)}
\label{eqslpm}
\end{equation}
and the $dN/dk\sim 1/k$ found by Bethe and Heitler (1934) changes
to $dN/dk\sim 1/\sqrt{k}$.  

\begin{figure}
\epsfig{file=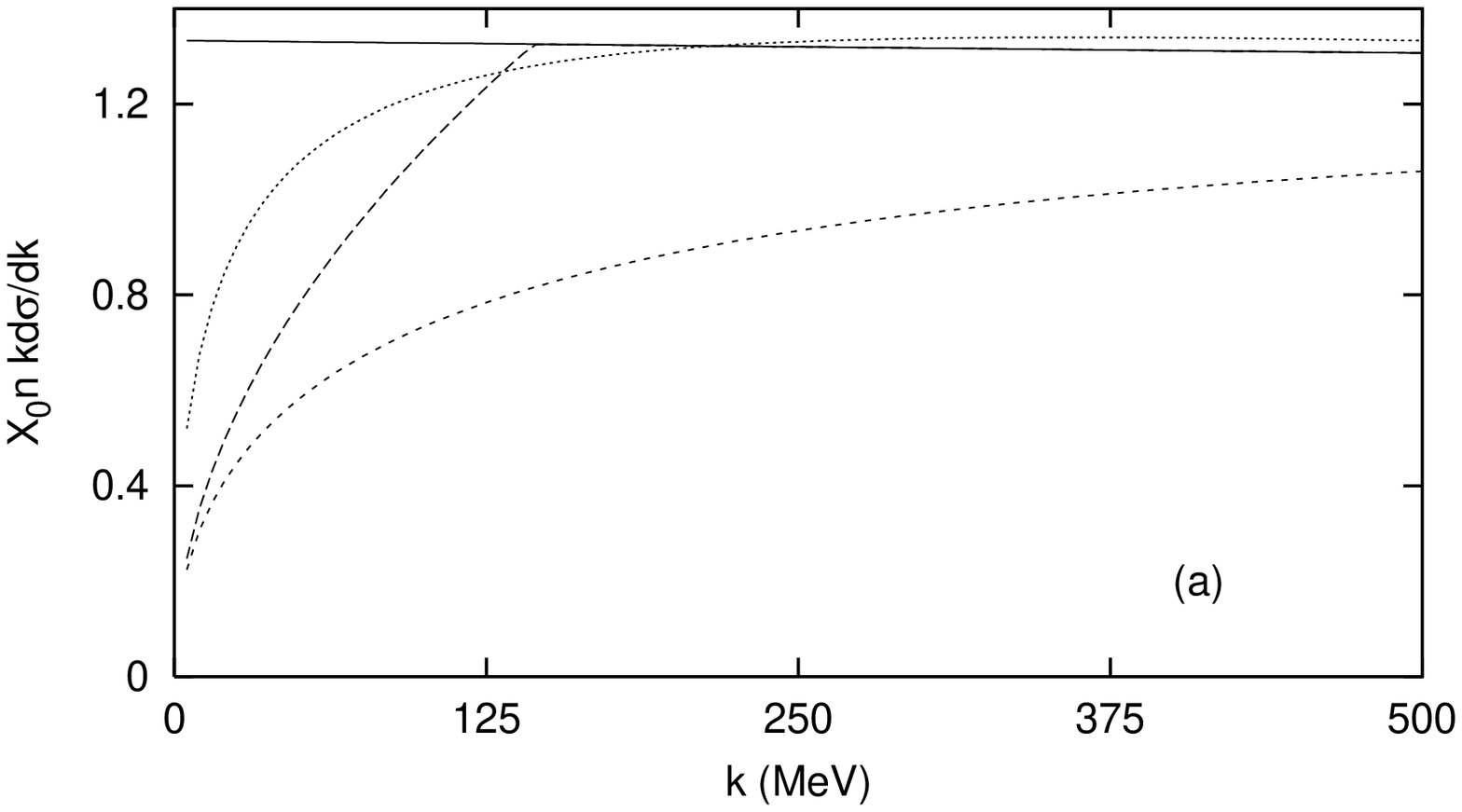,height=5in,width=2.8in,%
clip=,angle=270}
\epsfig{file=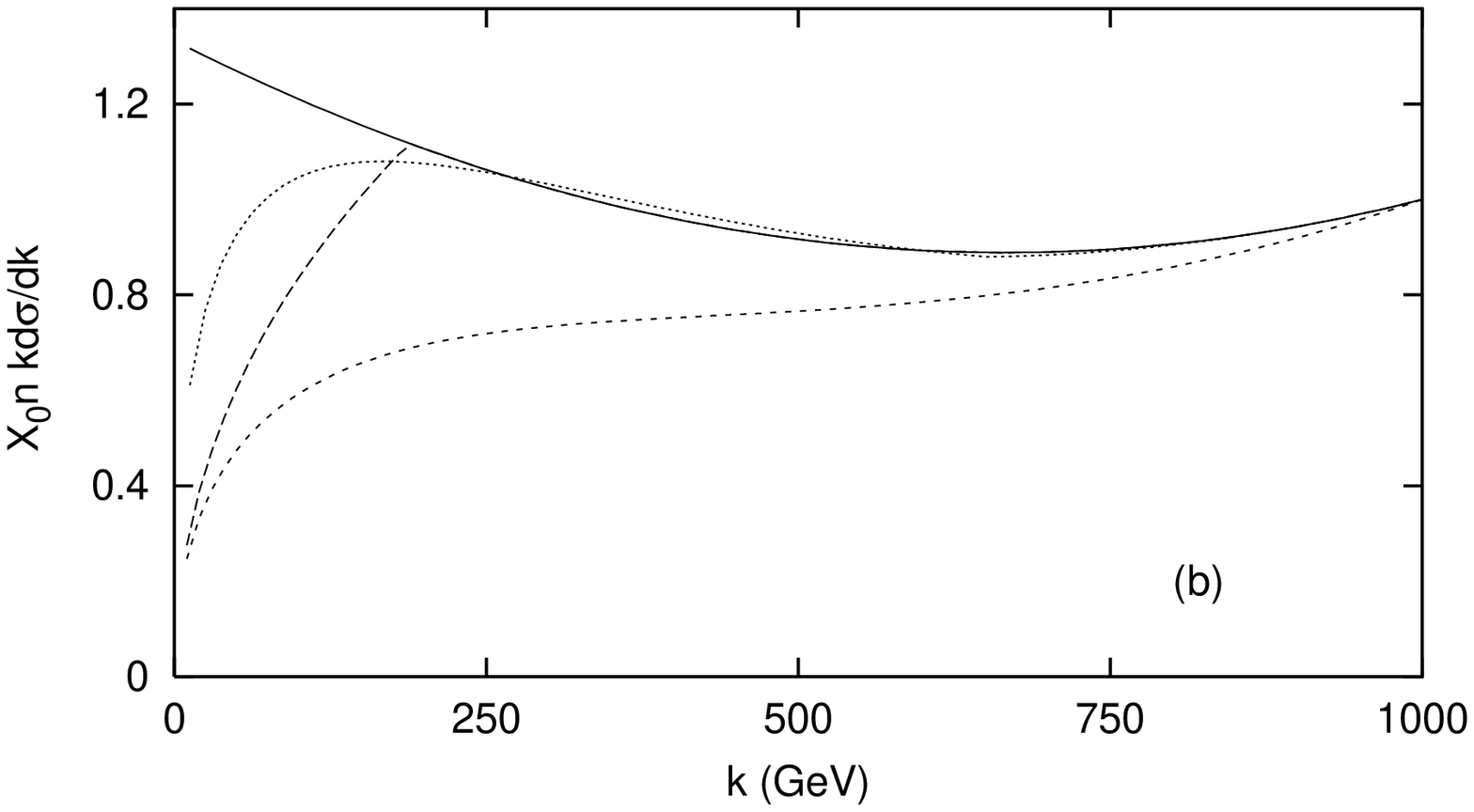,height=5in,width=2.8in,%
clip=,angle=270}
\vskip .1 in
\caption[]{Comparison of the differential energy weighted cross
sections per radiation length, $X_0n kd\sigma/dk$ for Bethe-Heitler
radiation (solid line), Migdal's detailed suppression calculation
(dotted line), quadratic suppression, Eq. (\ref{lflpm}) (short
dashes), and the strong suppression limit, Eq. (\ref{eqslpm}) (long
dashes), for (a) 25 GeV electrons and (b) 1 TeV electrons incident on
a lead target.}
\label{bremcompare}
\end{figure}

Figure \ref{bremcompare} compares bremsstrahlung cross sections for 25
GeV and 1 TeV electrons in lead.  The Bethe-Heitler cross section is
compared with three approaches to suppression: the full semi-classical
suppression Eq.~(\ref{lflpm}), the prevalent low energy limit
Eq.~(\ref{eqslpm}), and Migdal's calculation, included as a standard.
Equation~(\ref{lflpm}) predicts considerably more suppression than
Migdal.  Numerically, Eq. (\ref{eqslpm}) is closer to Migdal's
results, but the required approximation is unjustified in the
transition region $k\sim E(E-k)/E_{LPM}$.  Better agreement would be
found with a larger $E_{LPM}$, as would be given by the minimum
uncertainty principle.  With only $S$ considered, it is impossible to
separately determine $E_{LPM}$ and the overall cross section
normalization.  The onset of suppression must be seen to separate
these two factors.

\begin{figure}
\epsfig{file=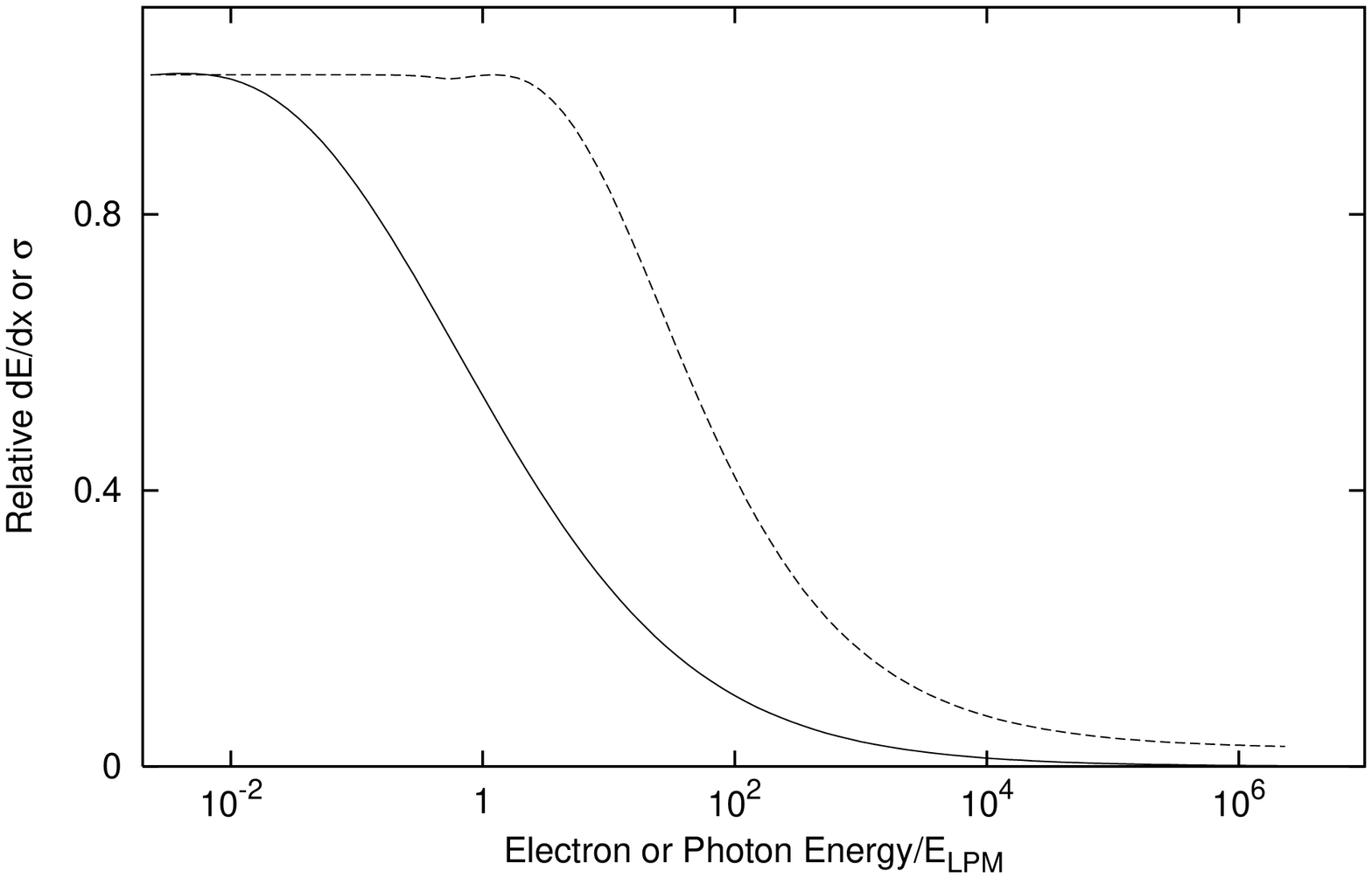,height=6in,width=3.5in,%
clip=,angle=270}
\vskip .1 in
\caption[]{The relative electron energy loss
$dE/dx$(Migdal)/$dE/dx$(Bethe-Heitler) from bremsstrahlung (solid
line) and the relative photon conversion cross sections
$\sigma$(Migdal)$/\sigma$(Bethe-Heitler).  Both curves are for lead,
with electron/photon energy in units of $E_{LPM}=4.3$ TeV.}
\label{suppressionvse}
\end{figure}

Nevertheless, Eq. (\ref{eqslpm}) demonstrates some implications of
bremsstrahlung suppression (Landau and Pomeranchuk, 1953b).  With
suppression, the number of photons emitted per radiation length is
finite, scaling as $\sqrt{E/E_{LPM}}$ for $E>E_{LPM}$.  Electron
$dE/dx$ due to bremsstrahlung is also reduced; instead of rising
linearly with $E$, it is proportional to $\sqrt{EE_{LPM}}$.  Figure
\ref{suppressionvse} shows the relative bremsstrahlung energy loss,
$(dE/dx)_{LPM}$ for Migdal compared with the Bethe-Heitler prediction.

Besides the photon spectrum, LPM suppression also affects the photon
angular distribution.  With the photon emission angle $\theta_\gamma$
included,
\begin{equation}
q_\parallel = p_e - p_e' - k/c \cos{\theta_\gamma}
=\ \sqrt{(E/c)^2\ -\ (mc)^2}
-\sqrt{[(E-k)/c]^2\ -\ (mc)^2} - k/c \cos{\theta_\gamma}.
\label{eqlongangle}
\end{equation}
The change in electron direction is assumed to be negligible.  Then,
\begin{equation}
q_\parallel\sim {m^2 c^3 k \over 2E (E - k) } + 
{k \theta_\gamma^2 \over 2c} 
\label{eqsangle}
\end{equation}
For $k\ll E$, $l_f=l_{f0}/ (1 + \gamma^2\theta_\gamma^2)$.  This
formula may be used to derive the angular distribution of
bremsstrahlung photons (Landau and Pomeranchuk, 1953b).  When
$\theta_\gamma$ increases $q_\parallel$, the multiple scattering term
becomes less important, so there is less suppression for
$\gamma\theta_\gamma>1$.  With both multiple scattering and a finite
$\theta_\gamma$,
\begin{equation}
l_f = { 2\hbar E (E-k) \over km^2c^3} \bigg[1+ {E_s^2l_f\over
2m^2c^4X_0} + {\theta_\gamma^2E(E-k)\over m^2c^4} \bigg]^{-1}
\label{lfwithms}
\end{equation}
Suppression is large when the multiple scattering term is larger than
the other terms.  Then
\begin{equation}
S(\theta)= \sqrt{k E_{LPM} (1+\gamma^2\theta_\gamma^2) \over E(E-k)}.
\label{eslpmangle}
\end{equation}
Suppression disappears rapidly as $\theta_\gamma$ rises.  

When $S(0)\ll 1$, the angular distribution is broadened.  It
follows from Eq. (\ref{eslpmangle}) that multiple scattering broadens
the angular distribution from $\langle\theta_\gamma\rangle \sim
1/\gamma$ to $\langle\theta_\gamma\rangle \sim 1/\gamma\sqrt{S}$
(Landau and Pomeranchuk, 1953b).  Starting from a different tack,
Galitsky and Gurevich (1964) found that $\langle\theta_\gamma\rangle$
is determined by the electron multiple scattering over the distance
$l_f$, giving the same algebraic result.

This increase in $\langle\theta_\gamma\rangle$ is difficult to
observe.  The angular distribution of bremsstrahlung photons in a
thick target is dominated by changes in electron direction due to
multiple scattering before the bremsstrahlung occurs.  However, for
sufficiently thin targets, multiple scattering will be small, so that,
if $S$ is small enough, then the photon emission angles dominate over
multiple scattering.  For thin targets, the photon spectrum measured
at angles $\theta_\gamma\gg 1/\gamma$ should exhibit less suppression
than at smaller angles.

\subsection{Photon Interactions with the Medium}

Ter-Mikaelian (1953a,b) pointed out that photon interactions can also
induce suppression.  Photons can interact with the medium by coherent
forward Compton scattering off the target electrons, producing a phase
shift in the photon wave function.  If this phase shift is large
enough, it can cause destructive interference, reducing the emission
amplitude.  Ter-Mikaelian used classical electromagnetism in his
analysis, calculating suppression in terms of the dielectric constant
of the medium,
\begin{equation}
\epsilon(k) = 1 - (\hbar\omega_p)^2/k^2
\label{eepsilson}
\end{equation}
where $\omega_p$, the plasma frequency of the medium, is $\sqrt{4\pi
nZe^2/m}$.  This is equivalent to giving the photon an effective mass
$\hbar\omega_p/c^2$.  The relationship between $k$ and $p_\gamma$
becomes $p_\gamma c=\sqrt\epsilon k$, so
\begin{equation}
q_\parallel = p_e - p_e' - k\sqrt\epsilon/c = 
{k \over 2c\gamma^2} +{(\hbar\omega_p)^2\over 2ck}.
\label{eqlongdiel}
\end{equation}
The formation length is then
\begin{equation}
l_f = { 2\hbar ck\gamma^2 \over k^2 + k_p^2 }
\label{lfdiel}
\end{equation}
where $k_p=\gamma\hbar\omega_p$.  When dielectric suppression is
strong, $q_\parallel$ is dominated by the photon interaction term and
$l_f$ becomes independent of $E$: $l_f=2ck/\hbar\omega_p^2$.  As with
LPM suppression, the cross section is proportional to the path length
that can contribute coherently to the emission, so $S$ is the ratio of
the in-material to vacuum formation lengths:
\begin{equation}
S =  { k^2 \over k^2 + k_p^2 }.
\label{sdiel}
\end{equation}
For $k< k_p$, bremsstrahlung is significantly reduced.  This happens
for $y=k/E<y_{die}$, where
\begin{equation}
y_{die}=\hbar\omega_p/mc^2
\end{equation}
is a material dependent constant.  For lead, $\hbar\omega_p= 60$ eV,
so $y_{die}\sim10^{-4}$.  Table I lists $y_{die}$ for a variety of
materials.

The same result can be obtained classically by including the
dielectric constant in Eq.~(\ref{cphase}), so
$\Phi=\exp{\{ikt[1-|{\bf v}|\sqrt{\epsilon}
\cos(\theta_\gamma)/c]/\hbar\}}$.  The case
$\cos{(\theta_\gamma)}=c/|{\bf v}|\sqrt{\epsilon}$, which gives an
infinite formation length, corresponds to \v{C}erenkov radiation
(Ter-Mikaelian, 1972, pg. 196).

Because photon emission angles are determined by the kinematics, a
finite $\theta_\gamma$ affects dielectric suppression the same way as
it does LPM suppression.  Including $\theta_\gamma$,
\begin{equation}
l_f = { 2\hbar ck\gamma^2 \over k^2(1+\gamma^2\theta_\gamma^2) + k_p^2 }.
\label{elfdielangle}
\end{equation}
For large suppression and $\gamma\theta_\gamma>1$,
$S=(k\theta_\gamma/\hbar\omega_p)^2$ is independent of $E$.  In this
limit, the angular spread is $<\theta_\gamma>\sim \hbar\omega_p/k$
(Galitsky and Gurevich, 1964).

Ter Mikaelian (1972, pg. 127) pointed out that the dielectric constant
also affects $q_{min}$ in the form factor logarithm.  The complete
cross section for $k\ll E$ is then
\begin{equation}
{d\sigma \over dk} =
{16Z^2\alpha r_e^2 k\over 3k_p^2}
\ln{\bigg(184Z^{-1/3} \sqrt{1+({ k_p\over k})^2}\ \bigg)}.
\label{sigmatern}
\end{equation}

Because $k_p^2\sim n$, except for the logarithmic term, photon
emission is independent of the density!  As the density rises,
increasing the number of scatters, suppression rises in tandem,
leaving the total photon production constant.

This suppression is sometimes known as the longitudinal density
effect, by analogy with the transverse density effect (Jackson, 1972,
pg. 632) which reduces ionization $dE/dx$.  It is also known as
dielectric suppression. Unfortunately, a quantum mechanical
calculation of dielectric suppression has yet to appear, nor has
dielectric suppression been described in terms of Compton scattering.

Because dielectric suppression and the LPM effect both reduce the
formation length, the effects do not merely add; the total
$q_\parallel$ must be calculated, and from that $l_f$ and the
suppression can be found.  Feinberg and Pomeranchuk (1956) showed that
when $k_p>k_{LPM}$ (i.e. $E < y_{die}E_{LPM}$), then dielectric
suppression overwhelms LPM suppression, and only the former is
observable.  For higher electron energies, LPM suppression is visible
for
\begin{equation}
k>k_{cr} = (k_p^4/k_{LPM})^{1/3}.
\end{equation}

\subsection{Bremsstrahlung Suppression due to Pair Creation}

Landau and Pomeranchuk (1953a) pointed out that, at the highest
energies, $l_f$ can approach a radiation length.  Then, the partially
created photon can pair create part way through the formation zone.
This destroys the coherence between different parts of the formation
zone, reducing the amplitude for photon emission.  Unfortunately,
there has been little attention to this problem, and the available
results are quite crude.

\begin{figure}
\center{\epsfig{file=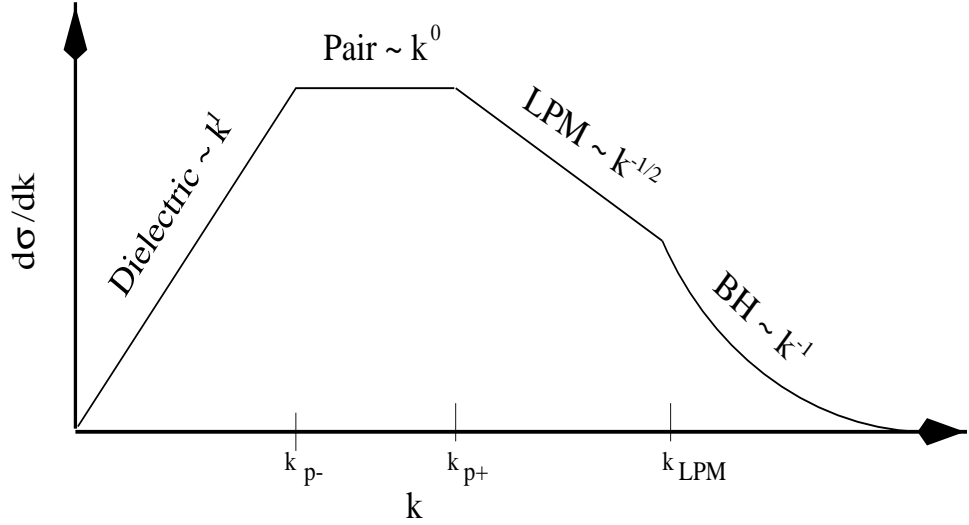,height=3in,width=5in,%
bbllx=0,bblly=0,bburx=520,bbury=270,%
clip=,angle=0}}
\vskip .1 in
\caption[]{Schematic view of bremsstrahlung $d\sigma/dk$ when several
suppression mechanisms are present.  For $E< E_p$, the pair creation
suppression disappears and LPM suppression connects with dielectric
suppression.}
\label{bremspectrum}
\end{figure}

For $k\gg mc^2$, the pair creation cross section is independent of
energy: $\sigma_\gamma= (28/9) \alpha r_e^2
Z^2\ln{(184Z^{-1/3})}$. This constant cross section limits the
formation length to roughly $X_0$.  Neglecting other suppression
mechanisms, $l_{f0}> X_0$ when $k < 2 \hbar E(E-k)/ X_0 m^2c^3$.
However, dielectric suppression and LPM suppression limit the range of
applicability.  With both these mechanisms considered, pair creation
further reduces photon emission when (Galitsky and Gurevitch, 1964):
\begin{equation}
{2E^2 \over E_s^2} \bigg({2 \hbar c\over X_0 k} - {\hbar^2\omega_p^2
\over k^2} \bigg) > 1.
\label{eforpair}
\end{equation}
The coefficients given here differ slightly from the original because
Galitsky and Gurevitch used a slightly different approach from that
presented here.  With other factors considered, this mechanism
is visible for
\begin{equation}
E > E_p = {X_0 \omega_p E_s\over \sqrt{2}c}
\end{equation}
is a material dependent constant. For $E \sim E_p$ this mechanism
dominates in a narrow window around $k= \hbar\omega_p^2 X_0/c$; for
$E\gg E_p$, the range is $k_{p-}= \hbar\omega_p^2 X_0 / 2c < k <
k_{p+}=4 \hbar cE(E-k)/(X_0E_s^2)$, as is shown in
Fig. \ref{bremspectrum}.  In this region, the suppression factor is
\begin{equation}
S= {X_0 \over l_{f0}} = {m^2c^3k X_0 \over 2\hbar E(E-k)} 
\label{sforpair}
\end{equation}
and $d\sigma/dk$ is independent of $k$. For $k<k_{p-}$, dielectric
suppression dominates, while for $k>k_{p+}$, LPM suppression is
dominant.  $E_p$ ranges from 25~TeV for lead to 15~PeV for sea level
air; other values are given in Table I.  For lead, when $E\gg E_p$, the
photon 'window' is $1.2\times 10^{7}$ eV$< k < 1.6\times10^{-19}
E^2$(eV), while for air it is $8.3\times10^7$ eV $<k<
3.0\times10^{-24} E^2$(eV).

Similarly, bremsstrahlung of a sufficiently high energy photon
can suppress pair production.  The bremsstrahlung can affect the
overall pair production rate if the emitted photon contributes
significantly to $q_\parallel$ of the entire reaction.

These formulae are only rough approximations.  At high enough
energies, the pair creation cross section is itself significantly
reduced because of LPM suppression and $X_0$ in Eqs.~(\ref{eforpair})-
(\ref{sforpair}) should be increased to account for this.  As
Fig.~\ref{suppressionvse} shows, for $k\gg E_{LPM}$, the pair
conversion length rises significantly.  Then, bremsstrahlung and pair
creation suppress each other and the separation of showers into
independent bremsstrahlung and pair creation interactions becomes
problematic.  For this, a new, unified approach is needed.

\subsection{Surface effects and Transition Radiation}

The discussion far has only considered infinitely thick targets.  With
finite thickness targets, the effects of the entry and exit surfaces
must be considered.  At first sight, this appears straightforward: the
only effect being a reduction in the multiple scattering when the
formation zone sticks out of the target, and hence there is less
suppression.  However, in addition to reduced suppression, multiple
scattering produces a new kind of transition radiation.

Conventional transition radiation occurs when an electron enters a
target, and the electromagnetic fields of the electron redistribute
themselves to account for the dielectric of the medium.  In the course
of this rearrangement, part of the EM field may break away, becoming a
real photon. Because transition radiation has been extensively
reviewed elsewhere (Artru, Yodh and Mennessier, 1975), (Cherry, 1978),
(Jackson, 1975), it will not be further discussed.  However, the
formula for radiation, neglecting interference between nearby edges,
is given here for future use.  The emission is  (Jackson, 1975, pg. 691):
\begin{equation}
{dN \over dk} = {\alpha \over \pi k}\bigg[\bigg(1+ {2k^2\over
k_p^2}\bigg)\ln{(1+{k_p^2\over k^2})} - 2\bigg]
\label{trconv}
\end{equation} 
photons per edge.  

\begin{figure}
\center{\epsfig{file=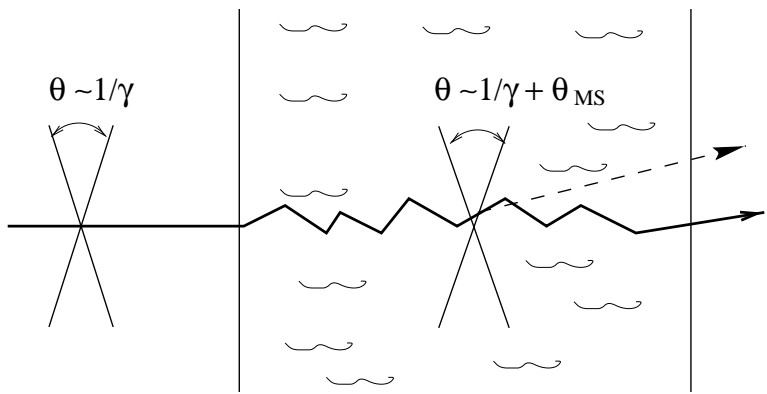,height=2in,%
bbllx=0,bblly=110,bburx=500,bbury=360,%
clip=,angle=0}}
\vskip .1 in
\caption[]{Diagram of transition radiation caused by multiple
scattering.  The pancaked electromagnetic fields are broadened by
multiple scattering from their free space width $1/\gamma$ to
$\sqrt{1/\gamma^2 + \theta_{MS/2}^2}$, where $\theta_{MS}$ depends on
$l_f$, and hence on $k$.}
\label{transitdiagram}
\end{figure}

The additional transition radiation occurs because multiple scattering
changes the trajectory of the electron.  The variation in electron
direction widens the directional distribution of the electromagnetic
fields carried by the electron, as is shown in
Fig. \ref{transitdiagram}.  Scattering broadens the EM fields from
their free space width $1/\gamma$ to a $l_f$ (and hence $k$) dependent
value.  As the EM field enters the target and realigns itself, it can
emit transition radiation.

Classically, this new form of transition radiation is closely related
to the old. Both depend on the difference in $l_f$ in the two
materials (Ter-Mikaelian 1972, pg. 233), with the complete radiation
\begin{equation}
{dN\over dkd\theta_\gamma} =
{2\alpha k \theta_\gamma^3 \over \pi\hbar^2 c^2 } (l_f-l_f')^2
\label{lftt}
\end{equation}
where $l_f$ and $l_f'$ are the formation lengths in the two media.  If
$l_f=l_f'$, then there is no transition radiation.  Conventional
transition radiation can be derived from this formula by focusing on
the dielectric constant of the media, while multiple scattering
transition radiation can be calculated by focusing on $\theta_{MS}$.
Of course, the complete spectrum includes both contributions.

The energy spectrum is given by integrating over $\theta_\gamma$; the
integral is complicated because the maximum $\theta_\gamma$ depends on
$k$. For $E>E_{LPM}$ and $k>k_p$, (Ter-Mikaelian, 1972, pg. 235)
\begin{equation}
{dN \over dk} \sim {\alpha \over\pi k}
\bigg( \ln{\big({1 + \sqrt{1+4k_{LPM}/k} \over 2}\big)}
+ {2\over 1+\sqrt{1+4k_{LPM}/k}} - 1\bigg)
\label{transms}
\end{equation}
per edge.  The total radiation may be found by integrating $dN/dk$ up
to $k_{LPM}$.  Since $k_{LPM}\sim E^2$, the total energy lost by the
electron $\Delta E$ rises as $E^2$, in contrast to conventional
transition radiation, where the loss is proportional to $E$.  When
$\Delta E$ becomes a significant fraction of $E$, quantum effects must
become important.

These calculations assume that the transition is instantaneous,
neglecting the sharpness of the surface.  They also neglect coherence
between nearby edges. More quantitative estimates are discussed in
Sec. IV.C.  For pair creation, one expects similar surface effects;
unfortunately these have yet to be worked out.

\subsection{Thin Targets}
\label{sthint}

In extremely thin targets, neither dielectric effects nor multiple
scattering produce enough of a phase shift to cause suppression.
Suppression due to multiple scattering disappears when the total
scattering angle in the target is less than $1/\gamma$.  This occurs for
targets with thickness $T<(mc^2/E_s)^2 X_0 \approx X_0/1720$.  With
dielectric suppression, $l_f\rightarrow 0$ as $k\rightarrow 0$, so
some dielectric suppression remains for any $T$, for photon energies
$k < T\hbar \omega_p^2 /2c$.  Of course, for extremely thin targets
transition radiation will dominate over bremsstrahlung.

For multiple scattering, intermediate thickness targets with $X_0/1720
< T < l_{f0}$ are of intererest because the entire target acts as a
single radiator.  The emission can be found from the probability
distribution for the total scattering angle in the entire target,
either classically (Shul'ga and Fomin, 1978) or with quantum
mechanical calculations (Ternovskii, 1960).  These calculations are
more complex than those presented earlier, because they use a
distribution of scattering angles rather than a single average
scattering angle.  The total radiation from the target is (Ternovskii,
1960)
\begin{equation}
{dN_{T}\over dk} = {2\alpha\over\pi k} \int_0^\infty
d^2\theta f(\theta) \bigg( {2\zeta^2+1 \over \zeta\sqrt{\zeta^2+1}}
\ln{(\zeta+\sqrt{\zeta^2+1})}-1\bigg)
\label{esigmashulga}
\end{equation}
where $\zeta=\gamma\theta/2$, and $\theta$ is the scattering angle.
Since $\theta$ is independent of $k$, this formula has the same $k$
dependence for $k\ll E$ as the Bethe-Heitler calculation.  The
scattering can be treated as Gaussian distribution, with rms
scattering angle $\theta_0$, where, for thin targets (Barnett \etal,
1996)
\begin{equation}
\theta_0= {E_s \over E} \sqrt{T\over X_0}\ \bigg(1+0.038 \ln{T\over
X_0}\bigg).
\label{ethetamsthin}
\end{equation}
For the relevant range of thicknesses, neglecting the logarithmic term
changes the suppression factor by at most a few percent.

\begin{figure}
\epsfig{file=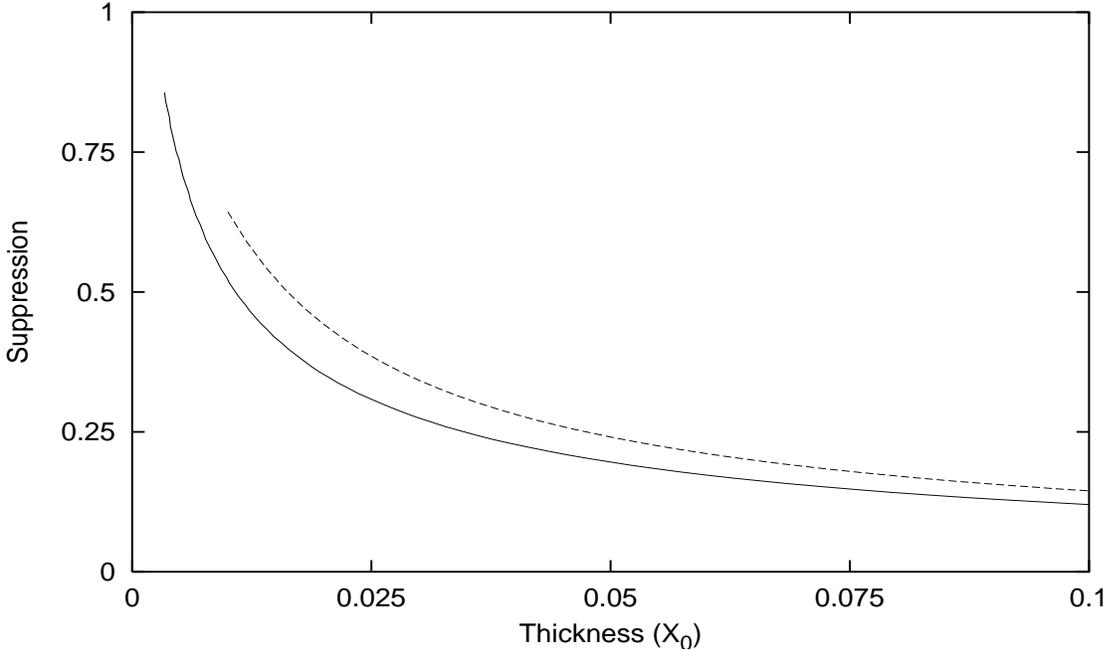,height=6in,width=3.5in,%
clip=,angle=270}
\vskip .1 in
\caption[]{Suppression factor $N_{T}/T\sigma_{BH}$ for thin targets.
The solid line follows Eq.\ (\ref{esigmashulga}) for E=25 GeV.  The
factor is only slightly energy dependent, as long as $T>X_0/1720$ and
$T<l_f$.  The dashed curve is the logarithmic approximation, Eq.\
(\ref{esigmashulgalimit}).}
\label{figshulga}
\end{figure}

Eq.\ (\ref{esigmashulga}) can be evaluated numerically.  For very thin
targets, it matches the Bethe-Heitler spectrum, except for a factor of
4/3.  For thicker targets, the suppression factor,
$S=N_{T}/T\sigma_{BH}$ is shown in Fig.\ \ref{figshulga}.

In the limit $T \gg (mc^2/E_s)^2 X_0$, (but with $T<l_{f0}$), $dN/dk\sim
\ln T$ (Ternovskii, 1960).  The intensity varies logarithmically with
the target thickness!  A slightly more detailed calculation finds
(Shul'ga and Fomin, 1996)
\begin{equation}
{dN\over dk} = {2\alpha \over \pi k }\bigg(\ln{E_s^2T \over m^2c^4
X_0}-1\bigg).
\label{esigmashulgalimit}
\end{equation}
This approximation, shown by the dashed line in Fig.\ \ref{figshulga},
overestimates Eq.\ (\ref{esigmashulga}) by 10-20\%. 

Shul'ga and Fomin (1998) found that the use of a screened Coulomb
potential, rather than a Gaussian scattering distribution, leads to an
equation like Eq. (\ref{esigmashulgalimit}), but with additional
terms.  This calculation shows the effects of scattering on an
electron by electron basis.  The more an electron scatters in a
target, the higher the average radiation.

A similar expression should apply for pair creation - the formation
length is the same.

\subsection{Suppression of Pair Creation}

\begin{figure}
\epsfig{file=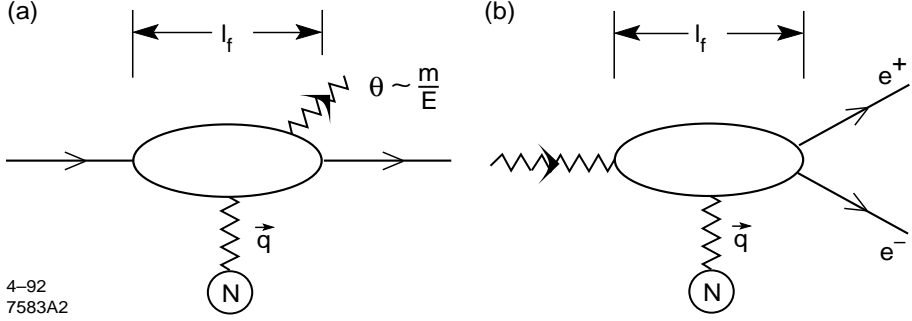,height=1.7 in,width=5in,%
clip=,angle=0}
\vskip .1 in
\caption[]{Schematic representations of $l_f$ for bremsstrahlung and
pair conversion, showing their relationship.}
\label{schematic}
\end{figure}

Multiple scattering can also reduce the cross section for
$\gamma\rightarrow e^+e^-$.  The relationship between pair creation
and bremsstrahlung, Fig. \ref{schematic}, is clear, and the two
Feynman diagrams easily map into each other.  The crossing does change
the kinematics of the process.  Since it is the electron and positron
that multiple scatter, and they must have energies lower than that of
the initial photon, suppression occurs only at higher incident
particle energies.  To avoid confusion, $k$ will continue to refer to
photon energy, with the produced pair having energies $E$ and $k-E$.

Pair production is not possible classically.  However, Landau and
Pomeranchuk (1953b) give some simple arguments why pair creation
should be sensitive to its environment.  The momentum transfer for
pair production is (Feinberg and Pomeranchuk, 1964)
\begin{equation}
q_\parallel = k/c - \sqrt{(k-E)^2/c^2-m^2c^2}  - \sqrt{E^2/c^2-m^2c^2}
= {m^2c^3k  \over 2 E(k-E)}.
\label{eqlongpair}
\end{equation}
Because $q_\parallel$ is unchanged when $E$ and $k-E$ are
interchanged, either can represent the electron or positron. The
formation length can be expressed in terms of the two final state
momenta or in terms of the invariant mass of the created pair.  Not
surprisingly, $l_{f0}$ for pair production is similar to the
bremsstrahlung case:
\begin{equation}
l_{f0} = {2 \hbar E  (k-E) \over m^2 c^3 k}.
\label{elfpair}
\end{equation}
It might seem surprising that $k$ is in the denominator of
Eq. (\ref{elfpair}).  But, $l_{f0}$ becomes a maximum for $E=k-E=k/2$;
then $l_{f0}= \hbar k/2m^2c^3$, and $l_{f0}$ rises with $k$. If $E\ll
k$, then this equation reduces to $l_{f0}=2\hbar E/m^2c^3$ and
$l_{f0}$ is very short.  This asymmetric energy division corresponds
to a pair with a large invariant mass.  In terms of pair mass, $M_p$,
\begin{equation}
l_{f0} = { 2\hbar k \over M_p^2c^3}.
\label{elfpairmass}
\end{equation}
One difference between pair creation and bremsstrahlung is that the
multiple scattering now applies to two separate particles.  The lower
energy particle scatters more, and so dominates the additional
$q_\parallel$.  The scattering is taken over $l_f/2$, as if the
charged particles are produced in the middle of the formation zone,
and the result is very similar to Eq.\ (\ref{lflpm}):
\begin{equation}
l_f= l_{f0} \bigg[ 1 + {E_s^2 l_f \over 2m^2c^4 X_0}\bigg]^{-1}
\label{lflpmpair}
\end{equation}
where $l_{f0}$ now refers to Eq.\ (\ref{elfpair}).  When the second
term is dominant,
\begin{equation}
l_f =  l_{f0} \sqrt{k E_{LPM} \over E(k-E)}
={2\hbar k \over M_pmc^3} \sqrt{E_{LPM}\over k},
\label{lflpmpair2}
\end{equation}
similar to Eq.\ (\ref{lflpm}). Then,
\begin{equation}
S=\sqrt{E_{LPM}k\over E(k-E)} = {M_p\over m} \sqrt{E_{LPM}\over k}.
\label{espair}
\end{equation}
For a given $k$, the suppression is largest when $E\approx
(k-E)\approx k/2$.  There is no suppression for $E\approx 0$ or
$E\approx k$.  For $E\gg E_{LPM}$, the total cross section scales as
$\sqrt{E_{LPM}/k}$. Figures \ref{suppressionvse} and \ref{comparepair}
show how $S$ drops as $k$ rises.

As with bremsstrahlung, the emission angles can affect suppression.
The relevant angular variables are $\theta_{e^+}$ and $\theta_{e^-}$,
the angles between the outgoing particles trajectories and the
incoming photon path.  If either angle is larger than $k/mc^2$, then
the formation length is shortened and suppression reduced.

Because of the high photon energy, there is no apparent analogy to
dielectric suppression for pair creation.

\subsection{Muons and Direct Pair Production}

Electromagnetic processes involving muons can also be suppressed.
However, because the muon mass $m_\mu\gg m_e$, the effects are much
smaller.  For a fixed energy, the formation length is reduced by
$(m_e/m_\mu)^2\sim 1/40,000$.  For muons, Eqs. (\ref{eqy}) and
(\ref{eqslpm}) hold, but with $E_{LPM}$ replaced by
\begin{equation}
E_{LPM(\mu)}= { m_\mu^4 c^7 X_0 \over \hbar E_s^2 } \approx
1.38\times10^{22}{\rm eV/cm}\cdot X_0.
\label{eelpmmu}
\end{equation}
This energy is high enough that LPM suppression is generally
negligible for muon bremsstrahlung and pair creation.

For muons, dielectric suppression still occurs for
$y<\hbar\omega_p/m_\mu c^2$, about $10^{-7}$ in solids.  This is very
small, but perhaps not unmeasurable.

Unlike electrons, high energy muons have a significant cross section
for direct pair production, $\mu^-N\rightarrow \mu^-e^+e^-N$.  This
process is similar to bremsstrahlung followed by pair creation, except
that the intermediate photon is virtual.  Both the $\mu$ and the final
state electrons are subject to multiple scattering.  The formation
length can be calculated by treating the pair as a massive photon,
starting from
\begin{equation}
q_{||} = \sqrt{E^2-m_\mu^2c^4} - \sqrt{(E-k)^2-m_\mu^2c^4} -
\sqrt{k^2-M_p^2c^4}  = { m_\mu^2c^3k\over 2 E(E-k) } + { M_p^2c^3
\over 2 k}
\label{elfdirect}
\end{equation}
where here $E$ is the muon energy, $k$ is the virtual photon energy
and $M_p$ is the pair mass.  Compared to bremsstrahlung, $l_f$ is only
decreased for $k/ E < M_p/ m_\mu $.  The final state has three
particles that can multiple scatter.  Since the incoming muon is very
energetic, it exhibits little multiple scattering.  The electron and
positron are less energetic, and multiple scatter more; the
contribution to $q_{||}$ due to (both) their multiple scattering is
$E_s^2 M_p^2 l_f/ 4m^2 k X_0$.  This is significant (the cross section
is reduced) for $l_f> (E_s/mc^4)^2 X_0/2$ when $k/E<M_p/m_\mu$.
Suppression is easiest for symmetric pairs because they lead to the
longest $l_f$, but, even then, energies above $10^{17}$ eV are
required to observe suppression.

Although it is much less probable, electrons can also lose energy by
direct pair production, $e^-N\rightarrow e^+e^-e^-N$.  For electrons,
$M_p>m$, so the second term in Eq.~(\ref{elfdirect}) always dominates,
and $l_f = 2\hbar k/M_p^2c^3$.  As with muons, suppression occurs for
$l_f > (E_s/mc^2)^2 X_0/2$, albeit without the restriction on $k/E$.
Still, suppression requires energies almost as high as the muon case.

The lack of suppression for direct pair production is an initial
demonstration that higher order diagrams typically involve larger
$q_\parallel$ than simpler reactions.  Therefore, higher order
processes are less sensitive to their environment, and, when
suppression is large, higher order diagrams become more important.

\subsection{Magnetic Suppression}
\label{smagsup}

External magnetic fields can also affect the electrons trajectory, and
hence its radiation.  This section will consider the effect of the
change in electron trajectory on bremsstrahlung emission, neglecting
the closely connected synchrotron radiation emitted by the same field.

An electron will be bent by an angle
\begin{equation}
\theta_{B/2} = {\Delta p \over p} = {eB l_f \sin{\phi_B} \over 2E}
\label{thetab}
\end{equation}
in a distance $l_f/2$ in a uniform magnetic field $B$.  Here, $\phi_B$
is the angle between the electron trajectory and the magnetic field.
As with multiple scattering, if $\theta_{B/2} > 1/\gamma$, then
bremsstrahlung is suppressed.  This happens when (Klein, 1993)
\begin{equation}
y < y_B = {\gamma B\sin{\phi_B} \over B_c}
\label{yb}
\end{equation}
where $B_c$ is the critical magnetic field,
$B_c=m^2c^3/e\hbar=4.4\times10^{13}$ Gauss. 

\begin{figure}
\epsfig{file=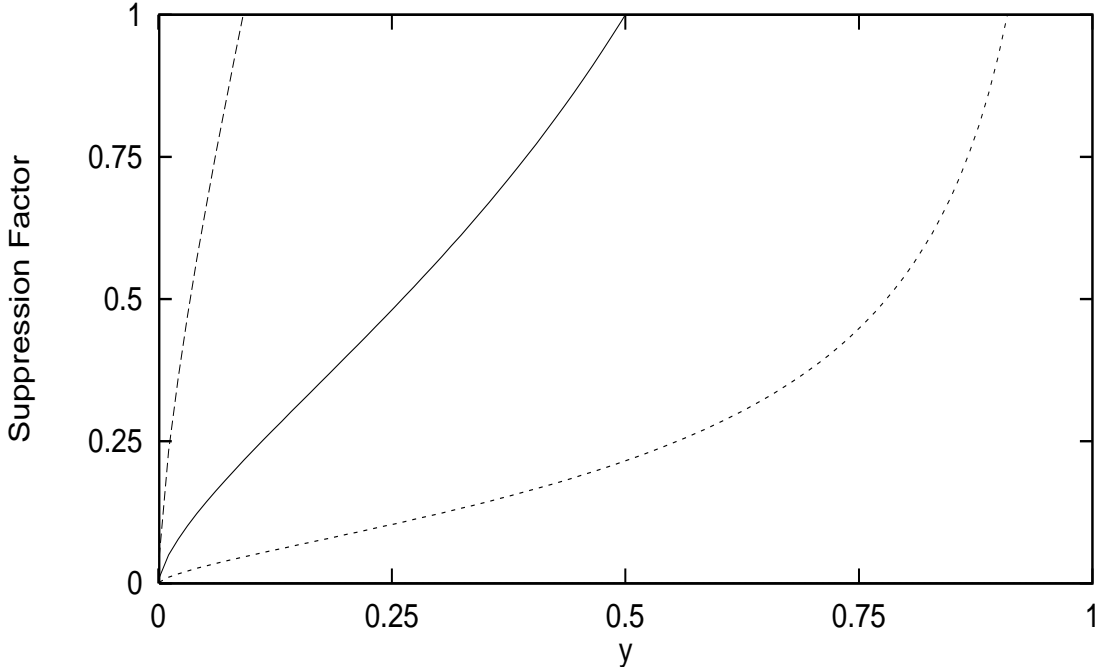,height=6in,width=3.5in,%
clip=,angle=270}
\vskip .1 in
\caption[]{Suppression factor $S$ for magnetic suppression, for
$E=0.1E_B$ (long dashes), $E=1E_B$ (solid line) and $E=10E_B$ (short
dashes).}
\label{magsup}
\end{figure}

The bending angle $\theta_{B/2}$ accumulates linearly with $l_f$, in
contrast to the LPM case where $\theta_{MS/2}\sim l_f^2$; this leads to a
stronger $k$ dependence than with LPM scattering.  If $\theta_{B/2}$ is
treated in the same manner as $\theta_{MS/2}$ in Eq.\ (\ref{eqlonglpm}),
then
\begin{equation}
l_f=l_{f0} \bigg[1+ \bigg({mcB\sin\phi_B l_f \over
\hbar B_c}\bigg)^2\bigg]^{-1}.
\end{equation}
This is a quartic equation for $l_f$, compared with the quadratic
found with multiple scattering.  In the limit of strong magnetic
suppression ($l_f \ll l_{f0}$), the suppression factor $l_f/l_{f0}$
has a form similar to the LPM effect (Klein, 1997):
\begin{equation}
S = \bigg({kE_B\over E(E-k)}\bigg)^{2/3}
\label{smag}
\end{equation}
where $E_B=mc^2B_c/B\sin{\phi_B}$.  Figure. \ref{magsup} shows the
suppression for three different values of $E/E_B$.

Because magnetic suppression has a weaker $k$ dependence than
dielectric suppression, it is only visible when dielectric suppression
does not apply, i.e. for $E>y_{die}E_B$, and then for $y_B > y >
y_{die}$.  For 25 GeV electrons in saturated iron ($\sim20$ kG),
$E_B\approx 1.5$ PeV and $y_B\sim 2.5\times 10^{-5}$, so the magnetic
effect will be hidden.  At higher electron energies, it becomes quite
visible.  For a 1 TeV electron in a 4T field, as will be found at in
the CMS detector at LHC, photons with energies below 900 MeV are
suppressed.

Suppression should also occur for pair production.  A similar
calculation finds
\begin{equation}
S = \bigg({kE_B\over E(k-E)}\bigg)^{2/3}.
\label{smagpair}
\end{equation}
For symmetric pairs (maximum suppression), $S= (4E_B/k)^{2/3}$.
Because the magnetic bending is quite deterministic, in contrast to
multiple scattering which is statistical, this semi-classical
calculation may be more accurate than that for multiple scattering.

Baier, Katkov and Strakhovenko (1988) considered bremsstrahlung
suppression in a magnetic field, for both normal matter (screened
Coulomb potentials) and $e^+e^-$ colliding beams.  They used kinetic
equations to find the radiation to power law accuracy, in both strong
and weak field limits.  These results are similar to Eq. (\ref{smag}).

When the magnetic field is confined to the material, magnetic
suppression should also produce transition radiation.  This should be
most visible with ferromagnetic materials.  Equation (\ref{lftt})
could be used to find the spectrum.

Both the semi-classical calculation and the more accurate result
neglect synchrotron radiation.  Because the formation length scales
for bremsstrahlung and synchrotron radiation are similar, they are
important in the same kinematic regions.  A complete calculation
should treat them together, calculating the electron trajectory due to
the combined field, and then calculating the radiation for that
trajectory.

\subsection{Enhancement and Suppression in Crystals}
\label{scrystals}

So far, we have considered only amorphous materials. In crystals,
however, the regularly spaced atoms produce a huge range of phenomena,
because the interactions with the different atoms can add coherently
(Williams, 1935).  When the addition is in phase, enhanced
bremsstrahlung or pair production results, while out of phase addition
results in a suppression.

Electrons can interact with atoms with impact parameters smaller than
$\hbar/q_\perp$.  The relative phase depends on the spacing between
atoms measured along the direction of electron motion.  The phase
difference for two interactions is $\Phi=\exp{\{i[\omega t-{\bf k\cdot
a}(t)]\}}$ where ${\bf a}$ is the atomic position.  If $\Phi$ has the
same phase for two nuclei, then the emission amplitudes add
coherently.

Including $\theta_\gamma$, the phase difference between two
adjacent sites is
\begin{equation}
\Phi=\exp{\{i[\omega - {\bf k}\cdot {\bf v}]a \cos(\theta_\gamma)\} }
\label{expcrystal}
\end{equation}
where $a$ is the spacing between two atoms along the direction
of electron motion. 
If the nuclei are spaced so that 
\begin{equation}
ak\cos(\theta_\gamma)= {4\pi n\hbar E(E-k)\over m^2 c^3}
\label{anglecrystal}
\end{equation}
where $n$ is an arbitrary integer, then $\Phi=1$ and the addition is
coherent.  As with \v{C}erenkov radiation, for certain
$\theta_\gamma$, the phase is always zero, and $l_f\rightarrow
\infty$, implying infinite emission (from an infinite crystal).
Conversely, if $ak\cos(\theta_\gamma)= (2n+1)2\pi\hbar E(E-k)/m^2
c^3$, there is complete destructive interference.

The large set of variables in Eq. (\ref{anglecrystal}) gives rise to a
variety of effects. As the incident electron direction (affecting $a$)
and/or $\theta_\gamma$ vary, the interference will alternate between
constructive and destructive, producing peaks in the photon energy
spectrum for most sets of conditions.

Although Eq.\ (\ref{expcrystal}) predicts infinite coherence, in a
real crystal several factors limit the coherence length.  One of
these is the thermal motion of the atoms. When the rms thermal
displacement of the atoms is larger than $1/q_\parallel$, the
coherence is lost.  When $\theta_\gamma$ is large, the transverse
separation between the electron and photon can limit the coherence.

Changes in the electron trajectory can also reduce the coherence
length.  The crystalline structure can generate very high effective
fields, causing strong bending, known as channeling (S\o rensen,
1996); this bending can limit the coherence length.  Multiple
scattering can also change the electron direction, and limit the
coherence (Bak \etal, 1988).  Finally, crystal defects and
dislocations can also limit coherence.  Because this is a vast
subject, with several good reviews available (Palazzi, 1968) (Akhiezer
and Shul'ga, 1987) (Baier, Katkov and Strakhovenko, 1989) (S\o rensen,
1996), this article will not consider regular lattices further.

\subsection{Summary \& Other suppression mechanisms}

The suppression mechanisms discussed so far are summarized in Table
II, in order of increasing strength.  Many other physical effects can
lead to suppression of bremsstrahlung and pair production. Many of
them involve partially produced photon interactions with the medium.
Some of the interactions that can affect bremsstrahlung are
photonuclear interactions, real Compton scattering, and, at lower
energies, a host of atomic effects including K and L edge absorption
and a variety of optical phenomena.

{\center{\epsfig{file=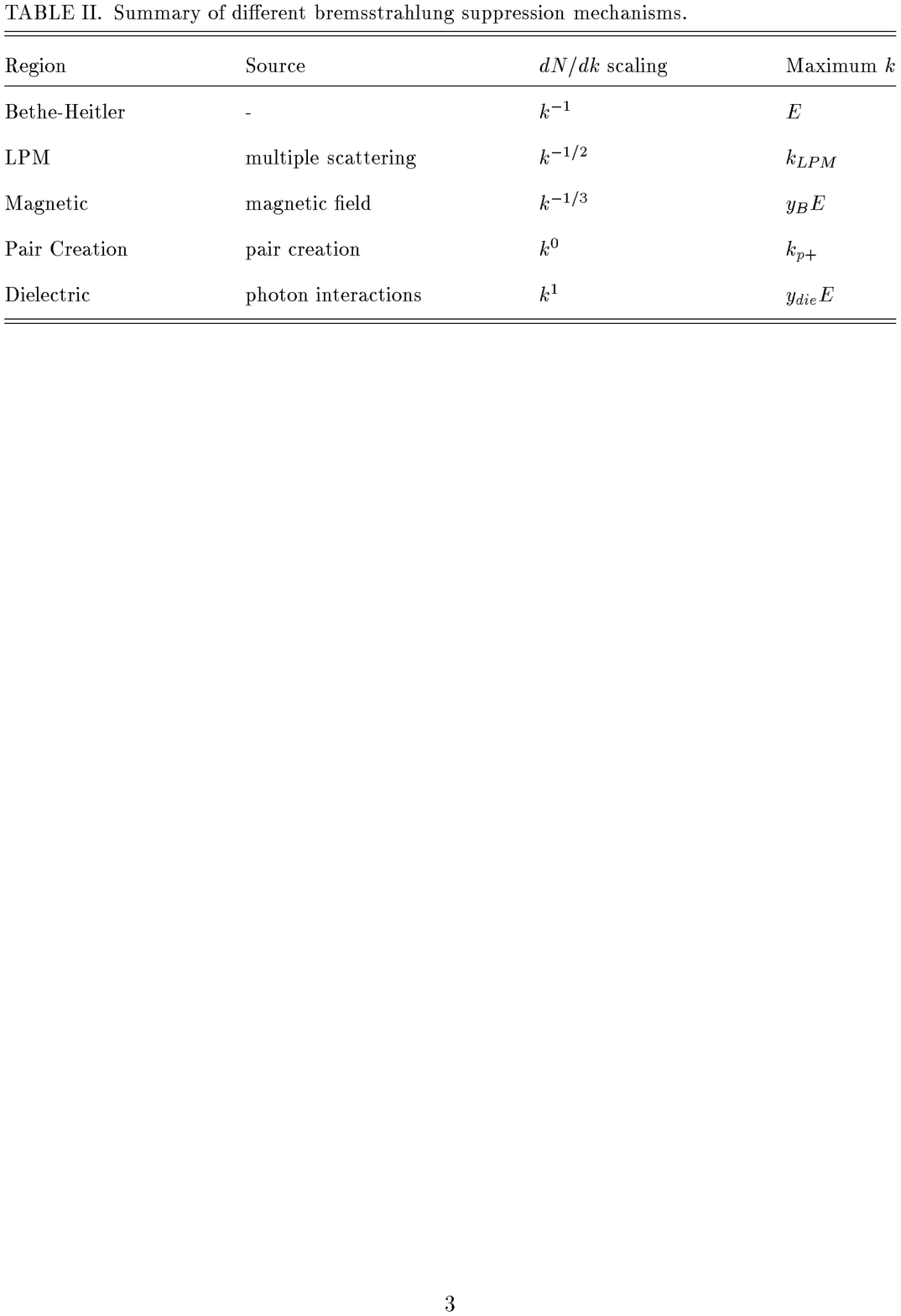,width=6.5in,%
bbllx=67,bblly=548,bburx=545,bbury=730,clip=}}}

Photonuclear interactions can have an effect similar to that of pair
conversion - the partly created photon is destroyed.  Because the
photonuclear cross section is much smaller than the pair conversion
cross section, this is a small correction.  Real Compton scattering
can also effectively destroy the photon.  Toptygin (1964) treated
these reactions as imaginary (absorptive), higher order terms to the
dielectric constant of the medium. The bremsstrahlung plus real
Compton scattering of the partially produced photon is in some sense a
new class of radiation, with its own Feynman diagram.  Because Compton
scattering involves momentum transfer from the medium, $l_f$ is short,
and so, in some regions of phase space (when dielectric suppression is
large), Toptygin found that this diagram can be the dominant remaining
source of emission.

Other photon absorption mechanisms occur at lower photon energies.
For example, K or L edge absorption produces a peak in the photon
absorption spectrum.  If, over a formation length, the absorption
probability due to these peaks is significant, then suppression can
occur.  These peaks are difficult to observe because of competition
from transition radiation, which is enhanced at the same energies (Bak
\etal, 1986).  For optical photons, a host of atomic effects can
affect the dielectric constant.  These variations can also introduce
suppression (Pafomov, 1967).

\section{Migdal formulation}
\label{smigdal} 

More quantitative calculations are more complex.  The first quantum
calculation, by Migdal (1956, 1957) is still considered a
standard. Migdal treated the multiple scattering as diffusion,
calculating the average radiation per collision, and allowing for
interference between the radiation from different collisions.  When
collisions occur too close together, destructive interference reduces
the radiation.

This approach replaces the average multiple scattering angle used
earlier with a realistic distribution of scattering, and, hence, of of
the electron path.  Migdal also allows for the inclusion of quantum
effects, such as electron spin and photon polarization.

Migdal treated multiple scattering using the Fokker-Planck technique
(Scott, 1963).  This technique is used to solve the Bolzmann transport
equation for $F(\theta,z)$, where $F$ is the probability distribution
of particles moving at an angle $\theta$ with respect to the initial
direction.
The equation is:
\begin{equation}
{\partial F(\theta,z) \over \partial z} = -\omega_o(z)F(\theta,z)
+ \int_0^\infty \theta' d\theta' \int_o^{2\pi} d\beta'
W(\chi,z) F(\theta',z)
\label{boltzmann}
\end{equation}
where $\omega_0$ is the scattering probability per unit thickness,
given by the integral of $W(\theta,z)$ over all angles.  Here,
$\theta$ is the scattering angle and $\beta'$ is the azimuthal angle.
$W(\chi,z)$ is the single scatter angular distribution. The angle
$\chi$ is the angular opening between vector representing $\theta$ and
$\theta'$: $\chi^2=\theta^2+\theta'^2-2\theta\theta'\cos\beta'$.  The
$W$ dependence on $z$ allows for inhomogeneity in the material;
otherwise $W$ is independent of $z$.  For each of the trajectories
allowed by the diffusion, Migdal calculated the photon radiation,
including electron spin and photon polarization effects.

The Fokker Planck method is valid if $W(\chi,z)$ is sufficiently
sharply peaked at $\chi=0$, so that it has a finite mean square
$\langle\chi^2\rangle_z$ and that $F(\theta',z)- F(\theta,z)$ can be
accurately approximated by a second order Taylor expansion in
$\theta'-\theta$.  Some calculations (Scott, 1963) lead to a Gaussian
distribution for $F$, with a mean multiple scattering angle
$\langle\chi^2\rangle_z$. Unfortunately, a Gaussian distribution
underestimates the number of scatters at angles larger than a few
times $\theta_0$ (Eq. (\ref{ethetamsthin})).  This problem limits the
accuracy of this calculation.  The problem is somewhat exacerbated
because $l_f$ is relatively short, so the number of scatterings is
fairly small.

\subsection{Bremsstrahlung}

\begin{figure}
\epsfig{file=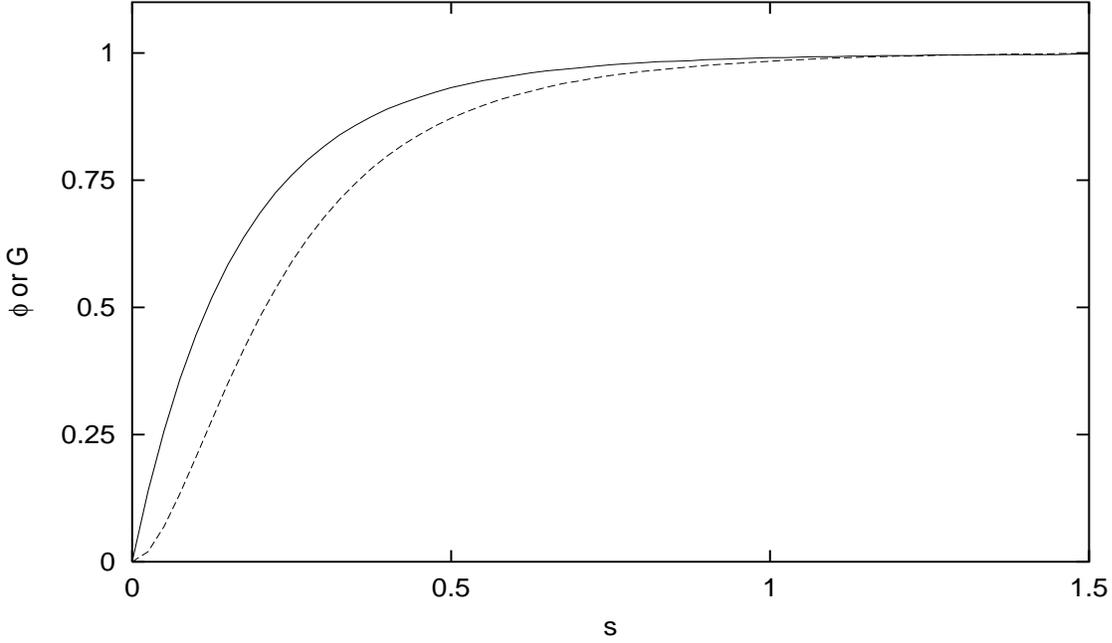,height=6in,width=3.5in,%
clip=,angle=270}
\vskip .1 in
\caption[]{Migdals $G(s)$ (dashed line) and $\phi(s)$ (solid line).
$\phi$ governs the suppression for $y\ll 1$.}
\label{gandphi}
\end{figure}

With these calculations, updated with a more modern form factor, the
Migdal cross section for bremsstrahlung is
\begin{equation}
  {d\sigma_{\text{LPM}} \over dk} = {4\alpha r_e^2\xi(s) \over 3k}
\{y^2 G(s) + 2 [1 +(1-y)^2 ] \phi (s) \} Z^2  \ln\bigg({184 \over Z^{1/3}}
\bigg)
\label{esigmamig}
\end{equation}
where $G(s)$ and $\phi (s)$ are the suppressions of the electron spin
flip and no spin flip portions of the cross section respectively,
\begin{eqnarray}
G(s) & = & 48s^2\bigg({\pi\over4} - {1\over2}\int_0^\infty e^{-st}
{\sin{(st)} \over \sinh{(t/2)}} dt \bigg) \\
\phi(s) & = & 12 s^2 \bigg(\int_0^\infty e^{-sx} \coth{(x/2)} \sin{(sx)} dx
\bigg) - 6 \pi s^2.
\label{ephimig}
\end{eqnarray} 
The factor $\xi(s)$ is
\begin{eqnarray}
\xi(s) & = & 2 \hskip 2.75 in(s<s_1) \nonumber\\
\xi(s) & = & 1 + \ln(s)/\ln(s_1) \hskip 1.725 in    (s_1<s<1) \nonumber\\
\xi(s) & = & 1 \hskip 2.75 in (s \ge 1)
\label{eximig}
\end{eqnarray}
with $s_1=Z^{2/3}/184^2$.  Here,
\begin{equation}
s= \sqrt{E_{LPM}k \over 8 E(E-k)\xi(s)}.
\label{esmig}
\end{equation}
Migdal gave infinite series solutions for $G(s)$ and
$\phi(s)$. However, they may be more simply represented by polynomials
(Stanev \etal, 1982):
\begin{eqnarray}
\phi(s) & = & 1-\exp{\bigg(-6s[1+(3-\pi)s]+s^3/(0.623+
0.796s+0.658s^2)\bigg)}\nonumber\\ 
\psi(s) & = & 1-\exp{\bigg(-4s-8s^2/(1+3.96s+4.97s^2-0.05s^3+7.5s^4)} \bigg)
\nonumber\\
G(s) & =  & 3\psi(s)-2\phi(s).
\label{estanev}
\end{eqnarray} 
These functions are plotted in Fig.\ \ref{gandphi}.

For $k\ll E$, $s\sim 1/\langle\gamma\theta_{MS}\rangle$.  For $s\gg1$,
there is no suppression, while for $s\ll1$, the suppression is large.
In the absence of suppression $G(s)=\phi(s)=1$, and Migdal matches the
Bethe-Heitler cross section.  For small suppression, where $s$ is
large, $G(s)= 1-0.22/s^4$ and $\phi(s) = 1-0.012/s^4$.  For strong
suppression, where $s\rightarrow 0$, $G(s)=12\pi s^2$ and
$\phi(s)=6s$.

If a Coulomb scattering distribution is fit with a Gaussian, the mean
of the Gaussian grows slightly faster than $\sqrt{l_f}$.  This
increase is reflected in the the logarithmic rise in $\xi$. For
sufficiently high energies, $q_{max}$ is limited by the nuclear radius
$R_A$, which Migdal approximated $R_A=0.5 \alpha r_e Z^{1/3}$.  The
nuclear radius limits $q_{max}$ to $\hbar/ R_A$; at higher momentum
transfers the nuclear structure is important and Eq.~(\ref{esigmamig})
loses accuracy.  The fortuitously simple `2' coefficient comes from
the chosen approximation for $R_A$. This cutoff is only reached for
extremely large suppression, and is hence of limited importance;
higher order terms are likely to dominate at this point.

One difficulty with these formulae is that $\xi$ depends on $s$, which
itself depends $\xi$, so the equations must be solved recursively.  To
avoid this problem, Stanev and collaborators (1982) developed simple,
non-iterative formulae to find $s$ and $\xi(s)$.  They removed
$\xi(s)$ from the equation for $s$, defining
\begin{equation}
s'= \sqrt{E_{LPM}k \over 8 E(E-k)}.
\label{esprime}
\end{equation} 
Then, $\xi$ depends only on $s'$:
\begin{eqnarray}
\xi(s') & = & 2 \hskip 2.75 in (s'<\sqrt{2}s_1) \nonumber\\
\xi(s') & = & 1 + h - {0.08(1-h)[1-(1-h)^2] \over \ln{\sqrt{2}s_1}}
\hskip .46 in (\sqrt{2}s_1<s'\ll 1) \nonumber\\
\xi(s') & = & 1 \hskip 2.75 in (s' \ge 1).
\label{exiprime}
\end{eqnarray}
where $h=\ln{s'}/\ln{(\sqrt{2}s_1)}$.  This transformation is possible
because $\xi$ varies so slowly with $s$.

\begin{figure}
\epsfig{file=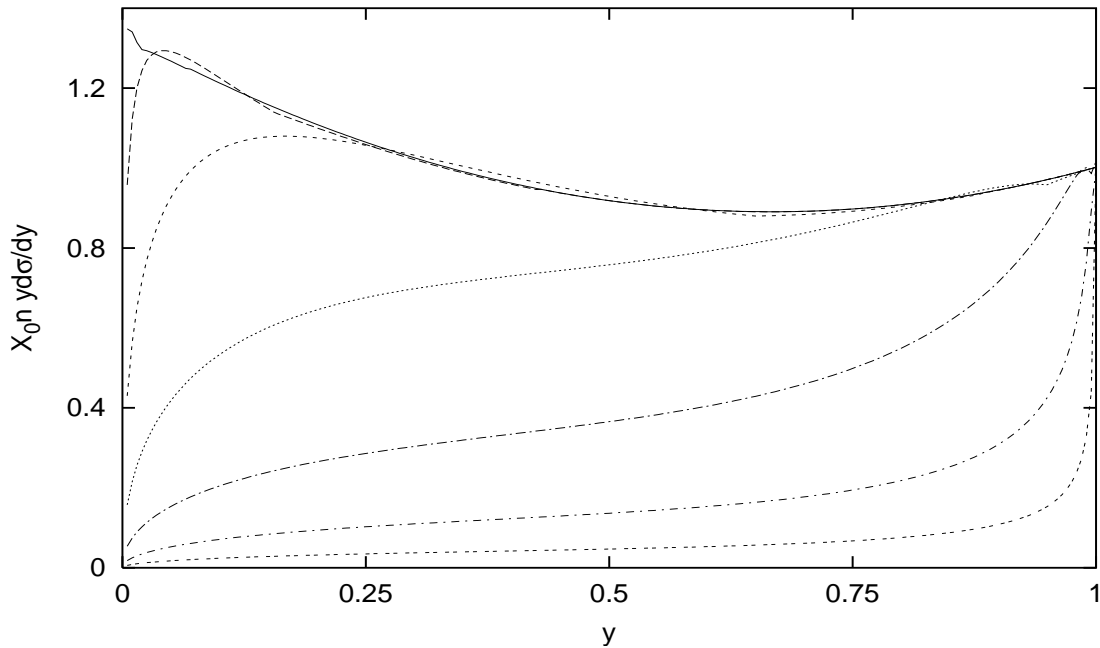,height=6in,width=3.5in,%
clip=,angle=270}
\vskip .1 in
\caption[]{The energy weighted differential cross section for
bremsstrahlung, $X_0n y d\sigma_{\text{LPM}}/dy$ for various electron
energies in a lead target, showing how the spectral shape changes.
Electrons of energies 10 GeV (top curve), 100 GeV, 1 TeV, 10 TeV, 100
TeV, 1 PeV and 10 PeV (bottom curve) are shown. The units are
fractional energy per radiation length.}
\label{bremcompare2}
\end{figure}

Fig. \ref{bremcompare2} compares the energy weighted cross section per
radiation length, $X_0nyd\sigma/dy$, Eq. (\ref{esigmamig}), for
several electron energies. As $E$ rises, the cross section drops, with
low energy photons suppressed the most.  The number of photons with
$k<E(E-k)/E_{LPM}$ is reduced.

Although it reproduces the main terms of the Bethe-Heitler equation,
Eq.~(\ref{esigmamig}) does not include all of the corrections that are
typically used today.  Without any suppression, Eq.~(\ref{esigmamig})
becomes
\begin{equation}
  {d\sigma_{\text{LPM0}} \over dk} = {4\alpha r_e^2 \over 3k}
\{y^2 + 2 [1 +(1-y)^2 ]\} Z^2F_{el},
\label{esigmamigbh}
\end{equation}
matching the Bethe and Heitler (1934) result.  In comparison a modern
bremsstrahlung emission cross section (Tsai, 1974) (Perl, 1994)
\begin{equation}
{d\sigma_{\text{BH}} \over dk} = {4\alpha r_e^2 \over 3k}\bigg[ \{ y^2
    + 2 [1 + (1-y)^2] \}\ \ [Z^2(F_{el}-f) + Z F_{inel}]+ (1 - y) {( Z^2 + Z
    )\over 3}\bigg]
\label{esigmatsai}
\end{equation}
includes several additional terms.  $F_{el}$ is the elastic form
factor from Sec. 2, $\ln{(184/Z^{1/3})}$, for interactions with the
atomic nucleus.  $F_{inel}$ is an inelastic form factor,
$\ln{(1194/Z^{2/3})}$ that accounts for inelastic interactions with
the atomic electrons.  Newer suppression calculations, discussed
later, treat inelastic scattering separately.  

The $f$ term accounts for Coulomb corrections because the interaction
takes place with the electron in the Coulomb field of the nucleus.
For lead $f=0.33$, while $F_{el}=3.7$.  The Coulomb correction may be
incorporated into suppression calculations by adjusting the form
factors (Sec. \ref{sbkatkov}; Baier and Katkov, 1997a).

These corrections may be accounted for with a simple assumption
(Anthony \etal, 1995). Since $q_{||} \ll mc$ the target mass is
irrelevant for suppression purposes, and the two form factors can be
lumped together.  The same can be done for the Coulomb corrections.
This is done by scaling the radiation length to include these
corrections; the standard tables of radiation lengths (Barnett, 1996)
include these factors.

The other difference, the final $1-y$ term, is problematic because of
the different $y$ dependence.  In a semi-classical derivation
(Ter-Mikaelian, 1972, pg. 18-20), this term only appears in the
no-screening limit, small impact parameter, high momentum transfer,
limit, where the formation length is short.  So, this term should
represent a part of the cross section that involves large momentum
transfers, and so is not subject to suppression.  Because it is only
about a 2.5\% correction for large $Z$ nuclei, current experiments
have limited sensitivity to this point.

In the strong suppression limit, for $y\ll 1$, the small $s$
approximations for $\phi(s)$ and $\xi(s)$ lead to the semi-classical
scaling
\begin{equation}
S=  3 \sqrt{kE_{LPM} \over  E(E-k)}.
\label{eslpm}
\end{equation}
Because $\xi(s)$ varies only logarithmically with $s$, this limit,
corresponding to $\xi(s)=2$, is only reached for very strong
suppressions, $S<10^{-3}$.  As Fig. \ref{bremcompare} shows, the
normalization differs from the semi-classical results, but it, as
previously mentioned, can depend on the how the absolute cross section
is treated.  Moreover, for a direct comparison, it might be fairer to
use $\xi=1$, since the semi-classical calculations neglect the
relevant non-linear scaling.  This changes the coefficient to
$3/\sqrt{2}\sim 2$. For lower energies, the strong suppression limit
is
\begin{equation}
S\sim \sqrt{{36 kE_{LPM} \over 8 E(E-k)}\bigg(1 - 0.13 \ln{S}\bigg)}
\label{elspm2}
\end{equation}
where the recursion for $\xi(s)$ has been removed, and $s_1$
calculated for lead.  It is clear that the correction for $\xi$ is
significant, particularly for small $S$.

These calculations contain some sources of error.  Because of the
Fokker-Planck method and the problematic mean scattering angle; these
results are of only logarithmic accuracy.  One numerical hint of
inaccuracy can be seen by comparing Migdal's formula with the
Bethe-Heitler limit.  For $k_{LPM} < k < 1.3 k_{LPM}$, $\phi(s)\xi(s)$
rises slightly above 1 and the Migdal result is slightly above Bethe
Heitler.  The maximum excess is about 3\%.  This can serve as a crude
estimate of the expected accuracy of Eq.~(\ref{esigmamig}).

Migdal also considered dielectric suppression.  Since it only occurs
for $y\ll 1$, only the $\phi$ term is relevant; Migdal replaced
$\phi(s)$ with $\phi(s\Gamma)/\Gamma$, where $\Gamma= 1 + k_p^2/k^2$,
to get
\begin{equation}
{d\sigma_{\text{LPM}} \over dk} = {16\alpha r_e^2\xi(s) \over 3k}
{\phi (s\Gamma)\over\Gamma} Z^2 \ln\bigg( {184 \over Z^{1/3}} \bigg).
\label{esigmamigdiel}
\end{equation}
The same substitution applies for the simplified polynomials (Stanev
\etal, 1982).  This is close to the Ter-Mikaelian result, although it
misses the modified form factor in Eq. (\ref{sigmatern}).

Equation~(\ref{esigmamig}) can also be found with a classical path
integral approach (Laskin, Mazmanishvili and Shul'ga, 1984).  The
radiation for a given trajectory, Eq, (\ref{longlp}), can be averaged
over all possible paths, In the limit $k\ll E$, required by the
classical nature of the calculation, this reproduces Migdal's result.
Dielectric suppression can also be included in the path integral
approach (Laskin, Mazmanishvili, Nasonov and Shul'ga, 1985).

Migdal notes that, for thick slabs, the photon angular distribution is
dominated by the electron multiple scattering.  However, nothing in
Migdal's calculation should change the semi-classical result that
suppression should be reduced for photons with $\theta_\gamma >
1/\gamma$; this effect may be visible in thinner targets.  Pafomov
(1965) discussed the energy and angular distribution of photons
emerging from thin slabs ($T\ll X_0$), producing a complex set of
results.

\subsection{Pair Creation}

The cross section for pair production may be found simply by crossing
the Feynman diagram for bremsstrahlung, as shown in
Fig. \ref{schematic}.  Migdal used this crossing to calculate the
cross section for pair production:
\begin{equation}
  {d\sigma_{\text{LPM}} (\gamma\rightarrow e^+e^-) \over dE} = {4\alpha
r_e^2\xi(\tilde{s}) \over 3k} \bigg( G(\tilde{s}) + 2 \big[{E^2 \over k^2 }
+ (1- {E\over k})^2\big] \phi(\tilde{s}) \bigg)
\label{sigmamigpair}
\end{equation}
where
\begin{equation}
\tilde{s} = \sqrt{E_{LPM} k \over 8 E(k-E) \xi(\tilde{s})}
\approx {mc^2\over k\gamma}
\label{smigpair} 
\end{equation} 
and $G$ and $\phi$ are as previously given.  In the limit
$\tilde{s}\gg 1$ there is no suppression, while $\tilde{s}\rightarrow
0$ indicates large suppression, both leading to the appropriate
semi-classical result. These equations can be simplified as was done
with bremsstrahlung (Stanev, 1982).

\begin{figure}
\epsfig{file=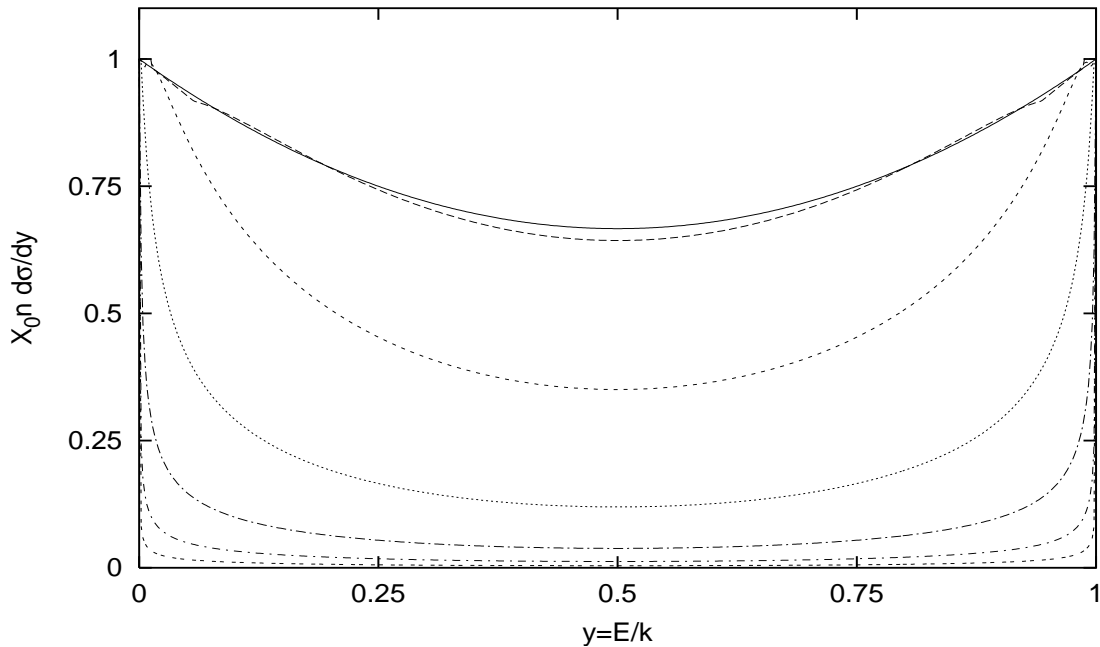,height=6in,width=3.5in,%
clip=,angle=270}
\vskip .1 in
\caption[]{The differential cross section for pair production, $X_0n
d\sigma_{\text{LPM}}/dy$ in lead for various photon energies, showing
how the spectral shape changes.  Cross sections are plotted for
photons of energies 1 TeV (top curve), 10 TeV, 100 TeV, 1 PeV, 10 PeV,
100 PeV and 1 EeV (bottom curve).}
\label{comparepair}
\end{figure}

Figure \ref{comparepair} gives the pair production cross section in
lead for a number of photon energies.  As $k$ increases above
$E_{LPM}$, the cross section drops, with asymmetric pairs increasingly
favored.  For $E_{LPM} < E(k-E)/k < 1.3 E_{LPM}$, $\phi(\tilde{s})
\xi(\tilde{s})>1$ and the Migdal cross section rises above the
Bethe-Heitler case, matching the bremsstrahlung overshoot.  It also
causes the 'hiccup' present at $k=E_{LPM}$ in
Fig. \ref{suppressionvse}.

\subsection{Surface Effects}
\label{ssedgeffects}

Migdal's calculations only apply for an infinite thickness target.
Several authors have calculated the transition radiation due to
multiple scattering, based on a Fokker-Planck approach.  Although
these calculations begin with the same approach, the details differ
significantly, as do the final results.  Gol'dman (1960) added
boundaries to Migdal's Fokker-Planck equation, calculating the emission
inside and outside the target, and an interference term.  Outside the
target, there is no emission, while inside he reproduced Migdal's
result.  For the surface terms, the photon flux for $y\ll 1$ and $s\ll
1$ is
\begin{equation}
{dN\over dk} ={\alpha\over\pi k} \ln{1\over s}\approx 
{\alpha\over 2\pi k} \ln{({8k_{LPM}\over k})}
\label{goldman}
\end{equation}
per surface. This result is simlar to the semi-classical result in
Eq.\ (\ref{transms}).

Ternovskii (1960) considered the dielectric effect as well as multiple
scattering, again starting from Migdal's kinetic equation.  He
considered the entire range of target thicknesses, including
interference between closely spaced boundaries.  By comparing the
radiation inside the target with an interference term, he found that
for $T<\alpha X_0/2\pi$, multiple scattering is insignificant, and the
Bethe-Heitler spectrum is recovered.  This is a much broader range
than the limit $T<X_0/1720$ presented in section \ref{sthint}.  His
results for intermediate thickness targets, Eq. (\ref{esigmashulga}),
apply for $\alpha X_0/2\pi < T <\alpha X_0/2\pi s$, also a wider range
than in Sec. \ref{sthint}.

For thicker targets, with $T\gg \alpha X_0/2\pi s$, if the dielectric
dominates ($s>1$ or $s(k_p/k)^2 >1$), he obtains Eq.\
(\ref{trconv}). Where multiple scattering dominates ($s<1,
s(k_p/k)^2\ll 1$), then
\begin{equation}
{dN\over dk} = {\alpha\over 2\pi k} \bigg({k^2\over E^2} + 2{E^2+(E-k)^2
\over E^2}\bigg) \ln{{\chi\over\sqrt{s}}}
\label{fternovskii}
\end{equation}
where '$\chi\approx 1$'.  This result is similar to
Eq.~(\ref{goldman}), but covers the complete range of $y$.  Neglecting
the logarithmic term, the spectrum matches that of unsuppressed
bremsstrahlung for a target of thickness $\sim 0.5\alpha X_0$.  This
has a different $E$ and $k$ dependence than would result from $1\ l_f$
of Bethe-Heitler radiation.

Unfortunately, Eq. (\ref{fternovskii}) is difficult to use.  It only
applies for $s(k_p/k)^2\ll 1$; no solution is given for the
intermediate region $s(k_p/k)^2\sim 1$.  Also, $\chi$ is poorly
defined.  If the equation is extended to $s(k_p/k)^2=1$, then choosing
$\chi=1$ creates a large discontinuity.  The discontinuity disappears
for $\chi=0.42$, but this also eliminates transition radiation in the
region $0.17 < s < 1$.

Garibyan (1960) extended his previous work with transition radiation
to include multiple scattering.  He found that multiple scattering
dominates for $k>k_{cr}$, matching the semi-classical result for
infinite targets.  If $\xi(s)$ is neglected, this is the same cutoff
found by Ternovskii.  Where multiple scattering dominated, he
reproduced Eq.~(\ref{goldman}).  Elsewhere, he found the usual
transition radiation from the dielectric of the medium.

Equations (\ref{goldman}) and (\ref{fternovskii}) are negative for
$s>1$; common sense indicates that they should be cut off in this
region where no transition radiation is expected.  However, Pafomov
(1964) considered this evidence that these works were wrong.  He
stated that they incorrectly separated the radiation into transition
radiation and bremsstrahlung.  He calculated the transition radiation
as the difference between the emission in a solid with and without a
gap, again starting with the Gol'dman's initial formulae. Pafomov
found that multiple scattering dominated transition radiation for
$k>k_{cr}$.  Surprisingly, he found that multiple scattering affected
transition radiation even for $E< y_{die}E_{LPM}$, with an additional
term $dN/dk=\alpha/\pi k (k_{LPM}/k)^2$ added to the conventional
result when $k>k_p$. For $k_{LPM} > k_p$, he found different formula
for different $k$.  For $k<k_{cr}$
\begin{equation}
{dN\over dk} = {\alpha\over\pi k} = \ln{ k_{cr}^2 \over
 k^2}
\label{pafamovc}
\end{equation}
while for $k_{cr}< k < k_{LPM}$,
\begin{equation}
{dN\over dk} = {\alpha\over\pi k} \ln{{2\over 3}\sqrt{k_{LPM} \over k}}.
\label{pafomova}
\end{equation}
Counterintuitively, Pafomov predicted that for $k> k_{LPM}$, there is
still transition radiation, with
\begin{equation}
{dN\over dk} = {\alpha\over\pi k} \ \ {8k_{LPM}^2 \over 21k^2}.
\label{pafomovb}
\end{equation}
These two equations do not match for $k = k_{LPM}$.  However, a
numerical formula, not repeated here, covers the entire range
smoothly.  It is worth noting that Eq.\ (\ref{pafomovb}) is quite
close to the semi classical result for $k<k_{LPM}$.

\begin{figure}
\epsfig{file=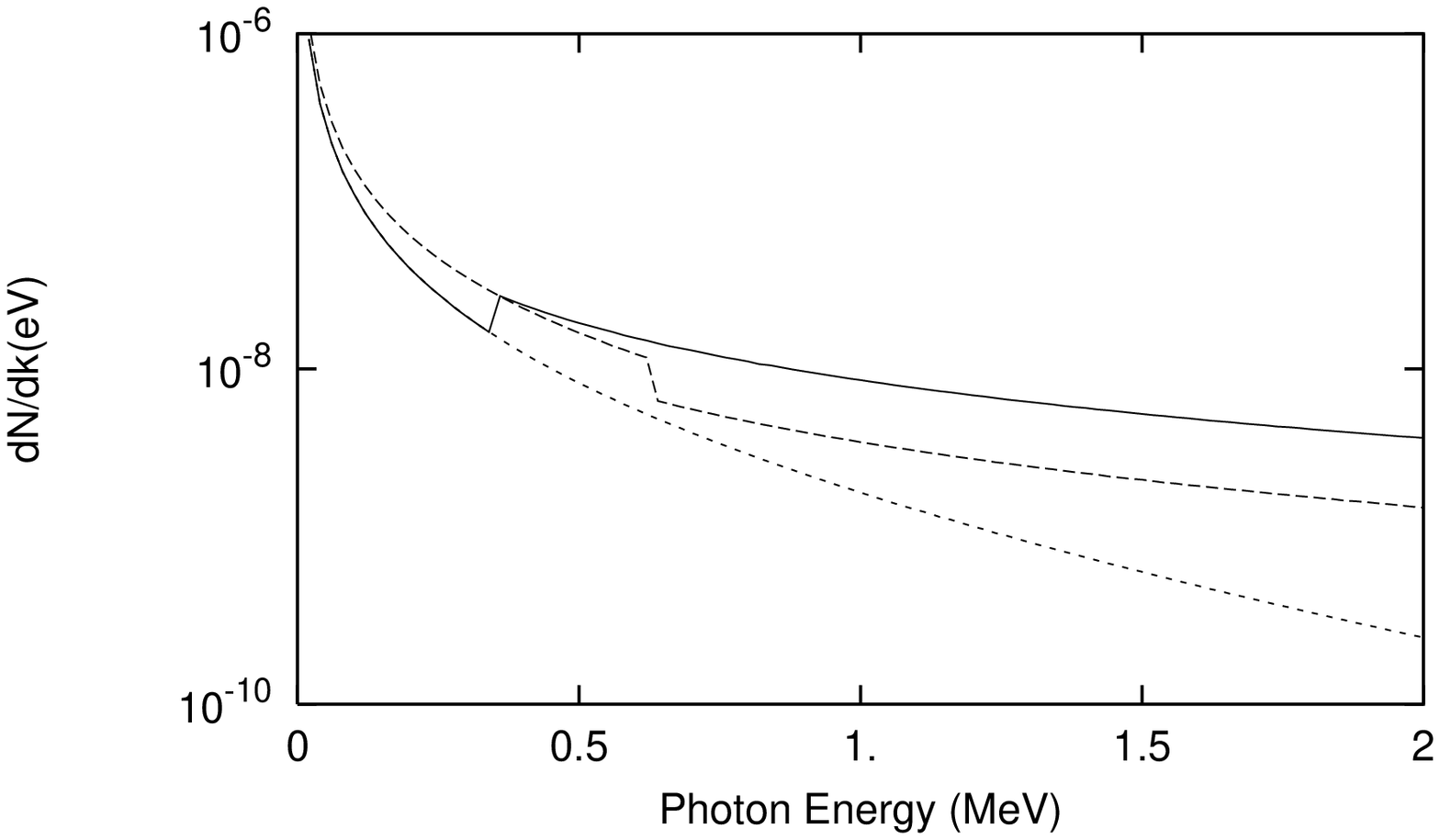,height=5in,width=2.8in,%
clip=,angle=270}
\epsfig{file=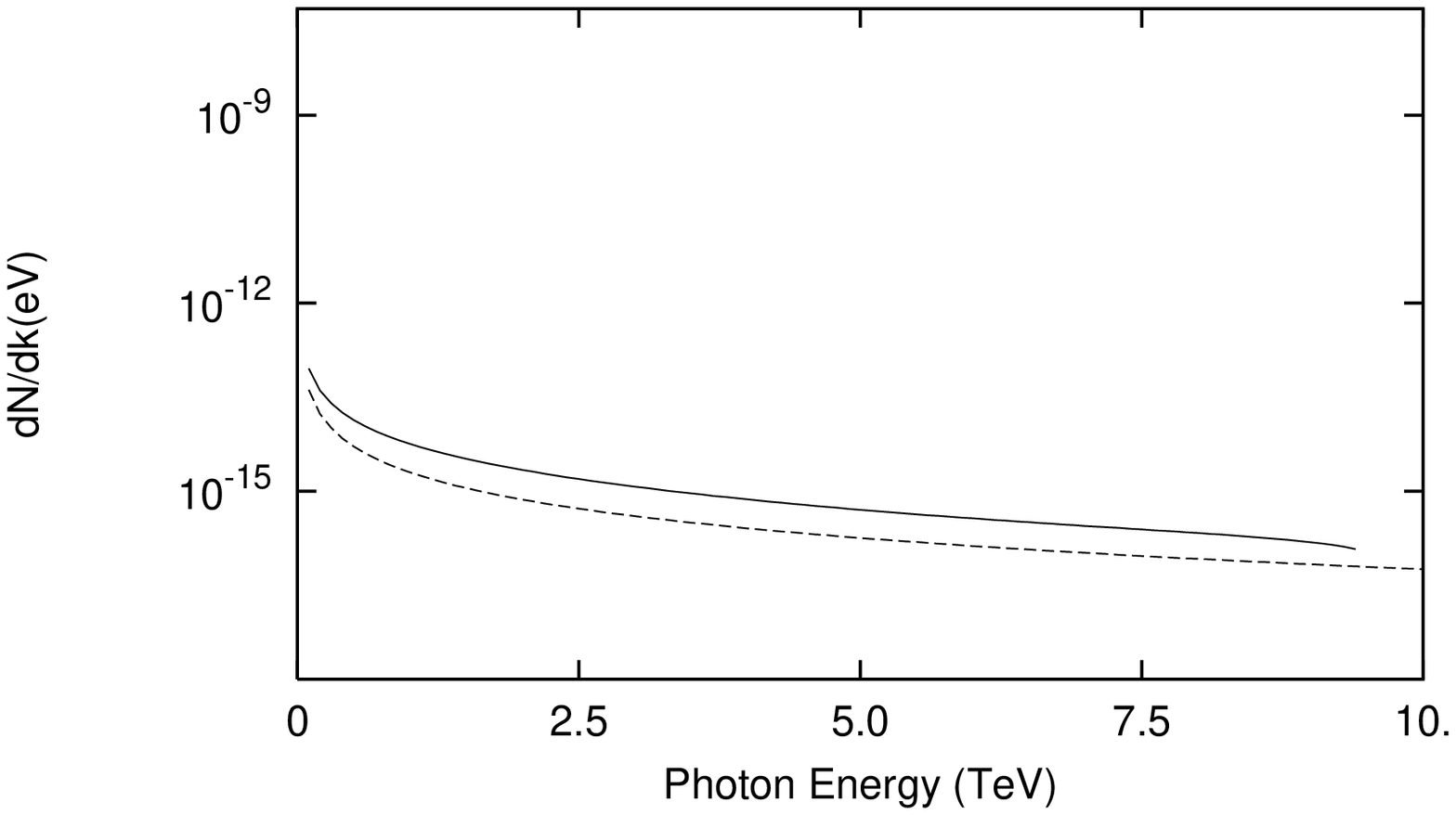,height=5in,width=2.8in,%
clip=,angle=270}
\vskip 0.1 in
\caption[]{Comparison of the edge effects predicted by Ternovskii
(with $\chi=1$) (solid line) and Pafomov (long dashes), for a single
independent edge, compared with conventional transition radiation
(short dashes).  (a) shows the radiation from a 100 GeV electron in
lead.  The jump in the Ternovskii curve at $k=350$ keV is at
$s(k_p/k)^2)=1$, while the drop in the Pafomov curve at 630 keV is at
$k=k_{cr}$.  Panel (b) shows the Pafomov and Ternovskii predictions
for a 10 TeV electron in lead; both curves are much smoother.
Conventional transition radiation is negligible in (b).}
\label{edgecompare}
\end{figure}

Fig.\ \ref{edgecompare} compares the transition radiation from a
single surface predicted by Gol'dman/Garibyan, Ternovskii (using
$\chi=1.0$), and Pafomov.  Except for Pafomov, these calculations
predict radiation up to $k=k_{LPM}=E^2/E_{LPM}$.  For a flat $kdN/dk$
spectrum, the total energy radiated per surface rises as $E^2$, as
with the semi-classical approach.  For large enough $E$, electrons
lose most of their energy to radiation when crossing an interface.

However, these equations fail before this point.  For high enough
electron energies, these calculations predict that each electron
should emit several photons per edge traversed.  Since the formation
lengths for the various photon emissions will overlap at the edge,
this is really a higher order process, not yet treated properly by
calculations.

Unfortunately, these calculations appear to do a poor job of
fitting the data. Fig. \ref{uranium} shows that they predict
transition radiation considerably above the data from SLAC E-146.

\section{Blankenbecler and Drell formulation}
\label{sbland}
Blankenbecler and Drell (1996) calculated the magnitude of LPM
suppression with an eikonal formalism used to study scattering from an
extended target.  The approach was originally developed to study
beamstrahlung.  One major advantage of their approach is that it
naturally accommodates finite thickness slabs, automatically including
surface terms.

They begin with a wave packet that scatters while moving through a
random medium.  For each electron path, they calculate the radiation,
based on the acceleration of the electron.  The radiation is
calculated for all possible paths and averaged.  One clear conceptual
advantage of this calculation is that it does not single out a single
hard scatter as causing the bremsstrahlung; instead all of the
scatters are created equal.  This differs from Landau and Pomeranchuk
and Migdal, who found the rate of hard scatters which produced
bremsstrahlung, and then calculated the multiple scattering in the
region around the hard scatter to determine the suppression.  For a
thick target, in the strong suppression limit $s\ll 1$, they predict
that the radiation is $\sqrt{3\pi/8}$ (about 8\%) larger that of
Migdal.

One advantage of this calculation is that it treats finite thickness
media properly.  There are three relevant length scales: $l_{f0}$
(Eq.\ (\ref{lfzero})),\ $T$ and $\alpha X_0$. Blankenbecler and Drell
(1996) called $\alpha X_0$ the mean free path for elastic scattering,
based on counting vertices in Feynman diagrams.  This correspondence
does not stand up to more detailed examination.  However, the
arguments in the paper are not affected.  These variables are
combined into two ratios:
\begin{equation}
N_{BD}= {\pi l_{f0} \over 3\alpha X_0}
\label{blandn}
\end{equation}
and
\begin{equation}
T_{BD}= {\pi T \over 3\alpha X_0}
\label{blandtt}
\end{equation}
where $T_{BD}$ is the target thickness, in mean free paths, while
$N_{BD}$ is the number of formation lengths per mean free path; when
$N_{BD}$ is large suppression is strong, while $N_{BD}<1$ corresponds
to a single interaction per electron, {\it i.e.} the Bethe Heitler
regime.

The eikonal approach finds the wave function phase and momentum
difference between different points on the electron path.  These
differences are then used to find the radiation for that length scale.
The probability of emission from the target is
\begin{equation}
{dP_{BD} \over dk} = {\alpha \over 2 kE(E-k)\hbar^2c^2}
\int {d^2k_\perp \over 4\pi^2}
\int_{-\infty}^{+\infty} dz_2
\int_{-\infty}^{z_2} dz_1 S(z_2,z_1)
\cos\bigg(\int_{z_1}^{z_2} dz {d\Phi(z,0)\over dz}\bigg).
\end{equation}
Here, $k_\perp=k\cos(\theta_\gamma)$ is the photon perpendicular
energy, $S(z_2,z_1)$ is a sum over electron polarization (spin flip
and no spin flip) of the squared perpendicular momentum acquired by
the electron due to scattering, and $d\Phi/dz$ is the differential
phase difference due to the scattering.  The $d^2k_\perp$ integration
is equivalent to an integration over photon angles.  In the absence of
scattering, $d\Phi/dz$ is constant, and then the integrals must be
carefully evaluated; Blankenbecler and Drell introduce a convergence
factor to insure that the boundary conditions are correct.

\begin{figure}
\center{\epsfig{file=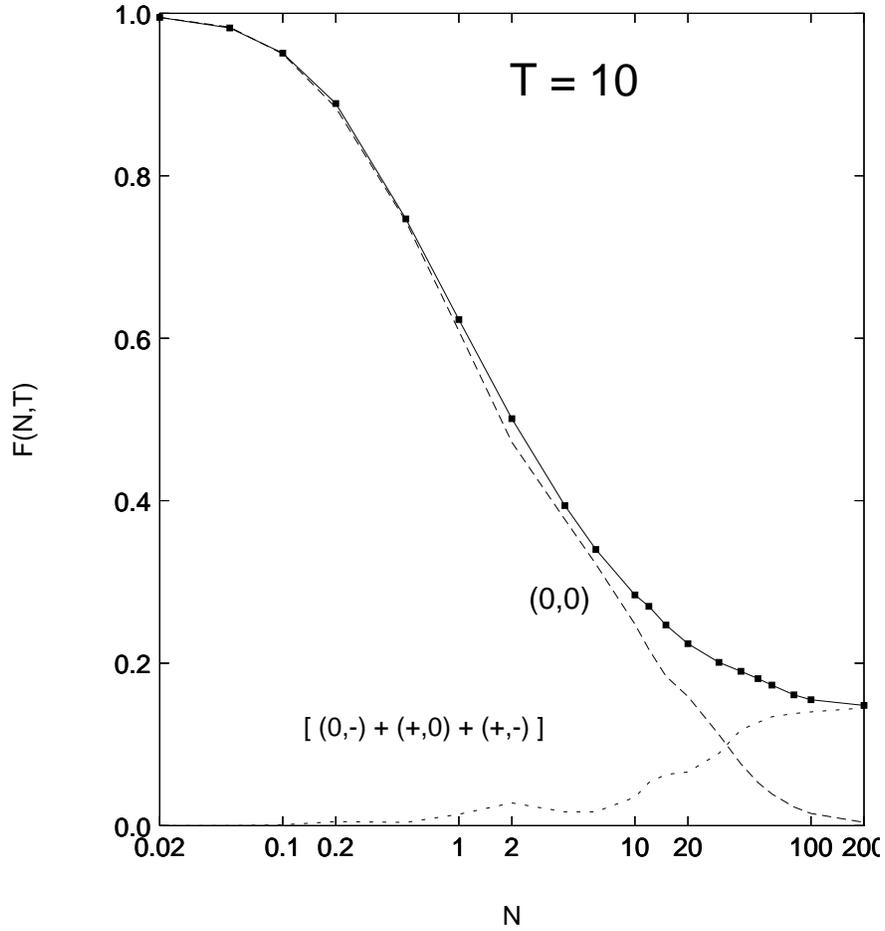,height=5in,%
bbllx=6,bblly=38,bburx=473,bbury=548,%
clip=,angle=0}}
\vskip .1 in
\caption[]{The Blankenbecler and Drell form factor $F(N_{BD},T_{BD})$
for a target with $T_{BD}=10$, showing the contributions for the bulk
emission ($00$) and the transition and interference terms $(0-)$,
($+0$) and ($+-$). For a fixed target thickness, $N_{BD}\sim
1/l_{f0}\sim k/E^2$; the edge effects are largest for small $N_{BD}$
corresponding to small $k$. From Blankenbecler and Drell (1996).}
\label{blandcomponents}
\end{figure}

Because the integrals include all possible values of $z$, three
regions must be considered: before the target (denoted $+$), inside
the target (denoted $0$), and after the target ($-$).  For the double
integral, there are 9 possible combinations, of which two ($++$ and
$--$) are clearly zero.  Time ordering eliminates ($-+$), ($-0$) and
($0+$), leaving the bulk term ($00$), two single surface terms ($+0$)
and ($0-$) and an interference term ($+-$).  For thick targets with
$T_{BD}\gg1$, the interference term vanishes, and the bulk term
dominates over the single surface terms.  In this case, their
calculations reproduce the Migdal results, with the 8\% higher cross
section.  For thinner targets, the surface terms are more important.
Figure~\ref{blandcomponents} compares these terms, using a suppression
form factor $S=F(N_{BD},T_{BD})$ relative to Bethe-Heitler.

For thin targets, $T_{BD}<1$, where $N_{BD}>1$, the interference term
($+-$) dominates, demonstrating a large transition radiation.
$N_{BD}/T_{BD}<1$ reduces to the Bethe Heitler free particle case.

\begin{figure}
\center{\epsfig{file=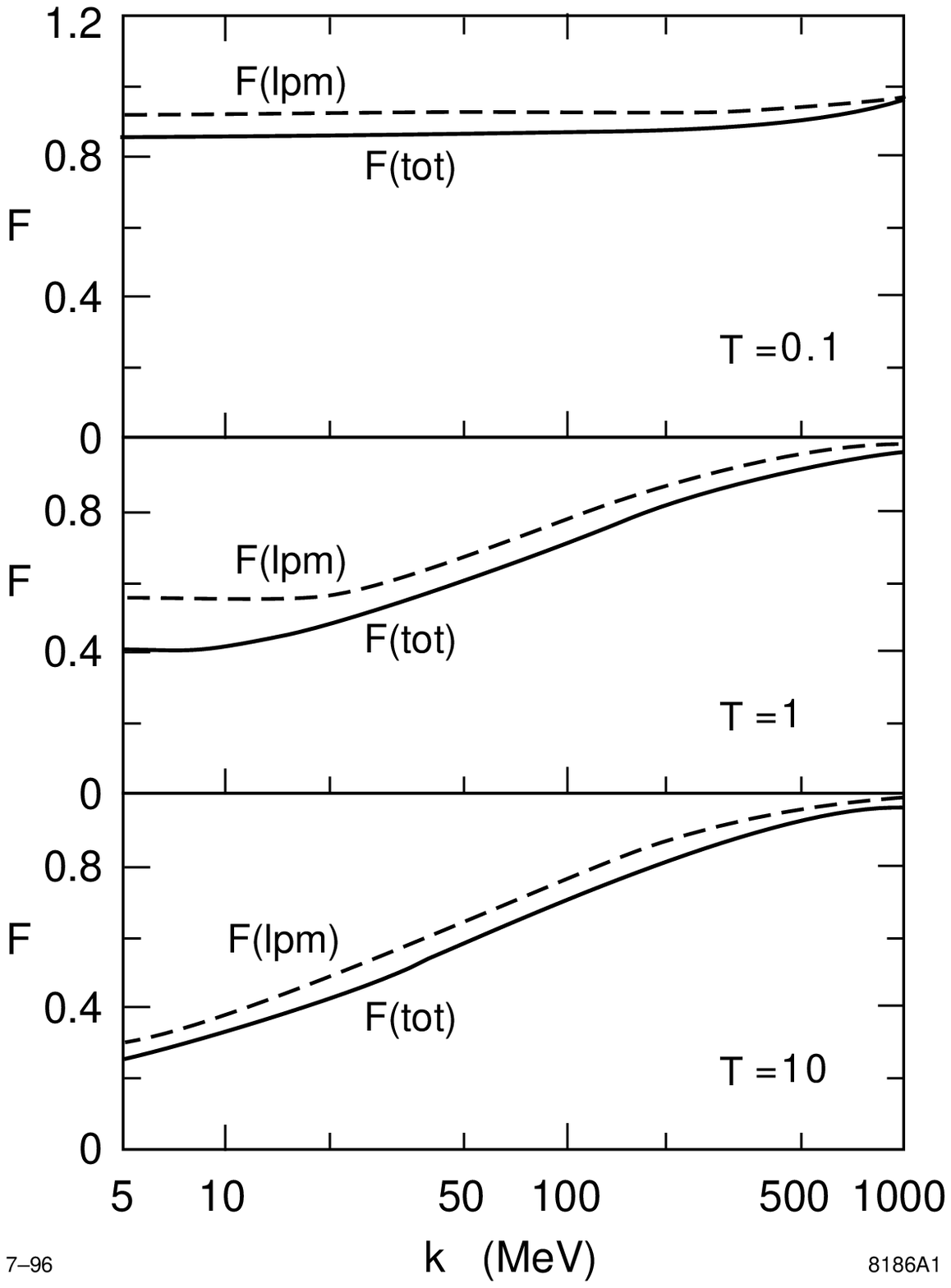,height=6in,%
clip=,angle=0}}
\vskip .1 in
\caption[]{The Blankenbecler and Drell form factor $F(N_{BD},T_{BD})$
for 25 GeV electrons in 3 thicknesses of gold targets.  F(lpm),
assumes that the wave function amplitude and phase fluctuate
independently, while F(tot) includes the correlation between amplitude
and phase, further reducing the emission.  From Blankenbecler
(1997b).}
\label{blandformfactor}
\end{figure}

For a thick target, $T_{BD}>1$, the central interaction ($00$) region
is important.  For large $N_{BD}$, their results are similar to those
of Migdal.  For very large $N_{BD}$, with $N_{BD} > T_{BD}$, the
formation length is longer than the target, and both the ($00$) and
mixed regions contribute.  Fig.\ \ref{blandformfactor} compares the
suppression as a function of $N_{BD}$ (which is proportional to
$1/k$), for a variety of $T$.  For large $T_{BD}$, suppression
increases with $N_{BD}$.  When $T_{BD}$ decreases, suppression drops,
eventually becoming almost independent of $N_{BD}$, at a value similar
to that predicted by Shul'ga and Fomin (1996).

There are a few caveats in this calculation.  The eikonal approach
assumes that the potential is smooth enough.  Blankenbecler and Drell
used a Gaussian scattering potential, which underestimates the rate of
large angle scattering.  Second, they assumed that the wave function
phase and amplitude fluctuate independently.  Blankenbecler (1997b)
showed that the correlation between amplitude and phase reduces the
emission by a further 5-15\%.  Figure~\ref{blandformfactor} compares
curves with and without the correlation.  The calculation has also
been extended to include multiple slabs separated by a gap
(Blankenbecler, 1997a).

Calculating the emission from a slab as a whole is problematic, These
results assume that there is either zero or one interaction per
incident electron.  But, for a typical bremsstrahlung cross section of
10 photons per $X_0$, the relative probability of getting 2
interactions compared with 1 interaction is $\sim 20 T/X_0$, so single
interactions are only prevalent for slabs with $T< 0.05 X_0$.  This
problem makes it difficult to apply these results in many real-world
situations.

\section{Zakharov calculation}
\label{szakharov}

Zakharov (1996a,b, 1998a,b) transformed the problem of multiple
scattering into a two dimensional (impact parameter and depth in the
target) Schr\"oedingers equation.  The imaginary potential is
proportional to the cross section for an $e^+e^-$ pair scattering off
of the atom, which is itself simply related to the bremsstrahlung
cross section.  This equation was solved using a transverse Greens
function based on a path integral. This impact parameter approach is
complementary to Migdal's momentum space approach.  Because this
approach allows for arbitrary density profiles, it naturally
accommodates finite thickness targets.

Zakharov (1996a) calculated the radiation due to a simple Coulomb
potential.  The calculation is keyed to the scattering cross section
for a $e^+e^-$ dipole of separation $\rho$,
$\sigma(\rho)=C(\rho)\rho^2$.  In the strong suppression limit, $C$
varies slowly with $\rho$.  Zakharov then found the frequency $\Omega$
for a harmonic oscillator in a potential which would reproduce this
scattering cross section.  This is roughly equivalent to describing
the multiple scattering with a Gaussian.  The radiation is governed by
a parameter $\eta=\Omega l_f$.  For an infinitely thick target, the
results are almost identical to Migdal (Zakharov, 1998a); functions of
$\eta$ that match Migdal's $\phi(s)$ and $G(s)$, for $\eta=1/\sqrt{8}s$.  The
only difference is the slowly varying part of the cross section:
$\xi(s)$ for Migdal and $C(y\rho_{\rm eff})/C(\hbar c/m)$ for Zakharov;
$\rho_{\rm eff}$ is the impact parameter where radiation is largest. In
the Bethe-Heitler limit, $\rho_{\rm eff}\sim \hbar c/ym$.

In the limit $y\rightarrow 0$, for strong suppression
\begin{equation}
{d\sigma_Z \over dk}= {2\alpha^2Z\over E} \sqrt{2\hbar c
\log{(2a/y\rho_{\rm eff})}\over \pi nk}
\label{zak2}
\end{equation}
where
\begin{equation}
\rho_{\rm eff} = \bigg({\pi Z^2\alpha^2 nEy^3 \log{(2/\alpha Z^{1/3})}
\over \hbar c} \bigg)^{1/4}
\label{zakr}
\end{equation}
where $a$ is the Thomas-Fermi screening radius Numerically,
$\rho_{\rm eff}({\rm m})\sim 10^{-9}/[E {\rm (eV)}y^3]^{1/4}$ for lead.
These equations are valid for $200/Z^{1/3} < \log(2a/y\rho_{\rm eff}) <
1.5\times 10^{5}/Z^{2/3}$; corresponding to Migdal's $\xi(s)=1$ and
$\xi(s)=2$.

Except for the `1' in $\xi(s)$, this equation has the same form as
Migdal's Eq. (\ref{elspm2}).  Although unimportant to the theory, the
`1' greatly reduces the effect of the slowly varying term.
Numerically, $2a/y\rho_{\rm eff}=6.8\times10^{-7} E(eV)/y$ for lead;
other solids aren't too different.  For a 1~TeV electron beam, as $y$
varies from $10^{-4}$ (a typical $y_{die}$) to $10^{-2}$ (an arbitrary
upper limit to 'low $y$'), the Zakharov logarithm varies by about
20\%, while the change in $\xi(s)$ is much smaller.  This variation
should be measureable.

For moderate suppression, a more detailed treatment is required.
Because $C(\rho)$ varies more quickly with $\rho$, the harmonic
oscillator approximation fails and the actual potential must be used
(Zakharov, 1996b).  Zakharov used separate screened elastic ($a=0.83
a_0 Z^{-1/3}$) and inelastic ($a=5.2 a_0 Z^{-2/3}$) potentials,
reproducing the appropriate unsuppressed bremsstrahlung cross
sections.  The separate potentials are most important for low $Z$
nuclei.  However, because of the small recoil, the separate form
factors have a small effect on suppression. The more complicated
potential requires additional integrations.  Because of this, this
approach can so far be used only for finite thickness targets.

\begin{figure}
\center{\epsfig{figure=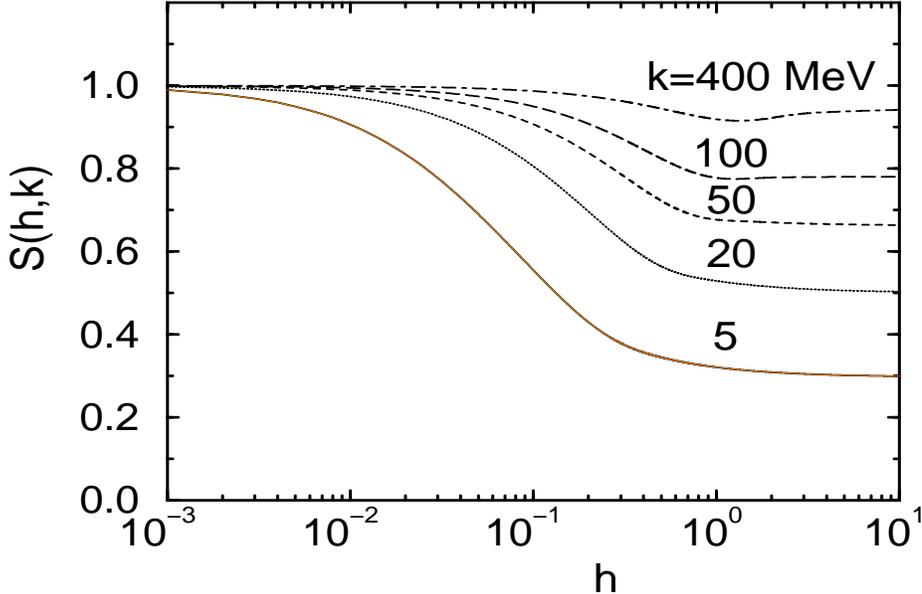,%
bbllx=127,bblly=285,bburx=509,bbury=589,%
height=8.5cm,width=12cm}}
\caption[]{Suppression factor $S(h,k)$ found by Zakharov.  The curves
are for a 25 GeV electron, with $h=T/l_{f0}$.  The $h=10$ results are
very close to the infinite thickness limit.  The dip around $h=1$ for
the highest photon energies is due to interference between the two
target surfaces. From Zakharov (1996b).}
\label{zakharov}
\end{figure}

Figure \ref{zakharov} shows Zakharov's suppression factors for finite
target thicknesses in a 25 GeV electron beam.  This demonstrates how
suppression increases with target thickness. The thickest target, with
$h=10$ is very close to the infinite thickness limit. Zakharov (1998b)
added a correction to allow for multiple photon emission, and did a
detailed comparison with SLAC E-146 data.  Except for the carbon
targets in 25 GeV electron beams, he found good agreement with the
data for photon energies $k>5$ MeV (above the region of dielectric
suppression).

Zakharov (1997b) presents a few results for multiple-slab
configurations, finding a smaller interference term than Blankenbecler
(1997a).  The two results would agree better if Blankenbecler included
the amplitude-phase correlation in his multiple slab calculations.

Zakharov's results for gluon bremsstrahlung from a quark will be
discussed in Sec.\ \ref{sqcd}.

\section{BDMS Calculation}
\label{bdms}

The BDMS group (R. Baier \etal, 1996) started with the Coulomb field
of a large number of scatterers, with the atomic screening modelled
with a Coulomb potential cut off with a Debye screening mass $\mu$.
An eikonal approach is used to account for the large number of
scatters.

The critical variable in this calculation is the (dimensionless)
phase difference between neighboring centers, 
\begin{equation}
\kappa = {X_0 \mu^2 c^3 k \over 2 \hbar E^2 }
\label{eqkappa}
\end{equation}
For QED calculations, $\mu=\hbar Z^{1/3}/a_0c$ reproduces Thomas-Fermi
screening. The authors also define a coherence number $\nu=\alpha
l_{f0}/X_0$, the number of scatters required for the accumulated phase
shift to equal 1. This $\nu$ is similar to Blankenbecler and Drells
$N_{BD}$.  A large phase shift between interactions, $\kappa>1$,
corresponds to the Bethe-Heitler limit.  Because this approach assumes
massless electrons, its 'Bethe-Heitler limit' does not completely
match the usual Bethe-Heitler formula.

In the factorization limit, $T< l_{f0}$, where the entire target
reacts coherently, their results match Ternovskii (1960):
\begin{equation}
{dN\over dk} = {2 \alpha \over \pi k }\langle\ln
({q_{tot}^2\over m^2c^2})\rangle
\end{equation}
where $q_{tot}$ is the total perpendicular momentum acquired by the
particle while traversing the target due to multiple scattering.
Here, the mass $m$ is introduced to remove a collinear divergence.

In the LPM regime, ($\kappa\ll 1)$, the bremsstrahlung cross section
is similar to Eq.\ (\ref{zak2}). The logarithmic term is due to
non-Gaussian large angle ($\theta > 1/\gamma$) Coulomb scatters.
Because of these large scatters, the mean squared momentum transfer is
poorly defined, and, in fact, diverges logarithmically; this logarithm
appears in the suppression formula.

\begin{equation}
{d\sigma\over dk} \sim {\alpha \over \pi k X_0 } \sqrt{\kappa
\ln{(1/\kappa)}}.
\end{equation}
They comment that they neglect a logarithmic factor under the
logarithm; if the corresponding term is removed from Zakharov's
formula, the two results have the same functional dependence.  If one
identifies $E_{LPM}=\kappa \mu^2/2$, as their paper indicates, then
the radiation takes the form
\begin{equation}
{d\sigma\over dk} \sim {\alpha \over \pi k } \sqrt{{kE_{LPM} \over E^2}
\log{E^2 \over kE_{LPM}}}.
\end{equation}

\section{Baier \& Katkov}
\label{sbkatkov}

Baier and Katkov (1997a) also studied suppression due to multiple
scattering, trying to reach an accuracy of a few percent, by including
several corrections omitted previously. They begin with the scattering
from a screened Coulomb potential, in the same impact parameter space
used by Zakharov.  Coulomb corrections are included to account for the
motion of the screening electrons, along with separate potentials for
elastic and inelastic scattering.  Finally, they allow for a nuclear
form factor, with an appropriately modified potential for impact
parameters smaller than the nuclear radius.

They find the electron propagator for a screened Coulomb potential, in
the Born approximation. This assumes Gaussian distributed scattering,
and reproduces Migdal's result.  The electron propagator is then
expanded perturbatively, with a correction term which accounts for
both large angle scatters and Coulomb corrections to the potential.
Without the Coulomb corrections, the first order result is similar to
that of Zakharov (V. Baier, 1998).

Coulomb corrections are incorporated by adjusting the parameters in
the potential.  With Coulomb corrections, the screening radius becomes
$a_2=0.81Z^{-1/3}a_0\exp[0.5-f(Z\alpha)]$.  while the characteristic
scattering angle changes from $\theta_1=\hbar c/Ea_0$ to
$\theta_2=\hbar c/Ea_0\exp[f(Z\alpha)-0.5)]$.  Here, $f(Z\alpha)$ is
the standard Coulomb correction (Tsai, 1974). This scaling accounts
for the extra terms in Eq.\ (\ref{esigmatsai}) compared with Eq.\
(\ref{esigmamigbh}).  For heavy nuclei, $a_2$ is about 20\% larger
than the Thomas-Fermi screening radius.

Higher order terms may also be calculated for the electron propagator.
The ratio of the first two terms of the expansion is $0.451$ divided
by a logarithmic term, so the series should converge reasonably
rapidly.

Inelastic scattering can be included by adding a term to the
scattering potential, changing the charge coupling from $Z^2$ to
$Z^2+Z$ and further modifying the characteristic angle, to
$\theta_e=\theta_1\exp{[Z/(1+Z)(f(\alpha Z)-1.88)-0.5]}$.

Dielectric suppression is included by modifying the potential, with a
replacement similar to Migdal's.  For $k\ll k_p$, the spectrum is
similar to Ter-Mikaelian's, but with power law and Coulomb
corrections:
\begin{equation}
{d\sigma\over dk} = {16Z^2\alpha r_e^2 k \over 3 k_p^2}
\bigg[\ln{\bigg({184 k_p\over k Z^{1/3}}\bigg)} + {1 \over 12}
-f(Z\alpha)\bigg]
\label{bkdiel}
\end{equation}
where $f(Z\alpha)$ is the standard Coulomb correction (Tsai, 1974).

Baier and Katkov considered the extremely strong suppression regime,
neglected by Migdal, where the finite nuclear radius becomes
important.  This region is reached at the lowest $E$ for $y\ll 1$,
where dielectric suppression otherwise dominates.  There, limiting $q$
(i.e. $q_\parallel$) to $\hbar/R_A$ changes the form factor for $k<k_p
R_A\lambda_e$.  In lead, this corresponds to $y<2\times 10^{-6}$.
Then,
\begin{equation}
{d\sigma\over dk} = {16Z^2\alpha r_e^2 k \over 3 k_p^2}
\bigg(\ln{({a\over R_A})}-0.02 \bigg).
\label{bkmag}
\end{equation}
For $y=10^{-6}$ in lead, this is about 25\% larger than
Eq. (\ref{bkdiel}).  This is probably measurable, although transition
radiation and backgrounds will be very large.

Baier and Katkov then considered targets with finite thicknesses,
breaking down the possibilities in a manner similar to Blankenbecler
and Drell, with a similar double integral.  For relatively thick
targets, $T>l_{f0}$, the results are consistent with Ternovskii
(1960), but with additional terms for the Coulomb correction.  For $T<
l_{f0}$ and strong suppression,
\begin{equation}
{dN\over dk} = {\alpha \over \pi k} \bigg[{k^2 \over E^2} + 
\bigg(1 + {(E-k)^2 \over E^2}\bigg)\bigg((1 + {1\over 2A})[\ln{(4A)}-0.578] +
{1 \over 2A} -1 + {0.578\over L_t}\bigg)\bigg]
\end{equation}
where $A=\pi Z^2\alpha^2nT\hbar^2/m^2c^2(L_t+1-2\times0.578)$ and
$L_t\sim 2\ln{(2a_2/\lambda_e\rho_t)}$, with $\rho_t$ the (scaled)
minimum impact parameter that contributes to the form factor integral.
Unfortunately, this equation diverges as $T\rightarrow 0$.

V. Baier and Katkov (1997a) compared their photon spectrum with SLAC
E-146 data from a $0.02 X_0$ thick tungsten target in 25 and 8 GeV
electron beams, and find good agreement.  However, a target this thick
has a substantial (roughly 20\%) correction to the photon spectrum to
account for electrons that undergo two independent bremsstrahlung
emissions.  This correction is not included in their calculation, and
the good agreement with data is surprising.  A later work (Baier and
Katkov, 1999) included multiphoton effects, and also finds good
agreement with the data; the difference between the two calculations
is not explained.

V. Baier and Katkov (1997b) considered thinner targets with $T\sim
l_{f0}$. They compare their calculations with SLAC E-146 data, for 0.7
\% $X_0$ thick gold targets in 8 and 25 GeV beams, and also find good
agreement.  Because this target is much thinner, the multiple
interaction probability is greatly reduced and the agreement is not
surprising.

\section{Theoretical Conclusions}
\label{theoryconclusions}

\subsection{Comparison of Different Calculations}

The calculations discussed here used a variety of approaches to solve
a very difficult problem.  Because the underlying techniques are so
different, it is difficult to compare the calculations themselves.
However, some general remarks are in order.

All of the post-Migdal calculations are done for a finite thickness
slab, integrating both bulk emission and transition radiation.
Unfortunately, the finite thickness slab calculations are not easily
usable by experimenters, because they assume each electron undergoes
at most one interaction in the target.  Multiple interactions are
easily accounted for in a Monte Carlo simulation.  However, the
simulation must be able to localize the photon emission, while these
calculations are for the slab as a whole. Because of the edge terms,
it isn't correct to simply spread the emission evenly through the
slab.  This limits their direct applicability to thin slabs.

The newest calculations by Zakharov and Baier and Katkov include
multiple photon emission.  However, these calculations are still for
bulk targets, and are not very amenable to complex geometries.  For
example, these calculations could not be used to model an
electromagnetic shower.

Here, we will use the Migdal approach as a standard for comparison.
Although Zakharov's approach is very different from Migdal, he
reproduces Migdal's result in the appropriate limit.  However,
Zakharov incorporates some further refinements which should lead to
increased accuracy.  Baier and Katkov have a similar approach to
Zakharov, and include a number of additional refinements, especially
for bremsstrahlung at very low $y$.

The other works have very different genealogies.  Because the BDMS
result does not reproduce the Bethe-Heitler limit, it is more relevant
to QCD than QED.  However, if the formalism is extended to include
hard radiation, and some additional graphs are added, then BDMS
becomes equivalent to the Zakharov formalism (R. Baier \etal, 1998b).

For Blankenbecler and Drell, the only obvious point of comparison is
the potential; Zakharov (1996b) states that the Blankenbecler and
Drell potential does not match the Coulomb potential, and will not
show the logarithmic dependence given by the slow variation of
$C(\rho)$ or $\xi(s)$.  

The results can also be compared numerically.  Unfortunately, these
calculations are very complex and the descriptions frequently lack
adequate information for independent computation.  So, it is necessary
to rely on the results given by the authors.  One point of comparison
is defined by SLAC E-146 data on a thin target: 25 GeV electrons
passing through a 0.7~\%~$X_0$ thick gold target.  The E-146 data
agreed well with Migdal as long as $T > l_f$; at lower $k$,
Ternovskii's Eq. (\ref{esigmashulga}) matched the data. Blankenbecler
and Drell, Zakharov, and Baier and Katkov all showed good agreement
with this data.  Zakharov initially added in a 7\% normalization
factor to match the data.  Some correction is required, because, even
for a 0.7\% $X_0$ target, the multiphoton pileup `correction' is still
several percent.  With a correction for multiphoton emission Zakharov
(1998b) found normalization coefficients average 1, except for the
uranium targets.  Zakharov also changed his approach to Coulomb
corrections between the two works.  The other authors do not discuss
normalization.  Overall, in this energy range, the different
approaches agree with each other to within about 5\%.

Because of the different logarithmic treatment, Migdal, Blankenbecler
and Drell, and Zakharov will scale slightly differently with energy.
Changes depending on target thickness are more complicated; the
surprising agreement found by Baier and Katkov for the 2\%~$X_0$
tungsten data, where multiple interactions are a 20\% correction,
shows that there there are significant uncertainties in scaling the
results with target thickness.  It would also be interesting to
compare the calculations for a low $Z$ target, where the E-146 data
showed some disagreement with Migdal.

\subsection{Very Large Suppression}

One weakness of all of these calculations is that they only consider
the lowest order diagrams.  For fixed $y$, $l_f$ rises with $E$, even
with LPM suppression.  At high enough energies, the formation zones
from different emissions will overlap, and any lowest order
calculation will fail.  Dielectric suppression is strong enough that
$l_f$ decreases and localization improves with increasing suppression,
so it is less subject to this problem.  However, for multiple
scattering, a method of dealing with higher order terms is needed.
While the radiative corrections to bremsstrahlung are known (Fomin,
1958), they were not computed with suppression mechanisms in mind.

Even neglecting the overlapping formation zones, when suppression is
large, higher order terms are important, because higher order
processes involve larger $q_\parallel$, and hence are less subject to
suppression; Sec. II.H illustrated this for direct pair production.
So, when suppression is strong ($S\sim\alpha$), current calculations
are suspect.  For QCD calculations, discussed in Sec. \ref{sqcd} the
problem is much worse because of the large coupling constant,
$\alpha_s$.

It is worth noting that suppression can affect other processes, via
radiative corrections.  The total cross section for Coulomb
scattering, for example, is the sum of the elastic cross section plus
the cross section for bremsstrahlung where the photon is not
observable.  Since suppression can affect the latter process, it can
indirectly change the elastic scattering cross section (Ter-Mikaelian,
1972, pg. 135).  Inelastic processes, which leave the target atom in
an excited state can also contribute significantly to the cross
section when suppression of the lowest order diagrams is large.

\subsection{Classical and Quantum Mechanical Approaches}

There has been some controversy as to how well classical
electrodynamics can predict LPM suppression.  The original Landau and
Pomeranchuk (1953a,b) calculations were completely classical; they
failed at the same point that classical bremsstrahlung calculation
fail: when $k\sim E$.  The path integral approach (Laskin,
Mazmanishvili and Shul'ga, 1984) goes further, and reproduces Migdal's
result in the same limit. Of course, at least semi-classical
calculations are necessary for pair production.

The main advantage of quantum mechanical calculations is that it
covers the complete range of $y$.  The newest calculations include
features not found in the classical approaches.  However, some of the
advances, such as separate elastic and inelastic potentials could
certainly be included in a classical calculation.

In the future, quantum approaches appear necessary to calculate the
higher order terms that become important at extremely high energies,
when formation lengths from different emissions begin overlapping.
Unfortunately, these calculations seem extremely difficult, and are
probably a long way off.

However, the classical approach to LPM suppression was not universally
accepted.  Bell (1958) stated that Migdal's predictions conflicted with
classical electrodynamics, and concluded that ``any real effect of
this kind is of essentially quantal origin.''  He pointed out that the
classical radiation, Eq.\ (\ref{classicalb}), is positive definite,
and hence monotonically increasing for increasing pathlength.
However, Bell neglected to account for the fact that, in a dense
solid, the electron trajectory ${\bf r}(t)$ changes, and one cannot
simply sum the radiation due to interactions with separate nuclei.
Even classically, emission from the different pieces of electron
pathlength can interfere, in a manner similar to other classical
interference effects.

\section{Experimental Results}
\label{sexperiment}

Bremsstrahlung or pair creation suppression can be studied with high
energy electron or photon beams. Because pair creation suppression
requires photons with $k>E_{LPM}$, beyond the reach of current
accelerators, pair creation has been studied only with cosmic rays,
with consequently very limited statistics.  The best suppression
studies have used electron beams at accelerators.  Besides the LPM
effect, these beams have been used to study dielectric suppression.

\subsection{Cosmic Ray Experiments}

The first tests of LPM suppression came shortly after Migdal's paper
appeared.  These experiments used high energy ($k>1$ TeV) photons in
cosmic rays, and studied the depth of pair conversion in a dense
target.  The earliest experiments looked for a deficiency of low
energy electrons in electromagnetic cascades (Miesowicz, Stanisz and
Wolther, 1957).  This analysis is difficult because of uncertainties
in the incoming photon spectrum and electron energy measurement.

Fowler, Perkins and Pinkau (1959) studied isolated pair production in
an emulsion stack.  They measured the shower energy and the distance
between the initial conversion and the first daughter pair resulting
from the primary.  They observed 47 showers with $k>1$ TeV.  As $k$
increased, the distance between the conversions rose, as predicted by
Migdal, contrasting with with decrease predicted by Bethe- Heitler.

Varfolomeev and collaborators (1960) used a similar approach, albeit
on a smaller data set, with similar results.  They selected events
where the energy of the daughter pair was very small ($y\ll 1$), and
found qualitative evidence for suppression; they did not differentiate
between LPM and dielectric suppression.

Lohrmann (1961), studied emulsion exposed on high altitude balloon
flights.  Compared to earlier results, Lohrmann studied lower energies,
where the LPM and Bethe-Heitler cross sections are much closer.
However, the increased photon flux allowed for much better statistics,
making up for the smaller cross section difference.  Lohrmann used a
variety of analysis techniques, all of which supported Migdal's
results.  More recently, long duration balloon experiments have
gathered somewhat larger data samples, up to 120 events, with similar
results (Strausz \etal, 1991).

Kasahara (1985) studied the development of $\sim$100 TeV showers in
lead/emulsion chambers exposed to cosmic rays on Mt. Fuji. His
analysis was based on the fact that electromagnetic LPM showers with
$E \gg E_{LPM}$ penetrate considerably more material than
Bethe-Heitler showers.  Kasahara found that the shower development
profiles matched simulations based on Migdal, but differed from his
Bethe-Heitler simulations.  The study suffered from a background from
hadronic and/or multiple photon showers, both also subject to large
fluctuations.  Kasahara removed 5 obvious background events from his
initial 19 event sample, but warned that some of the remainder could
also be hadronic.  This study is of special interest as it is the only
study yet in the region $E\gg E_{LPM}$, where suppression affects
shower development.  

All of the air shower experiments suffered from some common problems,
by far the largest being the poor statistics; high enough energy
photons are not common.  In addition, because of the limited emulsion
thickness, they did not consider surface effects.  Finally,
uncertainties in the photon spectrum complicated the analysis. For
these reasons, these experiments are at best qualitative
verifications of LPM suppression.

\subsection{Early Accelerator Experiments}

By the 1970's, 40 GeV electron beams were available at Serpukhov, and
the first accelerator based test of LPM suppression was done there
(Varfolomeev \etal, 1975).  Bremsstrahlung photons from 40 GeV
electron beams striking dense targets were detected in an sodium
iodide calorimeter.  The electrons were magnetically bent away from
the calorimeter.  Photons with 20 MeV $<k<$ 70 MeV emitted from
carbon, aluminum, lead and tungsten targets were studied.  Below 20
MeV, synchrotron radiation from the bending magnets dominated the
measurement, while above 70 MeV, the experiment was insensitive to the
spectral change predicted by Migdal.

The experiment suffered from several limitations.  Because the
electron beamline was in air, and included several scintillation
counters used as triggers, there was a significant light element
bremsstrahlung background.  The experimenters also mention a
significant background due to muon contamination in their beam.

The collaboration presented their data in terms of ratios of photon
spectra: lead/aluminum and tungsten/carbon.  This may have been done
to account for events containing several bremsstrahlung photons from a
single electron.  The data showed suppression in the region that
Migdal predicted.  However, the degree of suppression was larger than
Migdal predicted, although within the large errors.

CERN NA-43 was an experiment dedicated to studying channeling
radiation from electrons and positrons in crystals (Bak \etal, 1988).
In channeling, electrons or positrons travel along the crystal rows,
and hence are strongly affected by the coherently adding fields from
the atom rows.  At large angles to the axes, the coherence disappears,
and normal bremsstrahlung occurs.  In this large angle regime, they
saw suppression which they attributed to LPM suppression.  They also
observed suppression consistent with multiple scattering for electrons
incident along one of the crystal rows.  This is slightly different
from suppression in an amorphous material; here the suppression is
really the loss of coherent enhancement.

The first experimental studies of dielectric suppression were part of
a larger study of transition radiation (Arutyunyan, Nazaryan and
Frangyan, 1971).  Although the experiment focused on studies of
emission from stacks of thin radiators, the experimenters also
measured the radiation from 0.25 and 2.8 GeV electrons traversing
relatively thick glass and aluminum targets.  Although few details
were given, they appeared to observe suppression in the expected
energy range.

\subsection{SLAC E-146}

In 1992, the E-146 collaboration at Stanford Linear Accelerator Center
(SLAC) proposed an experiment to perform a precision measurement of
LPM suppression, and to study dielectric suppression (Cavalli-Sforza
\etal, 1992).  The experiment was conceptually similar to the
Serpukhov experiment, but heavily optimized to minimize background.
To minimize the statistical errors, the experimenters collected a
very large data set.  The experiment was approved in December, 1992,
and took data in March-April, 1993.

\subsubsection{Experimental Setup}

Figure\ \ref{e146layout} shows a diagram of the experiment.  An 8 or
25 GeV electron beam entered SLAC End Station A and passed through a
thin target.  The targets used are listed in Table III.  The beam was
then bent downward by a 3.25 T-m dipole magnet, through six wire
chamber planes which measured electron momenta and into an array of
lead glass blocks which accurately counted electrons.  Produced
photons continued downstream 50 meters into a BGO calorimeter array.
To minimize backgrounds, the electron path visible to the calorimeter
and the photon flight path were kept in vacuum.

{\center{\epsfig{file=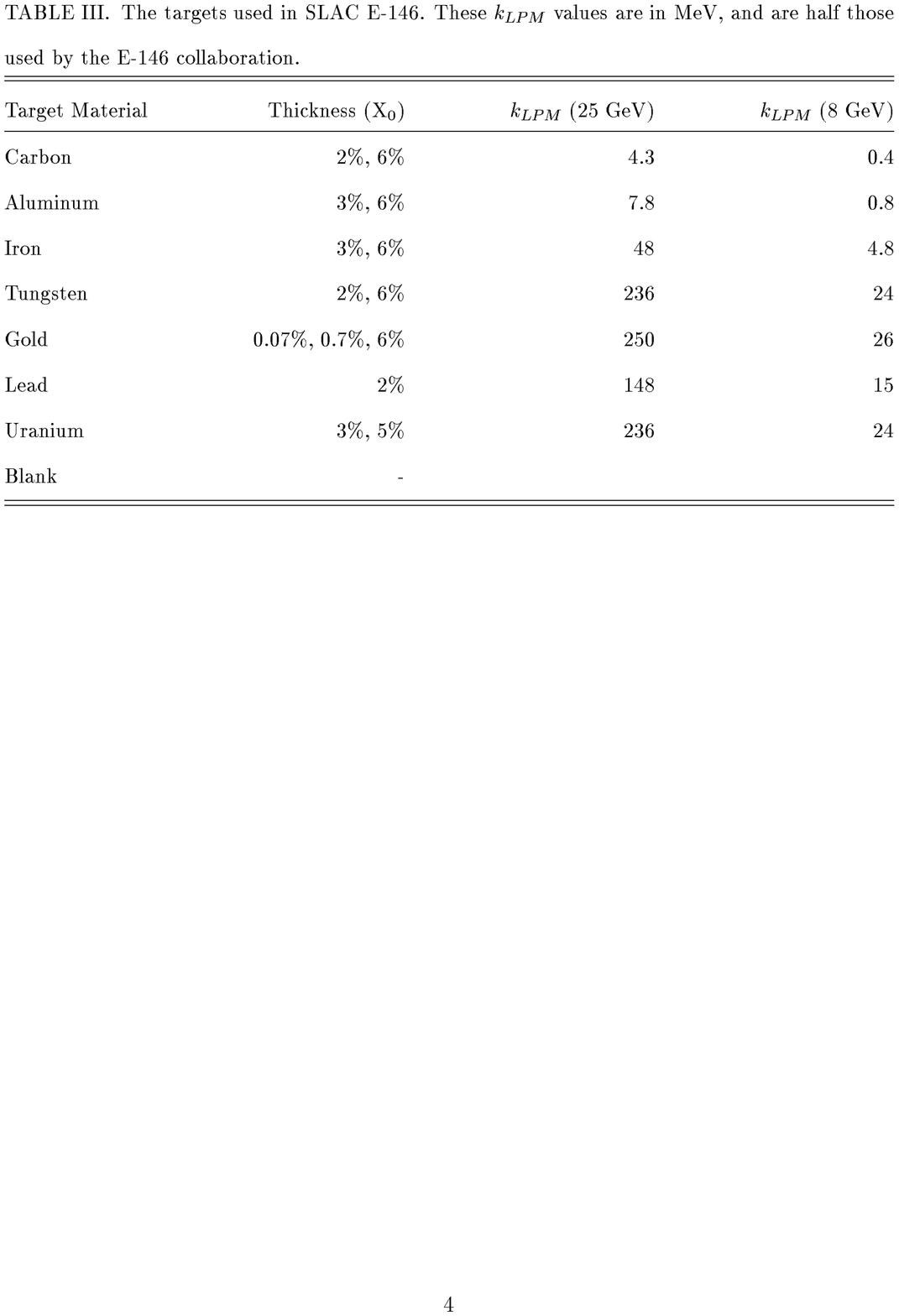,width=6.5in,%
bbllx=67,bblly=431,bburx=545,bbury=730,clip=}}}

\begin{figure}
\centering
\epsfig{file=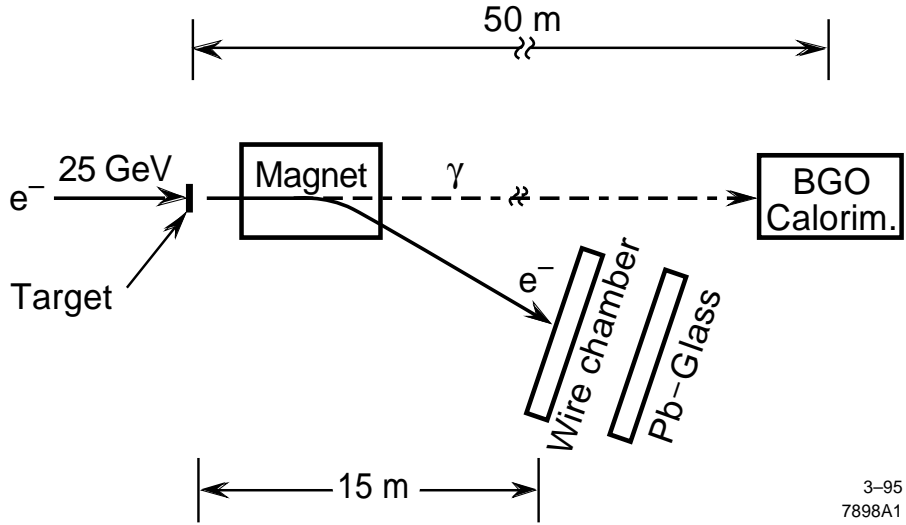,height=3in,width=5in,%
bbllx=145,bblly=310,bburx=420,bbury=470,%
clip=,angle=0}
\caption[]{Diagram of the SLAC-E-146 apparatus. From Anthony \etal\ (1995).}
\end{figure}

The calorimeter comprised 45 BGO crystals in a 7 by 7 array with
missing corners; each crystal was 2 cm square by 20 cm (18~X$_0$)
deep.  This segmentation provided excellent spatial resolution for
separating synchrotron radiation from bremsstrahlung photons.
Scintillation light from each crystal was measured separately by a
photomultipler tube (PMT).  The light yield was about 1 detected
photoelectron per 30 keV, providing good statistics down to 200 keV.
The calorimeter resolution was about 8\% (FWHM) at 100 MeV, with a
nonlinearity of about 3\%. The calorimeter temperature was monitored
throughout the experiment, and the data corrected using the measured
drifts.

The collaboration used several methods to calibrate the calorimeter,
to obtain both an absolute energy calibration and a crystal to crystal
intercalibration.  The primary tools for measuring the relative gain
were nearly vertical cosmic-ray muons, selected by a plastic scintillator
paddle trigger; the gain in each crystal channel was adjusted to
produce equal signals.
 
The absolute energy scale of the calorimeter was primarily determined
using a 500 MeV electron beam.  This calibration was checked with data
by comparing the electron energy loss, measured by the wire chamber,
with the calorimeter energy measurement.  Because of the steeply
falling photon spectrum and the non-Gaussian errors in the momentum
measurement, this was useful only as a cross-check.
 
Because of the large bremsstrahlung cross section, the experiment
required a beam intensity of about 1 electron per pulse.  Because it
would have been very uneconomical to use the SLAC linac to produce a
single electron per pulse, the collaboration developed a method to run
parasitically during SLAC linear collider (SLC) operations, by using the
off-axis electrons and positrons that are removed from the beam by
scrapers in the beam switchyard (Cavalli-Sforza \etal, 1994).

Normally, about 10\% of the SLC beam is scraped away by collimators in
the last 200 meters of the linac.  The collimators are only 2.2 X$_0$
thick, so a usable flux of high energy photons emerged from their back
and sides.  Some of these photons travel down the beampipe, past the
magnets that bend the electrons and positrons into the SLC arcs, and
into the beam switchyard, where a 0.7 $X_0$ target converted them into
$e^+e^-$ pairs.  Some of the produced electrons were captured by the
transport line optics, collimated, selected for energy, and
transported into the end station.
 
The beam worked well, with the size, divergence and yield matching
simulations.  At 8 and 25 GeV, the beam intensity was adjustable up to
about 100 electrons/pulse.  At 1 electron/pulse, the beam emittance
was limited by the optics, with a typical momentum bite of $\Delta p/p
< 0.5\%$.The beam optics were adjusted to minimize the photon spot
size at the calorimeter; spot sizes there were typically a few mm in
diameter. The beam spot was stable enough and small enough that beam
motion was not a major source of error.

Data was collected and written to tape on every beam pulse (120 Hz).
With the beam intensity averaging 1 electron/pulse, over 500,000
single electron events could be collected per eight hour shift.  The
experiment ran for a month, and good statistics were obtained with a
variety of targets.

\subsubsection{Data Analysis and Results}
 
The E-146 analysis selected events containing a single electron, as
counted by the lead glass blocks.  The photon energy was found by
summing the energies of hit BGO crystals using a cluster-finding
algorithm.  The cluster finding reduced the noise level by eliminating
random noise hits.

The experiment studied the photon energy range from 200 keV to 500
MeV, a 2500:1 dynamic range.  This exceeded the linear dynamic range
of the PMTs and electronics, so data was taken with two different PMT
gains, varied by changing the PMT high voltage.  The high-$k$ running
corresponded to 1 ADC count per 100 keV and the low-$k$ running was 1
ADC count per 13 keV, with the relative gains calibrated with the
cosmic ray data.  The high-$k$ data was used for 5 MeV $<k<$ 500 MeV
while the low-$k$ data covered 200 keV $<k< $ 20 MeV, with a weighted
average used in the overlap region.

These two sets of data differed in several significant ways.  There
were large differences in calorimeter behavior and background levels,
as well as the physics topics.  For $k> 10$~MeV, photons largely
interacted by pair conversion, producing an electromagnetic shower.
Showers typically deposited energy in 3-20 crystals in the
calorimeter.  For $k< 2$ MeV, the photons interacted by single or
multiple Compton scattering.  Usually, Compton scattering deposited
energy in a single calorimeter crystal.  Sometimes, the photon Compton
scattered once and then escaped from the calorimeter, taking some
energy with it.  This added a low energy tail to the energy deposition
curve.  While the high-$k$ data had very low backgrounds, the low-$k$
data had significant backgrounds due to synchrotron radiation, at
least for the 25 GeV beams.  Finally, the two data sets emphasize
different physics, with the high-$k$ data most relevant for LPM
suppression, with the low-$k$ data more useful for studying dielectric
suppression.  For these reasons, the two sets of data were analyzed
quite independently, and combined in the final histograms.

\subsubsection{Backgrounds and Monte Carlo}
 
One advance introduced by E-146 was the use of a detailed, high
statistics Monte Carlo. The main purpose of the Monte Carlo was to
understand multi-photon pileup.  This occurred when a single electron
passing through the target interacted twice, radiating two photons.
The Monte Carlo also simulated photon absorption in the target
(Anthony \etal, 1997) and modelled the detector.  Transition radiation
was treated as an integral part of the physics, rather than a
background.  In addition to conventional transition radiation, the
predictions of Ternovskii and Pafomov were included as options.

The Monte Carlo tracked electrons through the material in small steps,
allowing for the possibility of bremsstrahlung in the material and
transition radiation at each edge.  LPM suppression was implemented
using simple formulae (Stanev \etal, 1982) with dielectric suppression
incorporated using Eq.~(\ref{esigmamigdiel}). For consistency, the
Bethe-Heitler cross sections were included by turning LPM suppression
off from Migdal's formulae, rather than using a more modern formula.

The expected and measured backgrounds were both small.  The major
background was synchrotron radiation from the spectrometer magnet.
Synchrotron radiation was significant for $k<1$ MeV in the 25 GeV
data.  Because the magnet bent the beam downward, the synchrotron
radiation painted a stripe in the calorimeter, extending downward from
the center.  Because of the large lever arm, and because the bending
started in the fringe field of the magnet, where the field was low,
the synchrotron radiation was small near the center of the
calorimeter. Because of the good spatial resolution of the
calorimeter, most of the synchrotron radiation was removed with a cut
on the photon position in the calorimeter; photons within $45^o$ of a
line downward from the calorimeter midpoint were removed; this cut
removed most of the synchrotron radiation, along with 25\% of the
signal.

Non-target related backgrounds were measured with target empty runs.
The backgrounds were typically 0.001 photons with $k>200$ keV per
electron, much less than the $\sim10 T/X_0$ bremsstrahlung photons
with $k>200$ keV per electron.  Target related backgrounds were
expected to be small; photonuclear interaction rates are small, and
the events are unlikely to appear in the E-146 analysis.
 
\subsubsection{Results}

Because of the high statistics and low background, the E-146 data
allowed for detailed tests of the theory; photon spectra could
be easily compared with different predictions.

\begin{figure}
\epsfig{file=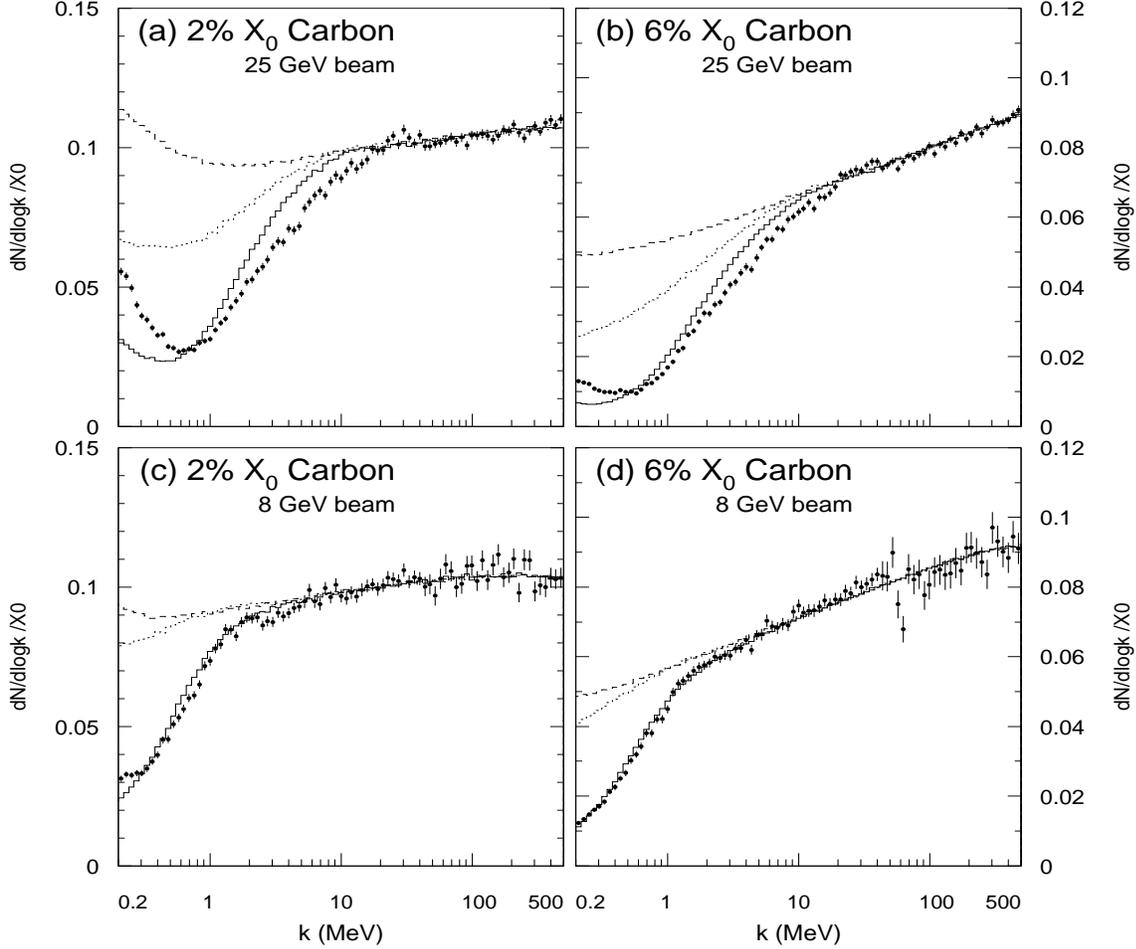,height=5in,width=6in,%
bbllx=30,bblly=165,bburx=570,bbury=670,%
clip=,angle=0}
\vskip .1 in
\caption[]{Comparison of data from SLAC-E-146 with MC predictions for
200 keV to 500 MeV photons from 8 and 25 GeV electrons passing through
2\% and 6\% $X_0$ carbon targets. The cross sections are given as
$dN/d(\ln k)/X_0$ where $N$ is the number of events per photon energy
bin per incident electron, for (a) 2\% \xo\ carbon and (b) 6\% \xo\
carbon targets in 25 GeV electron beams, while (c) shows the 2\% \xo\
carbon and (d) the 6\% \xo\ carbon target in an 8 GeV beam.  Three
Monte Carlo curves are shown.  The solid line includes LPM and
dielectric suppression of bremsstrahlung, plus conventional transition
radiation.  Also shown are the Bethe-Heitler plus transition radiation
MC (dashed line) and LPM suppression only plus transition radiation
(dotted line).  Adapted from Anthony \etal\ (1997).}
\label{carbon}
\end{figure}

\begin{figure}
\epsfig{file=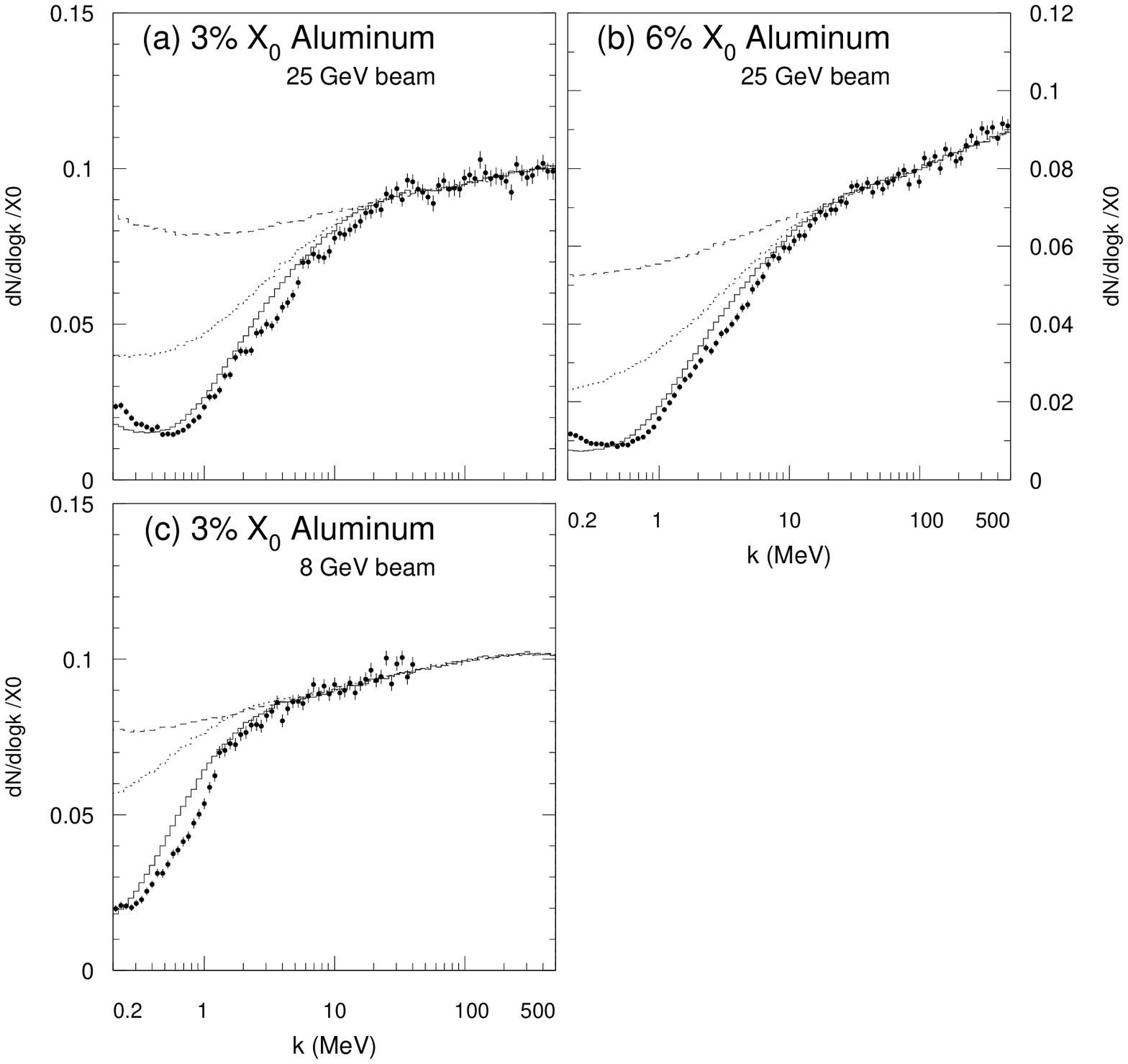,height=5in,width=6in,%
bbllx=30,bblly=165,bburx=570,bbury=670,%
clip=,angle=0}
\vskip .1 in
\caption[]{Comparison of data from SLAC-E-146 with Monte Carlo
predictions for 200 keV to 500 MeV photons from 8 and 25 GeV electrons
passing through 3\% and 6\% $X_0$ aluminum targets.  The format and
Monte Carlo curves are the same as in Fig. \ref{carbon}. Adapted from
Anthony \etal\ (1997).}
\label{aluminum}
\end{figure}

\begin{figure}
\epsfig{file=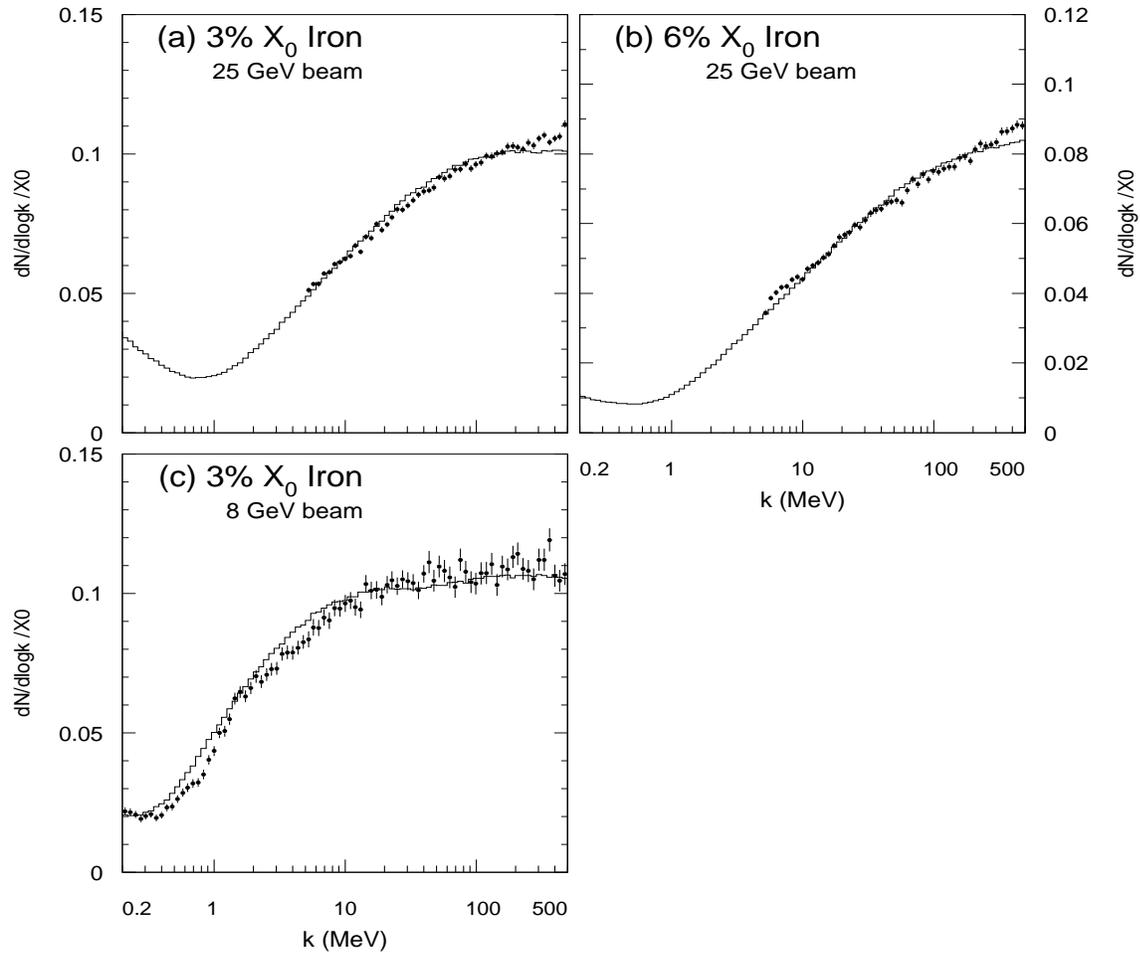,height=5in,width=6in,%
bbllx=30,bblly=165,bburx=570,bbury=670,%
clip=,angle=0}
\caption[]{SLAC-E-146 measurements and Monte Carlo predictions for 8 and
25 GeV electrons passing through a 3\% and 6\% $X_0$ iron targets.
The Monte Carlo curve is based on LPM and dielectric suppression, plus
conventional transition radiation. Adapted from Anthony \etal\ (1997);
Panel (c) is mislabelled as 6\% $X_0$ there.}
\label{iron}
\end{figure}

\begin{figure}
\epsfig{file=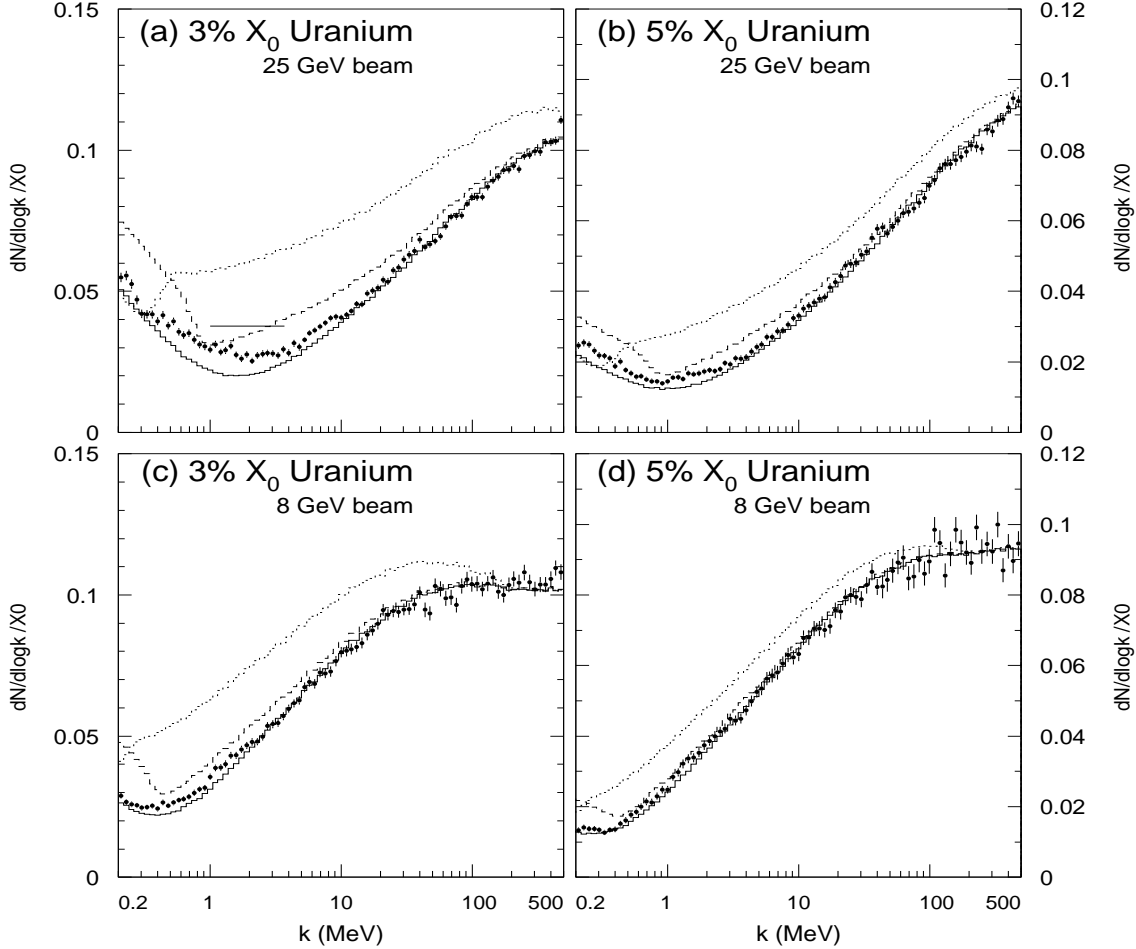,height=5in,width=6in,%
bbllx=30,bblly=165,bburx=570,bbury=670,%
clip=,angle=0}
\caption[]{SLAC E-146 measurements and Monte Carlo for 3\% \xo\ and 5\%
\xo\ uranium targets in 8 and 25 GeV electron beams.  The solid line
shows the LPM and dielectric suppression, conventional transition
radiation Monte Carlo prediction. The other lines include simulations
based on calculations of transition radiation due to Pafomov (dashed
line) and Ternovskii (dotted line), Eq. (\ref{fternovskii}), with
$\chi=1$.  The flat solid line in panel (a) is a calculation based on
Eq. (\ref{esigmashulga}).  Adapted from Anthony \etal\ (1997).}
\label{uranium}
\end{figure}

\begin{figure}
\epsfig{file=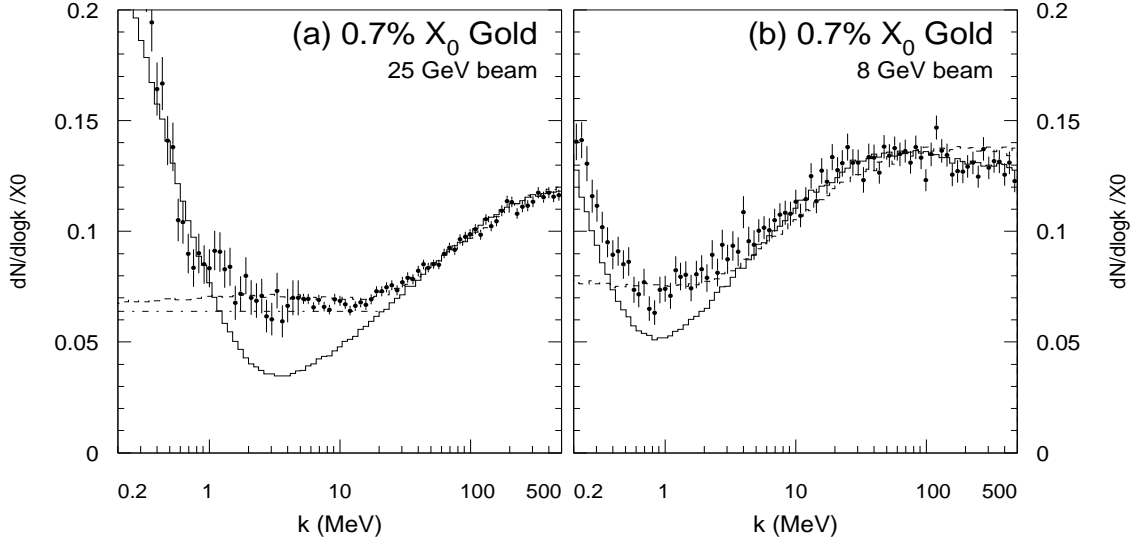,height=3.in,width=6in,%
bbllx=30,bblly=160,bburx=572,bbury=446,%
clip=,angle=0}
\caption[]{SLAC E-146 data on 8 and 25 GeV electrons hitting a
0.7\%~$X_0$ gold target.  Shown are calculations by Blankenbecler and
Drell (dashed line), and Shul'ga and Fomin (dot-dashed line). For
comparison, the Migdal MC is shown as the usual solid line.  Adapted
from Anthony \etal\ (1997).}
\label{gold}
\end{figure}

Figures \ref{carbon}-\ref{gold} show a sampling of E-146 results.
Photon energies were histogrammed logarithmically, using 25 bins per
decade of energy, so that each bin had a fractional width $\Delta
k\sim 10\% k$. The logarithmic scale is needed to cover the 2500:1
energy range.  The logarithmic binning $dN/d\ln k = kdN/dk$ was chosen
so that a $1/k$ Bethe-Heitler (BH) spectrum will have an equal number
of events in each bin, simplifying the presentation and statistical
analysis.

Figure\ \ref{carbon} shows the photon spectrum (points with error
bars) from 8 and 25 GeV electrons passing through 2\% and 6\% $X_0$
carbon targets. Also shown are 3 Monte Carlo (MC) histograms.  The top
histogram (dashed line) is a simulation of BH bremsstrahlung plus
conventional transition radiation; the transition radiation is
substantial below $k_p$, 1.4 (0.4) MeV at 25 (8) GeV.  Above $k_p$,
the spectrum is sloped because there is a finite probability of a
single electron interacting twice while passing through the target.
Because the calorimeter cannot separate single photons from multiple
hits, but instead measures total energy deposition, this depletes the
low energy end of the spectrum (shown here), while increasing the
number of calorimeter overflows.  In the absence of multiple
interactions, the Bethe-Heitler spectrum would be flat at $(1/X_0)
dN/d\ln k =4/3\ln{(k_{max}/k_{min})}=0.129$ for bins with logarithmic
widths $k_{max}/k_{min}= 10^{1/25}=1.096$.  The bin heights directly
scale with the bin fractional width $\Delta k/k$.

The dotted histogram is a simulation that includes LPM suppression,
but not dielectric suppression, plus conventional transition
radiation.  The solid line includes LPM and dielectric suppression,
along with conventional transition radiation.  This was 
the 'standard' E-146 choice for simulation.

Both suppression mechanisms are required to approach the data.
However, there are still significant discrepancies between the LPM
plus dielectric MC and the data.  Below 800 keV (350 keV) for 25 (8)
GeV beams, the upturn in the data may be residual background,
especially synchrotron radiation.  The difference at higher photon
energies is more complex, and will be discussed in the following
subsection.

Figure\ \ref{aluminum} shows the spectrum from 8 and 25 GeV electrons
passing through 3\% and 6\% $X_0$ aluminum targets. The same three
simulations are shown.  Because aluminum has twice the $Z$ of carbon,
LPM suppression is considerably enhanced, with $k_{LPM}$ 8 MeV and 800
keV at 25 and 8 GeV respectively.  Because the density is similar to
carbon, dielectric suppression is similar.  Because dielectric
suppression dominates for $k<k_{cr}$, the curve for both suppressions
is similar to that of carbon.  The agreement between the data and the
standard curve is better than with carbon.

Figure\ \ref{iron} shows the spectrum from 8 and 25 GeV electrons
passing through 3\% and 6\% $X_0$ iron targets, with just the
'standard' MC.  The general slope of the data matches the simulation,
but the behavior at higher $k$ for 25 GeV beams is quite different.

Figure\ \ref{uranium} shows the bremsstrahlung spectra in uranium
targets.  Uranium is dense enough that LPM suppression is dominant,
and the E-146 collaboration compared simulations with dielectric and
LPM suppression, plus conventional transition radiation (TR), or the
calculations of Ternovskii (1960) or Pafomov (1964) of transition
radiation due to multiple scattering.

The Pafomov curve jumps at $k=k_{cr}$, around 800 keV (400 keV) at 25
(8) GeV, corresponding from the switch from Eq.\ (\ref{pafamovc}) to a
numerical approximation for Eqs. (\ref{pafomova}) and
(\ref{pafomovb}).  For $k<k_{cr}$, Pafomov is considerably above the
data and the conventional TR curve.  Above the jump, the curve shows a
reasonable trend, but the TR appears to be several times too high.

The Ternovskii curve also jumps, at $sk_p^2/k^2=1$, about 500 keV
(below 200 keV) for the 25 (8) GeV data. Below the jump, Ternovskii
matches conventional TR.  Above it, Ternovskii is quite far above the
data.  Moreover, it extends to too high an energy, above $k_{LPM}$.
The Ternovskii radiation could be reduced by lowering $\chi$ below 1
in Eq.\ (\ref{fternovskii}).  However, a considerable adjustment would
be required.

Figure\ \ref{gold} shows the spectrum from 8 and 25 GeV electrons
passing through a 0.7\% $X_0$ gold target.  The solid histogram is the
standard simulation.  This target is especially interesting because,
at 25 GeV, for $k< 7$ MeV, $T < l_{f0}$.  Including LPM suppression,
$T<l_f$ for $k<3$ MeV, and the target should interact as a single
unit. The BH $1/k$ spectrum should be recovered, albeit at a reduced
intensity. The dot-dashed line shows a calculation based on
Eq. (\ref{esigmashulga}), in good agreement with the flattening
observed in the data.  For the E-146 2\% $X_0$ lead (not shown here)
and 3\% $X_0$ uranium 25 GeV data, Eq. (\ref{esigmashulga}) also
applies, but only for a very limited range of $k$.  However, as
Fig. \ref{uranium} shows, for 3\% $X_0$ uranium, Eq.\
(\ref{esigmashulga}) predicts $dN/d(\ln k)=0.038$ for 1.0 MeV $<k<$
3.7 MeV, considerably above the data. The lead data shows a similar
discrepancy.  Of course, these discrepancies may be because the
equation is does not apply so close to $T=l_f$.

Fig. \ref{gold} also shows the results of a calculation by
Blankenbecler and Drell.  Neither prediction has been corrected for
multiple interactions in the target or detector resolution.  Because
neither calculation includes dielectric suppression or transition
radiation, they fail for $k<k_{cr}$.

\begin{figure}
\epsfig{file=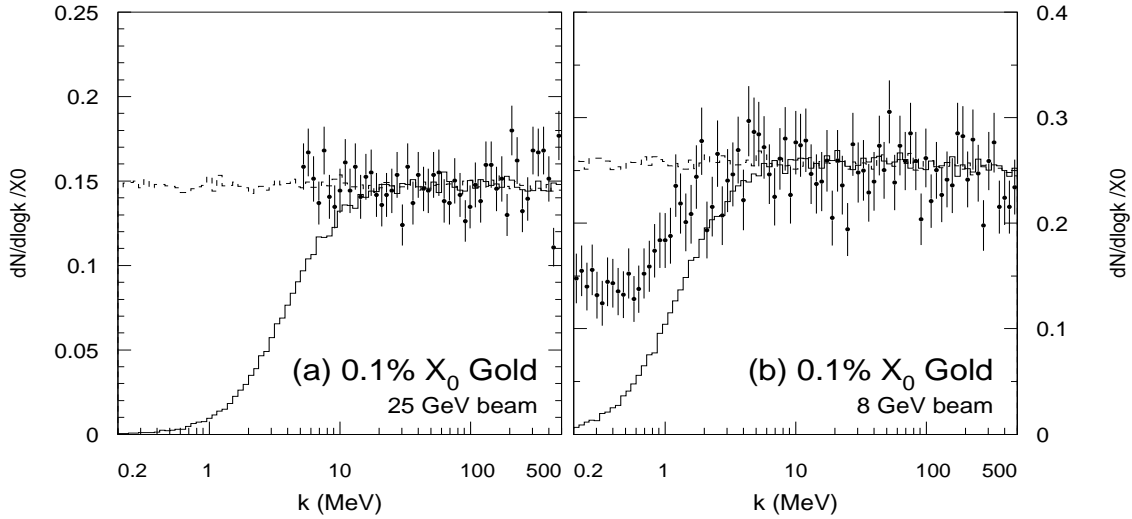,height=3in,width=6in,%
bbllx=30,bblly=160,bburx=572,bbury=460,%
clip=,angle=0}
\caption[]{Measurements and Monte Carlo for a 0.1\%~\xo\ gold target at
(a) 25 GeV and (b) 8 GeV.  The dashed line is the BH prediction (no
suppression), while the solid line is a Monte Carlo which includes
dielectric suppression, but not LPM suppression. Because the very thin
target should exhibit little transition radiation, no transition
radiation is included in these Monte Carlos. From Anthony \etal\
(1997).}
\label{thingold}
\end{figure}

Fig. \ref{thingold} shows the spectrum from 8 and 25 GeV electrons
hitting a 0.07\% $X_0$ target.  This target is only about $1.2\times
(X_0/1720)$ thick, so that multiple scattering should produce little
suppression, and the Bethe-Heitler simulation (dashed histogram)
should be a good match to the data.  Dielectric suppression should
also be small, because the total phase shift in the target,
$\hbar^2\omega_p^2 T/k^2 l_f$, is small for $k> 500$ keV.  Transition
radiation should also be strongly suppressed, by $(T/l_f)^2$.  Without
this suppression, transition radiation would completely dominate the
data.  As expected, the 25 GeV electron data matches the Bethe-Heitler
spectrum.  However, the 8 GeV data drops off for $k<2$ MeV.  For
comparison, a dielectric suppression only Monte Carlo simulation is
shown.  The data is considerably above this curve.

Blankenbecler and Drell (Blankenbecler, 1997c) predict that, at 25
GeV, emission is suppressed by 10\% compared to BH for k= 500 MeV,
rising to 13\% at k=100 MeV.  At 8 GeV, the difference is only a few
percent.  Unfortunately, these predictions differ from Bethe-Heitler
by less than the experimental errors.

For a target this thin, the collaboration has noted that the signal is
very small and the potential backgrounds are large.  Furthermore, the
target thickness and overall normalization between the signal and
simulations cannot be well determined.

In all of these these plots, the Monte Carlo curves were normalized to
the data by multiplication by a constant adjustment, chosen so the MC
best matches the data above 20 MeV (2 MeV) at 25 (8) GeV.  The
thresholds were chosen to avoid thin target corrections for $T <
l_{f0}$, backgrounds and transition radiation; for the 0.7\% $X_0$
target, higher limits were chosen, 30 (10) MeV at 25 (8) GeV. Overall,
the standard Monte Carlo curves had to be scaled up by an average of
5\% (2$\sigma$) to match the data.  This discrepancy would likely
disappear with an input cross section where the onset of suppression
was more gradual around $k\sim k_{LPM}$.

The errors shown on the plots are statistical only.  The E-146
collaboration has carefully studied the systematic errors on these
measurements.  The point-to-point systematic errors vary slowly with
$k$ and correspond to a 4.6\% uncertainty for $k>5$ MeV.  Below 5~MeV,
the systematic errors rise to 9\%, because of increased uncertainties
in photon energy cluster finding as Compton scattering takes over from
showering as the dominant energy loss.  The $\pm 3.5\%$ systematic
error on the normalization was determined separately.

\subsection{Bulk versus Surface Effects and Conclusions}

Although the data clearly demonstrates LPM and dielectric suppression,
the thinnest targets show excess radiation over the bulk LPM
predictions.  This can easily be ascribed to surface radiation.  As
long as interference between the two target edges is negligible, the
surface and bulk effects can be separated in a model independent way
by subtracting spectra from targets made of the same material, but
different thicknesses.

\begin{figure}
\epsfig{file=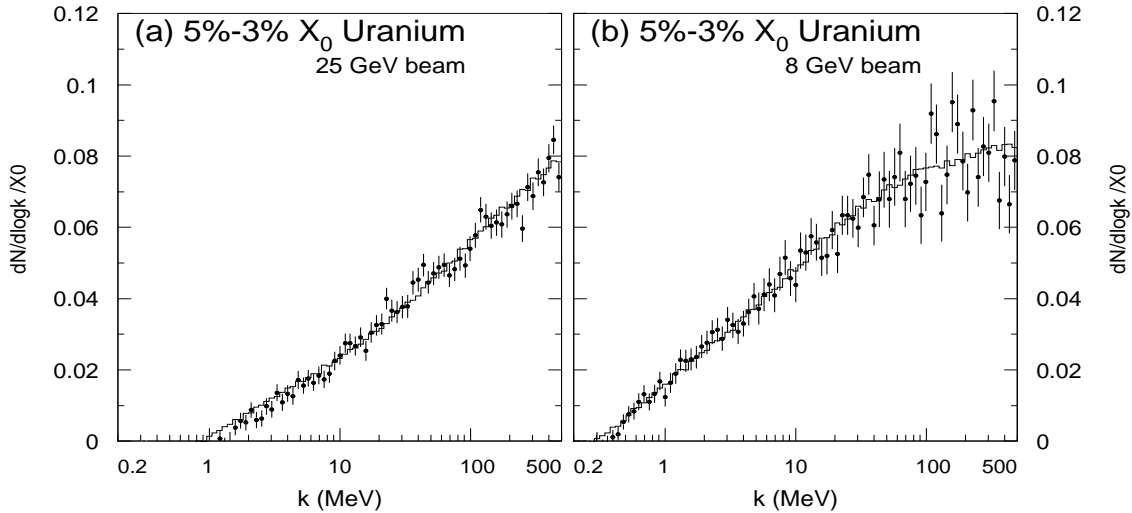,height=3in,width=6in,%
bbllx=30,bblly=160,bburx=572,bbury=460,%
clip=,angle=0}
\caption[]{Bin by bin subtraction of the 3\% $X_0$ uranium data from
the 5\% $X_0$ data by SLAC E-146, for 25 and 8 GeV beams.  The solid
line is the result when the same procedure was applied to their
standard Monte Carlo. From Anthony \etal\ (1997).}
\label{subtract}
\end{figure}

Figure \ref{subtract} shows the results of this subtraction for
uranium.  Because the subtraction exacerbates the effects of multiple
interactions in the target, it is necessary to compare the subtracted
spectra with similarly treated simulations.  For all of the E-146
materials, the subtracted spectra and simulations agree more closely
than the unsubtracted spectra, and, for most targets, the agreement is
within the statistical and systematic errors.  The notable exceptions
are the 25 GeV iron and carbon data.

The improved agreement indicates that the discrepancies in the
unsubtracted spectra are due to surface effects.  For the denser
targets, this fits the expectations for transition radiation.
However, for the lighter carbon and iron targets, the unsubtracted
data indicates more suppression than expected.  This implies that
there is negative emission from the surfaces, an extremely unlikely
result.

The collaboration (Anthony \etal, 1997) considered a few explanations
for the discrepancies.  For carbon, the increased suppression could be
due to the crystalline structure of the pyrolitic graphite target.  If
the target density profile varied, then the higher density regions
would show more suppression, and the aggregate would show somewhat
higher LPM and dielectric suppression.  For iron, magnetization of
individual domains could produce some magnetic suppression.
Unfortunately, the details depend on the (unknown) domain structure.

The density profile could also vary if the target surfaces were
oxidized.  For the $2\% X_0$ tungsten target, the collaboration saw
some evidence for this; thickness measurements based on weighing
(giving results in $g/cm^2$) were lower than those obtained with
calipers; an oxide layer could explain this.  The anomalous
normalization on the 2\% tungsten target could also be explained by an
oxide layer. For the lighter targets, a surface layer could explain
the shapes of the curves, but the required thickness seems very
unrealistic, and, in any case, the presence of such a layer is not
supported by the normalization and thickness measurements.

It is also possible that Migdal's theory may be inadequate for lighter
targets.  However, Zakharov (1998b) also found a very poor fit for the
E-146 carbon data, despite a careful treatment of inelastic
electron-electron interactions.  In comparison, his fits to the other
data were quite good.  It is possible, although probably unlikely,
that a different treatment, would find a better fit.  It is also
possible that a calculation taking into account the chemical structure
of the pyrolitic graphite and using screening based on the actual
electron density might better fit the data.

The subtraction procedure could be modified to measure the transition
radiation spectrum from a single surface.  However, the errors are
slightly too large for the result to be interesting.

Overall, the E-146 heavy target data is in good agreement with most of
the recent treatments of LPM suppression.  However, the shape of the
spectra from the lightest targets remain a bit of a mystery.

\section{LPM effect in Plasmas}
\label{splasmas}

So far, we have considered particles moving through a medium.
However, in a plasma, there is no separate incident particle, and each
particle must be treated on an even footing.  If the medium is
sufficiently dense, there can be suppression even for non-relativistic
particles.  Then, the formation time $t_f$ is easier to use than the
formation length.  If $t_f$ is longer than the mean time between
collisions $t_c=1/\Gamma$, $\Gamma$ being the collision rate, then
emission can be suppressed.

One hadronic example of a plasma is a supernova; with the high
temperature and density, $t_c$ is very short.  For reactions with
$t_f>t_c$, suppression may be present.  One interesting reaction is
the production of right handed neutrinos or axions through
$NN\rightarrow NN\nu\overline\nu$, $nn\rightarrow npe\overline\nu_e$
and $NN\rightarrow NNa$.  If these hypothetical particles exist and
are produced, they will carry energy away from the explosion,
increasing the cooling rate.  The measured cooling rate has been used
to put limits on these particles (Raffelt and Seckel, 1991).  

For axions or neutrinos, $t_f = \hbar/\Sigma E$, where $\Sigma E$ is
the sum of the neutrino energies or the axion energy.  When $t_c \ll
t_f$, then 'free' collisions are rare.  Instead, the interacting
nucleons are excited, This can be modelled by giving the nucleon an
effective mass, as may be done with photons in dielectric suppression.
Two complications come from the possibility of back reactions, and
from the degeneracy of the incoming particles.  The suppression is the
ratio of the emission $Q(\Gamma)$ to $Q(\Gamma=0)$:
\begin{equation}
S = {t_f \over t_{f0}} = 
\bigg\langle {\Sigma E^{n+2} \over \Sigma E^2 + \Gamma^2/4} \bigg\rangle 
\langle{1 \over \Sigma E^n}\rangle.
\label{sraffelt}
\end{equation}
Here, $n$ accounts for the multiplicity of the emitted particles, with
$n=2$ for axions and $n=4$ for neutrino pairs; for axions this
equation matches Eq.\ (\ref{sdiel}). Because $\Gamma$ depends on the
local density; finding the overall suppression factor for an entire
supernova requires detailed modelling.  However, Raffelt and Seckel
estimate that $S$ could be as small as 0.1.  So, far fewer axions and
neutrinos are emitted than if suppression were absent, and earlier
limits that neglect this reduction are invalid (Raffelt and Seckel,
1991).

The general solution for an non-equilibrium electromagnetic plasma has
been elegantly formulated using non-equilibrium quantum field theory
(Knoll and Voskresensky, 1995, 1996).  By careful classification of
diagrams, and appropriate resumnation, Knoll and Voskresensky avoided
infrared divergences, and reproduced both the classical
(non-suppressed) and quasi-particle (low density) limits by
appropriate choice of subsets of graphs.  In a dense plasma, the
quasi-free scattering approximation breaks down, and the reaction rate
is reduced by
\begin{equation}
S = { k^2 \over k^2 + \Gamma^2 }.
\label{sknoll}
\end{equation}
This equation also matches Eq. (\ref{sdiel}), with
$\gamma\hbar\omega_p$ of the medium replaced by $\Gamma$, the
relaxation rate of the source.

A plasma like this might surround a quantum black hole.  Before a
quantum black hole explodes, it emits a huge flux of charged
particles. This radiation forms a nearly thermal photosphere,
consisting of electrons, positrons and photons (Heckler, 1995).  The
charged particle emission rate and consequent density is high enough
that bremsstrahlung and pair production should be suppressed in this
plasma.  Since most of the emitted particles are hadrons, the black
hole may also generate a dense color plasma, akin to a quark gluon
plasma, with interactions suppressed as discussed in the following
section.

Similar plasma effects probably occurred during the big bang.
Unfortunately, we do not know of any calculations involving it.

\section{Electromagnetic Showers}
\label{sshowers}

Although the LPM effect is best studied using single interactions in
thin targets, most real world situations involve electromagnetic
showers in thick targets.  This includes natural processes like cosmic
ray air showers and neutrino induced electromagnetic showers, and
showers in man-made detectors.  Although modelling showers with
suppression effects requires complex analytic calculations or Monte
Carlo simulations, this section will present some simple calculations
that show when suppression effects can be important.

Suppression affects showers in several ways.  Besides the obvious
elongation when $E>E_{LPM}$ and the radiation length increases, the
shower changes form, with the low energy 'fuzz' disappearing, and
shower to shower fluctuations become much more important, because the
number of interactions drops greatly.

\subsection{Natural Showers}

Many calculations have concerned high energy shower development in
water or ice, an area relevant to high energy neutrino astronomy, or
in air, an application motivated by extremely high energy (EHE) cosmic
ray air showers. The latter case is complicated because the air
density varies exponentially with altitude.

Water and ice showers are of interest for detecting very high energy
astrophysical $\nu_e$ through resonant $W$ production, $\nu_e
e\rightarrow W\rightarrow e\nu_e$ and similar reactions.  This cross
section rises rapidly at the W pole, corresponding to a $\nu_e$
initial energy of 6.4 PeV, an average of 2.1 PeV of which goes to the
outgoing electron.  Since 2.1 PeV$\sim 7 E_{LPM}$, LPM suppression is
significant. As Fig. \ref{suppressionvse} shows, the electron
radiation length is increased by about 35\%.  Since $E\sim 4E_p$, pair
creation will also cause suppression for a range of $k$.

At similar energies, the search for $\nu_\tau N\rightarrow\tau
N\rightarrow e\nu_\tau\overline\nu_eN$ events (Learned and Pakvasa,
1995) in ice requires distinguishing the $\tau$ track from the
beginning of an electron shower; when suppression mechanisms reduce
the number of low energy bremsstrahlung photons, this separation
becomes more difficult.

Because both applications involve very high energy showers, direct
simulations have been limited by the available computer power; until
recently at least partially analytic calculations were needed.  These
analytic methods owe much of their history to earlier analytic calculations
of Bethe-Heitler shower development.

One of the first analytic calculations (Pomanskii, 1970) showed that
the penetrating power of electromagnetic showers rises as the LPM
effect becomes important.  At $10^{19}$-$10^{20}$ eV in earth,
electrons and photons become as penetrating as muons.  Misaki (1990)
used the matrix method to show that EM showers above $10^{15}$ eV in
water are elongated by the LPM effect.  In the tails of the shower, at
a given sampling depth, the LPM effect roughly triples the density.

Unfortunately, while analytic calculations can determine the average
shower shape, they have limited value for understanding shower to
shower variations; for this, simulations are needed.  To reduce the
computing load, hybrid Monte Carlos are often used.  A hybrid MC
simulates the initial shower development; shower tails are added on
based on a library of simulated complete showers. This reduces the
computational requirements significantly.  Stanev and collaborators
(1982) measured shower elongation in water and lead due to the LPM
effect using a hybrid MC.

Konishi \etal\ (1991) used a hybrid MC to study fluctuations in a
regime where suppression is very strong: $10^{17}$ eV showers in lead.
In contrast to BH showers, LPM showers showed large shower to shower
variations.  By eliminating soft bremsstrahlung, the LPM effect
greatly reduces the number of interactions per radiation length, so
the shower development depends on far fewer interactions, greatly
increasing the shower to shower variation.  The variation complicates
energy measurement, especially for detectors with limited
sampling. Misaki (1993), quantified the degree of fluctuation,
measuring the distribution of the depth of shower maximum, the depth
at which the number of shower particles is a maximum.  For $10^{17}$
eV showers in rock ($E_{LPM}=77$ TeV), he found that shower maximum
occurs at $18\pm13$ (FWHM) $X_0$ for BH showers, compared with
$152\pm176$ $X_0$ for LPM showers. Not only is the maximum deeper, but
its position varies much more.

Not considered here is the increased angular spreading caused by
suppression.  Although it may be not affect the general
shower shape, it is probably important in understanding radio
waves produced by showers in ice (Zas, 1997), where the spectrum
depends on the transverse separation between particles.

\subsection{Cosmic Ray Air Showers}

Cosmic ray air showers occur when EHE cosmic rays hit the earths
atmosphere and interacts to produce a cascade of particles (Sokolsky,
Sommers and Dawson, 1992). Showers up to $3\times10^{20}$ eV have been
seen (Bird \etal, 1994). Two techniques are used to study the highest
energy air showers. Large aperture telescopes, like the Flys Eye (Bird
\etal, 1994) observe the shower induced fluorescence of the $N_2$ in
air, measuring the shower development in the atmosphere.  Ground based
arrays of hundreds or thousands of small detectors, spaced up to a
kilometer apart, observe the remnants of the shower that reach the
ground (Takeda, 1998) (Auger, 1996).

Ground based detectors act as calorimeters with a single sampling
layer, located behind a 28 $X_0$, or 15 hadronic interaction lengths,
$\lambda$ thick atmospheric absorber.  For a vertical $10^{20}$ eV
shower, sea level is near shower maximum, so measurements are not too
sensitive to the position of the first interaction.  For obliquely
incident showers, the detector is considerably behind shower maximum,
and as a result, ground observations are subject to significant
fluctuations depending on the position of the first interaction.
Because of their limited ground coverage, most of their samples are
many Moliere radii from the shower core, in the tails of the angular
distribution.  This heightens the sensitivity of the detector to the
details of the initial interactions.

On the other hand, air fluorescence detectors measure the whole shower
profile, with the atmosphere acting as a fully active sampling medium.
They are thus less sensitive to the details of the shower development.

There has been considerable disagreement about whether the LPM effect
is important in ultra-high energy showers. Capdevielle and Attallah
(1992) found that LPM suppression had a significant effect on
$10^{19}$ to $10^{20}$ eV proton induced air showers. However,
Kalmykov, Ostapchenko and Pavlov (1995) found a much smaller change;
for a $10^{20}$ eV shower, they found a 5\% decrease in the number of
electrons at shower maximum, and a $15\pm 2$ g/cm$^2$ downward shift
in the position of shower maximum.  Although this shift is smaller
than the measurement resolution, it is systematic, and important in
studies of cosmic ray composition.  Kasahara (1996) found that the LPM
effect reduces the number of particles reaching the ground by about
10\% for a $5\times10^{20}$ eV proton shower.  None of these authors
considered the effect of other suppression mechanisms.

Although a complete Monte Carlo simulation is required to quantify the
effect of suppression, simple calculations can illustrate some
qualitative features of shower development (Klein, 1997).  The
calculations depend strongly on the identity of the incoming
particles, with protons (or neutrons) the most popular, although
photons or heavier nuclei cannot be excluded.  For heavy ions,
suppression is greatly reduced because of the lower per-nucleon
energy.  Because this model is very simple, the possibility of photons
pair converting in the earth's magnetic field will be neglected
(Kasahara, 1996)(Stanev and Vankov, 1997).  The probability depends on
the photon energy and the angle between the photon and the earths
magnetic field.  For primary photons with $k > 10^{20}$ eV, the
probability is large, partly because the magnetic field extends to
much higher altitudes than the atmosphere.

Because the atmospheric pressure, $1/X_0$ and $1/E_{LPM}$ decrease
exponentially with height, it is convenient to use an exponential
depth variable such as column density.  Column density is measured in
g/cm$^2$, with ground level, $A_0$ at 1030 g/cm$^2$.  Then,
$E_{LPM}=(A_0/A)$ 234 PeV, where $A$ is the depth in column density.
Similarly, $y_{die}=1.3\times10^{-6}(A/A_0)$ and $E_p=\sqrt{A_0/A}\
42$ PeV.  Because the atmosphere is much cooler at high altitudes,
these numbers underestimate suppression by about 25\%.  They also
neglect the fact that air is composed of diatomic $N_2$ and $O_2$
molecules; when an electromagnetic interaction involves one atom of
the molecule, the other atom will introduce additional multiple
scattering over that expected from a monoatomic gas. This could be a
significant effect, but it has yet to be studied.

Incoming photons react by pair production, while protons interact
hadronically.  A central hadronic collision will produce a shower of
several hundred pions; the neutral pions will decay to photons.  The
highest energy $\pi^0$ will have a rapidity near to the incoming
proton, and their decay photons will have energies up to
$2\times10^{19}$ eV.  Many diffractive processes, such as $\Delta$
production can also produce photons with similar energies.  Overall,
photons from central interactions will have an average energy of about
$2\times10^{17}$ eV.  Above $\sim 10^{18}$ eV, the finite $\pi^0$
lifetime becomes important (Kasahara, 1996); a $2\times10^{19}$ eV
$\pi^0$ travels about 5 km before decaying and is likely to interact
hadronically instead of decaying.  This will reduce the number of high
energy photons, reducing the significance of LPM suppression in the
earliest part of the shower development.

\begin{figure}
\epsfig{file=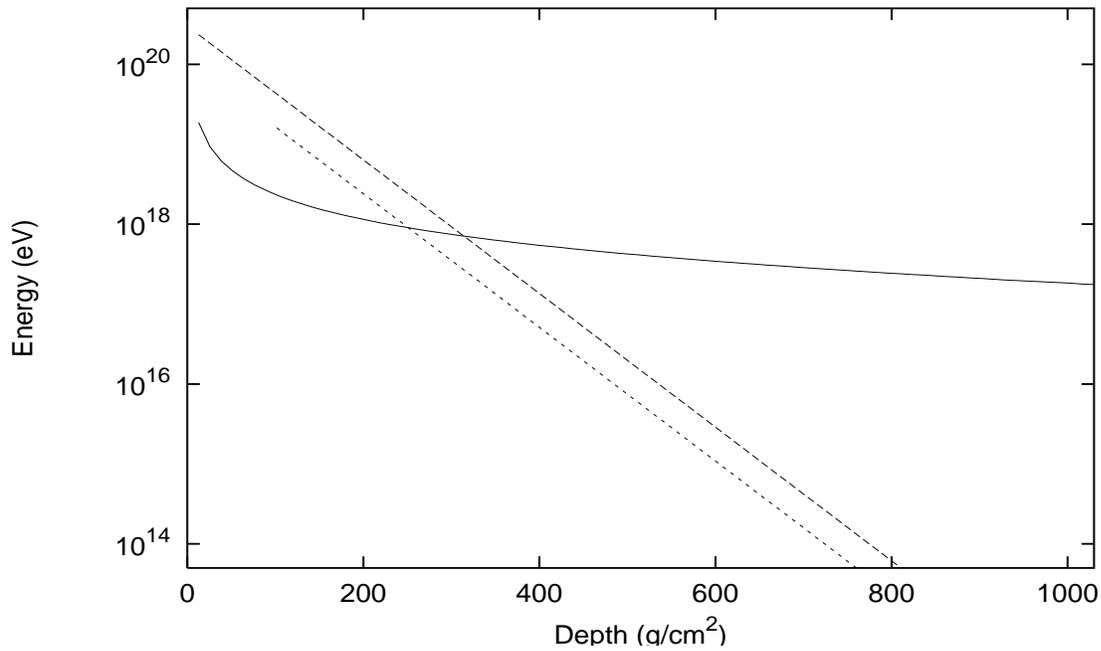,height=6in,width=3.5in,%
clip=,angle=270}
\vskip .1 in
\caption[]{$E_{LPM}$ (solid line), $\overline E$ for a
$3\times10^{20}$ eV photon shower (long dashes) and $\overline E$ for
a $2\times10^{19}$ eV photon created at $1\lambda$ (short dashes).
$\overline E$s is based on an idealized Bethe-Heitler shower where the
number of particles doubles each radiation length.  At altitudes where
$\overline E> E_{LPM}$, suppression is important; the ratio of the two
energies determines the degree of suppression. Here, a temperature
correction has been added to $E_{LPM}$.}
\label{airdepth}
\end{figure}

Because both $E_{LPM}$ and the average particle energy,
$\overline{E}$, decrease with depth, suppression mechanisms can
actually become stronger as a shower moves deeper in the atmosphere.
Fig.~\ref{airdepth} compares $E_{LPM}$ with $\overline{E}$ for an
idealized Bethe-Heitler shower from a $3\times10^{20}$ eV photon.  In
each successive radiation length, there are twice as many particles
with half the energy.  The dashed line shows a similar cascade, from a
$2\times10^{19}$ eV photon starting at $1\lambda$.  The
electromagnetic interactions at a given depth are determined by the
ratio $\overline{E}/E_{LPM}$.  For the photon shower, suppression is
largest around 75 g/cm$^2$, where $E\sim 40 E_{LPM}$.  Electron
$dE/dx$ is reduced about 80\%, and pair production cross section is
reduced by 60\%; $X_0$ has more than doubled. For hadronic showers,
the effect is smaller, and, of course, these high energy photons are
only a small portion of the total shower.  On the other hand, since
hadronic interactions are only partially inelastic, the proton may
carry a significant fraction of its momentum deeper in the atmosphere,
where suppression is larger.  In general, at least part of the shower
will be suppressed.  Because of the large shower to shower variations,
it is difficult to give more quantitative estimates.

This simple model underestimates the importance of suppression.  When
suppression slows shower development, $\overline E$ of the remainder
of the shower will increase, increasing the suppression in the next
radiation length.  However, the model shows that suppression is very
important in photon showers, and in at least parts of proton initiated
showers.  However, suppression is clearly less significant for proton
showers than predicted by Capdevielle and Attallah (1992).  These
estimates appear consistent with Kalmykov, Ostapchenko and Pavlov
(1995) and Kasahara (1996).

Fluctuations are also very important in air showers. Because the
cosmic ray energy spectrum falls as $dN/dE\approx 1/E^3$, it is
important to understand the tails of the energy resolution
distribution; without accurate simulations, showers whose energy is
overestimated can skew the measured spectrum.  Suppression mechanisms
exacerbate the fluctuations, by greatly decreasing the number of
interactions in the initial stage of the shower.  For example, at sea
level, a $10^{10}$ eV electron emits an average of 14 bremsstrahlung
photons per $X_0$, while a $10^{17}$ eV electron emits only 3 photons.
Pair creation is similarly affected; pairs become more and more
asymmetric.  The result is that shower to shower fluctuations are much
larger.

Suppression also greatly reduces the number of particles in the early
stages of a shower, eliminating most of the low energy 'fuzz.'  The
LPM effect, dielectric suppression and pair conversion suppression all
contribute to this reduction.  Although the LPM effect covers the
widest range of energies, the other effects are stronger in the
limited range in which they operate.  For $E>E_p$, dielectric
suppression reduces the number of bremsstrahlung photons with $k<331$
MeV by two orders of magnitude.  None of the current simulation
efforts include these other mechanisms. Even relatively late in the
shower, there will be a reduction in the number of low energy
particles.  For example, where $\overline E\sim10^{13}$ eV, the LPM
effect suppresses photons below $\sim$500 MeV.  Although a low energy
cutoff does not affect the overall shower development, it can
significantly reduce the number of particles early in the shower. This
reduction may affect the shower profile observed by future air
fluorescence detectors.

Because $E_{LPM}$ drops as altitude decreases, suppression greatly
increases the chance of photons penetrating deep in the atmosphere,
even if the average interaction depth isn't too different.  For
trajectories where the probability of pair conversion in the magnetic
field is small, Bethe-Heitler and LPM predict similar average
conversion depths for a $3\times10^{20}$ eV shower, 114 and 122
g/cm$^2$ respectively.  However, Migdal predicts that the photon has a
7\% chance of surviving to a depth of 6$X_0$, while Bethe and Heitler
predict only a 0.25\% survival probability at the same depth. These
occasional deep interactions produce showers quite different from
photons that convert higher in the atmosphere, and much more energy
will reach the ground.  Although the magnitude is smaller, similar
effects may be seen for the most energetic photons from proton
interactions.

There have also been searches for neutrino interactions with air deep
in the atmosphere, producing nearly horizontal air showers.  Because
the neutrino interaction cross section rises with energy, this becomes
especially attractive for very high energy neutrinos, $E\sim 10^{15} -
10^{21}$ eV (Capelle \etal, 1998).  Since these neutrino interactions
occur at low altitudes, where $E_{LPM}$ is a few hundred PeV,
suppression is extremely important, especially at the higher energies.
For example, a $10^{20}$ eV electron at sea level has a $dE/dx$ a
factor of 10 lower than the Bethe-Heitler prediction.

Although suppression mechanisms may be of little importance to current
cosmic ray studies, the next generation of experiments will reach both
higher energies and higher statistics (Auger, 1996, Krizmanic, Ormes
and Streitmatter, 1998).  Although careful, detailed simulations are
needed to reach definite conclusions, it appears that suppression
mechanisms will affect the data collected.

\subsection{Showers in Detectors}

Changes in shower energy deposition may also be important in the next
generation of high energy and astrophysics detectors.  This review will
consider two examples, with a few representative numbers.

Direct cosmic ray composition measurements depend on measuring the
cosmic ray charge and energy; the latter in a calorimeter.  Because
these experiments must be done above much of the atmosphere, either in
balloon flights or space based experiments, calorimeter thickness is
limited by the allowable weight.  For example, the JACEE collaboration
has flown $\sim 6 X_0$ thick lead + emulsion/X-ray film calorimeters
on long duration balloon flights (Parnell \etal, 1989). Protons with
energies up to 500 TeV were observed.  Since a proton can be
diffractively excited to a $\Delta$ particle, which can decay
$p\pi^0\rightarrow p\gamma\gamma$, photons with energies up to about
7\% of the proton energy, 35 TeV, can be produced.  Figure
\ref{suppressionvse} shows that the location of the photon conversion
point is not affected since 35 TeV $\sim 8 E_{LPM}$.  However, the
produced electrons and positrons will exhibit reduced energy loss, and
overall energy leakage out of the back of the calorimeter may
increase.  The next generation of space based experiments will detect
protons with energies well beyond 1 PeV, with a consequently increased
effect (Parnell \etal, 1989).

At the Large Hadron Collider, now under construction at CERN,
electrons with energies up to about 1 TeV will be produced.  Electron
identification will be aided by a pre-shower radiator, which will
separate electrons from hadrons by measuring the energy radiated in
the first 2 or 3 $X_0$ of a lead calorimeter (Aspell \etal, 1996).
Because suppression reduces the number of photons emitted by the
electron, electrons will behave more like heavier particles such as
muons and pions. Without suppression, the number of photons emitted by
a 1 TeV electron in lead is infrared divergent: $N_\gamma\sim T/X_0
\ln{(E/k_{min})}$.  For a detection cutoff $E_{min}=1$ MeV, this is 14
photons per radiation length.  Suppression reduces this to an average
of 3 photons per radiation length.  The small average number of
photons increases the probability that the early stages of an electron
shower will be indistinguishable from a muon or pion interaction.

\section{QCD Analogs}
\label{sqcd}

QCD analogs of the LPM effect involve quarks and gluons moving through
a strongly coupling nuclear medium.  Because the nuclear medium is so
dense, a quark or gluon moving through the medium typically undergoes
many soft color exchange interactions, analogous to multiple
scattering.  Because a nucleus is small, the formation length is often
larger than the nucleus, so the nucleus acts as a single radiator.

The literature on this subject is voluminous and this is not the place
for a detailed review.  Instead, we will discuss a few examples, to
illustrate the basic features of the interactions, and also to
emphasize the both the similarities and the differences between the
QED and QCD cases.  Because QCD couples strongly and is non-Abelian,
the theoretical calculations are much less robust, and the
experimental data is much harder to interpret.

\subsection{Hadron Level Calculations}

In hadronic interactions, particle masses play a larger role than they
do in electromagnetic interactions.  Suppression effects in hadronic
interactions can be seen easily by considering the photoproduction of
vector mesons, for example $\gamma N\rightarrow\phi N$.  This reaction
is similar to $\gamma\rightarrow e^+e^-$, with a quark pair produced
instead of a lepton pair. One difference is that the interaction
between the quark pair and the nucleus is hadronic, rather than
electromagnetic.

The formation length for vector meson production is the same as for
pair production, $2\hbar k/M_V^2 c^3$.  For a 100 GeV photon producing a
$\phi$, $l_{f0}= 40$ fm, far larger than the nuclear diameter. So, the
photon/vector meson interacts with the nucleus as if it is a single
particle, with the interactions determined by
Eq. (\ref{esigmashulga}). Unfortunately, QCD does not specify
$f(\theta)$, but, in principle, this equation could be used to explore
how the cross section depends on the nuclear thickness.

Another example where suppression effects can be important is low mass
Drell-Yan dilepton production, $q\overline q\rightarrow l^+l^-$.  In
the rest frame of the target nucleus, the incoming quark or antiquark
can be thought of as pair producing the leptons.  The momentum
transfer from the target quark or antiquark can be small, so the
formation zone can extend far outside the nucleus.

\subsection{Quark Level Calculations}

More fundamental calculations involve quarks and gluons moving through
a possibly very hot nucleus.  One focus of recent calculations is the
search for the quark gluon plasma; a fast quark in a hadron gas
(normal nucleus) may interact differently than one in a quark gluon
plasma in which protons and neutrons are replaced by a sea of quarks
and gluons.  The sea of quarks and gluons, will have a longer
screening length and a larger cross section, so quarks and gluons
should lose energy considerably faster than in medium consisting of
confined quarks and gluons.  Measurements of high energy jets (or
hadrons) energy loss has been proposed as a tool for detecting the
quark gluon plasma (Wang, Huang and Sarcevic, 1996).

It is instructive to consider bremsstrahlung energy loss ($dE/dx$) by
a quark with energy $E$ emitting gluons of energy $k$ in a nuclear
medium.  Including the quark and gluon masses $m_q$ and $m_g$
(S\o rensen, 1992)
\begin{equation}
l_{f0} = {2\hbar Ek(E-k) \over m_q^2 c^3 k^2 + m_g^2 c^3 E(E-k)}.
\end{equation}
Unfortunately, the quark and gluon masses critically affect $l_{f0}$.
Besides bremsstrahlung, gluons can 'pair create' quark anti-quark
pairs, a close analog of pair production.  Because of the possibility
of experimentally measuring jet or hadron energy loss discussed above,
most calculations have focused on bremsstrahlung.

S\o rsensen (1992) repeated the Landau and Pomeranchuk semi-classical
derivation, and also adapted Migdal's formulae to quarks and gluons.
Although the results depend on $m_q$ and $m_g$, and hence have large
systematic errors, he did show that, for reasonable mass choices,
suppression was important.  This mass problem can be avoided because,
for QCD, the gluon emission angle (equivalent to $\theta_\gamma$ is
large, and Eq. (\ref{lfwithms}) can be written as
\begin{equation}
l_{f0} = { 2\hbar k \over k_\perp^2}
\end{equation} 
where $k_\perp$ is the perpendicular energy of the emitted gluon.
This avoids any explicit mass dependence.

Brodsky and Hoyer (1993) used quantum mechanical arguments to find the
energy loss of a parton travelling through nuclear matter.  They
decomposed the incident hadron into a spectrum of Fock states of
varying masses, thereby avoiding the parton mass uncertainty.  The
resulting energy loss is limited by $\Delta E/E < \kappa A^{1/3}/
x_1^2 s$ where $\kappa\sim0.5$ GeV$^2$, $x_1$ is the fractional energy
of the produced parton and $\sqrt{s}$ is the center of mass energy.
Here, $x_1\sqrt{s}$ is analogous to $E$. The target thickness is given
by the nuclear radius, $A^{1/3}$.  However, surface effects are not
included in the calculation.  Numerically, this energy loss should be
negligible for reasonably high energy hadrons.

Wang, Gyulassy and Pl\"umer (1995) did a detailed calculation of
energy loss in a quark gluon plasma.  They found that the radiation,
while obeying the bound listed above, was very sensitive to the color
screening distance in the plasma.  The strength of the suppression
depends on the ratio of the mean free path $\lambda$ to the formation
time $t_f$, as with the QED plasma.  When $t_f <\lambda/c$,
\begin{equation}
{dE \over dz} = {C_2\alpha_s c \langle q_\perp^2\rangle \over \pi\hbar}
\ln{\bigg({2 r_2 E\hbar \over \mu^2c^3 \lambda}\bigg)}
\end{equation}
where $\alpha_s$ is the strong force coupling constant, $\langle
q_\perp^2\rangle$ is the square of the average momentum transfer from
a single scattering, which should be proportional to $\mu^2$, with
$\mu$ the color screening mass. For quarks, the color factors
$C_2=4/3$ and $r_2=9/8$, while for gluons $C_2=3$ and $r_2=9/8$.  So,
the energy loss only grows logarithmically with particle energy; a
typical value is $dE/dx\sim$ 3.6 GeV/fm.  In the opposite limit,
$t_f>\lambda/c$, $dE/dx\sim E$, as with Bethe Heitler.

More recent calculations have emphasized the finite size of nuclei.
Many of these works have also considered the QED case, useful as a
check of their results.  Zakharov (1997) applies the techniques of
Sec. \ref{szakharov} to finite thicknesses of both nuclear matter and
the quark gluon plasma.  For a quark travelling through a nuclear
medium, the energy loss is proportional to the distance travelled,
$\Delta E/E\sim T/10$ fm, considerably higher than Brodsky and Hoyer.
Zakharov found that surface effects are large, so LPM suppression
plays a limited role.  This probably explains the disagreement with
Brodsky and Hoyer, who neglected surface effects.

For high energy quarks ($l_{f0}> R_A$) created inside the nucleus,
$\Delta E\sim T^2$; this dependence comes about because the quark
takes some time to develop its gluon field; this is analogous to the
lower $dE/dx$ observed for newly created $e^+e^-$ pairs.  For quarks
created at lower energies, where $l_f < R_A$, the energy loss is
proportional to the distance travelled, $\Delta E/E\sim T/10$ fm.

R. Baier and collaborators (1995) considered a fast quark or gluon
propagating through nuclear matter.  In an infinite medium, the soft gluon
spectrum matches their soft photon QED result (R. Baier \etal\ 1996),
with $dE/dx\sim \sqrt{E}$. This differs from the logarithmic energy
dependence found by Wang, Gyulassy and Pl\"umer. The difference is
that this work included additional diagrams, such as the non-Abelian
3-gluon vertices; they note that these are the dominant source of
emission.  A later, expanded collaboration (R. Baier \etal, 1997)
considered a finite size quark gluon plasma and found the same energy
dependencies as earlier, along with the same $R_A^2$ dependence found
by Zakharov.  However, for lower energy emission, they find $\Delta
E/E\sim T/4$ fm, much higher than Brodsky and Hoyer or Zakharov.  This
is because Zakharov includes additional diagrams in his calculation,
and also treats hard gluons differently  (R. Baier \etal, 1998b).

The collaboration has recently studied energy loss in a longitudinally
expanding quark gluon plasma (R. Baier \etal, 1998a).  This is the
first attempt to model a realistic, time varying temperature and
density distribution that could occur when two heavy nuclei collide.

While the different calculations agree in many ways, there is still
some significant disagreement.  They generally agree about the
appropriate energy scaling in the Bethe-Heitler (no suppression) and
strong suppression regimes, and also show a good correspondence with
the QED calculations.  The disagreement is over where these two
regimes apply.  Some of this stems from differing treatment of surface
terms.  Some of it may stem from the details of the initial state
used.  Whatever the cause, the numerical results vary greatly.
Unfortunately, because of the difficulty of clear experimental tests,
it may be some time before data can choose the best result.

\section{Suppression in E$^+$E$^-$ Collisions}

Future high energy electron-positron colliders will collide extremely
dense beams of electrons and positrons.  Besides the desired hadronic
interactions, large numbers of photons will be created.  Photon
emission before the hadronic interaction lowers the average collision
energy.  Also, some of the photons will pair convert, introducing a
charged particle background into the detector. Photons can be produced
by coherent beamstrahlung, where a single particle interacts with the
other beam as a whole, or by incoherent bremsstrahlung.  Because of
the density of the plasma, multiple scattering, Compton scattering and
magnetic suppression can suppress some of these interactions.

Although suppression occurs for similar reasons as in bulk matter,
many of the details are different because of the very different
environment.  Both bremsstrahlung and pair creation are modified
because bare charges are involved; $q_{min}$ is governed by the size
of the beams.  For both, the magnetic fields produced by the opposing
beam can have strong effects, and both bremsstrahlung and pair
creation are significantly suppressed.  (Baier and Katkov,
1973)(Katkov and Strakhovenko, 1977), as is beamstrahlung.

In contrast, because of the low density, LPM suppression is small in
currently envisioned machines.  Dielectric suppression transfers over
rather directly, after adjusting for the differing electron densities.
It is in principle strong enough to cause significant suppression.
However, because the beams are much shorter than $l_{f0}$, dielectric
suppression is reduced, and of only marginal significance (Chen and
Klein, 1993).

\section{Open Problems and Future Possibilities}
\label{openprobs}

Theoretically, there has been much progress in the past few years, and
several new approaches to LPM suppression have appeared.  The problem
of radiation from slabs, sets of foils and the like have all been
considered for a wide range of kinematic factors.  Despite this, these
calculations have some limitations.  Most only consider multiple
scattering, and none cover the full range of suppression mechanisms.
None cover pair production; this may no longer be a trivial extension.

Finally, the high energy regime when bremsstrahlung and pair creation
suppress each other needs to be explored.  Even if QED tests are not
practical, this regime may be important for understanding the QCD
case.  More generally, none of the calculations consider higher order
diagrams.

Even in the absence of a unified approach, a better treatment of
dielectric suppression is of interest.  To treat multi-GeV
bremsstrahlung with classical electromagnetism concepts like the
dielectric constant of a medium is rather anachronistic.  The
dielectric constant comes from photons forward Compton scattering off
the electrons in the medium; it would be very nice to see a
calculation of dielectric suppression that uses Compton scattering as
a starting point; such a calculation might yield some interesting
surprises.

All of these recent results need to be made generally accessible.  A
public program library would let experimenters compare the different
approaches under a variety of conditions.

Many implications of the LPM effect are poorly appreciated, and there
is a need to learn more about them.  One `lesson' from SLAC-E-146 is
that there is very little data on small $y$ bremsstrahlung, and
consequently little theoretical attention to this area. Of particular
concern are radiative corrections to electromagnetic effects.  Besides
the effect that suppressed bremsstrahlung will have on elastic
scattering, little is know about when higher order terms become
important.

Experimentally, there is a need to accurately measure suppression at
higher energies, and to look at new phenomena.  Cosmic ray based
experiments have poor statistics, and current accelerators only reach
the semi-classical ($k\ll E$) regime.  Ideally, a study of quantum
bremsstrahlung would use electrons with $E>E_{LPM}\approx\ 2.5$ TeV
for gold.  An experiment at Fermilab (Jones \etal, 1993) with 250 GeV
electrons would be a significant step forward, but would not reach
this goal.  Unfortunately, a TeV electron beamline seems a long way
off.  However, because a relatively low flux is sufficient, it might
be possible to use electrons and or photons produced in 14 TeV $pp$
collisions at LHC to study bremsstrahlung or pair production.
Suppression effects are large enough that they may also be observable
in the LHC general purpose detectors.  This is particularly likely in
apparatus such as pre-shower radiators that look at early shower
development.

Any future test of LPM or dielectric suppression must at least match
E-146 in statistics, backgrounds and control of systematics.  The
statistics should be simple, but the backgrounds and systematics will
require some effort.  It will be necessary to minimize the magnetic
fields in the bending magnets, photon angular acceptance, and
backgrounds due to the beam transport system.  With a higher energy
beam, it should also be possible to reduce the systematic errors, by
avoiding the calorimetric 'transition region' between energy
deposition by Compton scattering and by showering.  It would also be
very useful to collect data with a wider range of target thicknesses,
and also more low $Z$ targets. With an experiment accurate to 1-2\%,
it should be possible to compare the data with the recent theoretical
predictions, assuming that the problems of multiple emission by a
single electron can be handled properly, either theoretically or
experimentally.  While experiments with higher accuracy than E-146 are
very desirable, improving the systematic errors by a large factor over
E-146 would require extreme attention to detail.

Any new experiment should also investigate other phenomena, such as
magnetic suppression.  With $E> 200 GeV$, $y_B>y_{die}$ and magnetic
suppression should be directly observable. Studies in a constant
magnetic field are probably the most interesting, but the effect of
randomly ordered domains might also be observable.  Synchrotron
radiation could be separated from bremsstrahlung by measuring targets
of different density in a fixed magnetic field.  A similar apparatus
could measure emission from targets consisting of stacks of foils, to
investigate transition radiation due to multiple scattering.  It is
desirable to try to push bremsstrahlung measurements to the smallest
$y$ possible.  As Eq. (\ref{bkmag}) demonstrates, this is a new
regime, and some surprises may await us.  One tangible goal would be
to reach the point where higher order corrections cause current
theories to fail.

There is currently considerable theoretical effort studying LPM-like
phenomena in QCD.  With the coming data on high energy heavy ion
collisions from RHIC, this work will probably continue, with an
increasing emphasis on ways to differentiate between suppression in a
hadron gas and that of a quark gluon plasma.  Also, the current
calculations are limited by their neglect of higher order terms, which
are likely to be very important.  Experimentally, there is a clear
need is for a convincing demonstration of suppression involving QCD.

One place where more theoretical work is needed is to investigate
suppression in astrophysical phenomena.  One down-to-earth need is a
detailed study of suppression in EHE cosmic ray air showers.  Other
topics, less tied to specific experimental techniques, include the
study of suppression in astrophysical regions of extreme density,
temperature, and/or magnetic field.  This review has discussed a few
such areas, but there are many more.

Between the new theoretical approaches and the expanding range of
applications in QCD, plasmas and astrophysics, suppression mechanisms
seem destined to be a growing area of study over the coming years,

\section{Acknowledgements}

This review would not have been possible without much help.  V. Baier,
R. Becker-Szendy, R. Blankenbecler, S. Drell, D. Schiff, N. F. Shul'ga
and B.G. Zakharov explained various theoretical aspects of
suppression.  V. Baier, R. Becker-Szendy, R. Blankenbecler, S. Drell,
R. Engel, L. Kelley, D. Loomba, A. S\o rensen, X. N. Wang and
D. Zimmerman contributed by reading and commenting on early drafts of
this manuscript.  I am also grateful for the useful comments of the
Reviews of Modern Physics reviewers and editors.  Needless to say, any
mistakes are my responsibility. I'd also like to thank my E-146
collaborators for continuing discussions of the experiment.  Don
Coyne, Hans Georg Ritter and Jay Marx provided bureaucratic and moral
support.  This work was supported by the U.S. D.O.E. under contract
DE-AC03-76SF00098.


\begin{references}

\bibitem{akhiezer} 
Akhiezer, A. I. and N. F. Shul'ga, 1987, ``Influence of multiple
scattering on the radiation of relativistic particles in amorphous and
crystalline media,'' Usp. Fiz. Nauk {\bf 151}, 385 [Sov. Phys. Usp.
{\bf 30}, 197 (1987)].

\bibitem{akhiezer2}
Akhiezer, A. I. and N. F. Shul'ga, 1996, {\it High Energy
Electrodynamics in Matter}, Gordon and Breach.

\bibitem{prl}
Anthony, P. \etal, 1995, ``An accurate measurement of the 
Landau-Pomeranchuk-Migdal effect,'' Phys. Rev. Lett. 
{\bf 75}, 1949.

\bibitem{dprl}
Anthony, P. \etal, 1996, ``Measurement of dielectric suppression
of bremsstrahlung,'' Phys. Rev. Lett. {\bf 76}, 3550.

\bibitem{prd} Anthony, P. \etal, 1997, ``Bremsstrahlung suppression due
to the LPM and dielectric effects in
a variety of targets,'' Phys. Rev. {\bf D56}, 1373.

\bibitem{Artru} Artru, X., G. B. Yodh and G. Mennessier, 1975,
``Practical theory of multilayered transition radiation detector,''
Phys. Rev. {\bf D12}, 1289.

\bibitem{Arutyunyan}
Arutyunyan, F. R., A. A. Nazaryan and A. A. Frangyan, 1972,
``Influence of the medium on the emission of relativistic electrons,''
Zh. Eksp. Teor. Fiz. {\bf 62}, 2044 
[Sov. Phys. JETP, {\bf 35}, 1067 (1972)].

\bibitem{aspell} Aspell, P., \etal, 1996, ``Energy and spatial resolution of
a shashlik calorimeter and a silicon preshower detector,''
Nucl. Instrum.  and Meth. {\bf A376}, 17.

\bibitem{Auger} Auger Collaboration, 1996, ``The Pierre Auger project
design report,'' Fermilab-Pub-96-024 (unpublished).

\bibitem{Baier2} Baier, R., Yu. L. Dokshitzer, A. H. Mueller and
D. Schiff, 1996, ``The Landau-Pomeranchuk-Migdal effect in QED,''
Nucl. Phys. {\bf B478}, 577.

\bibitem{Baier3} Baier, R., Yu. L. Dokshitzer, A. H. Mueller and
D. Schiff, 1998a, ``Radiative energy loss of high energy partons
traversing an expanding QCD plasma,'' Phys. Rev. {\bf C58}, 1706.

\bibitem{Baier4} Baier, R., Yu. L. Dokshitzer, A. H. Mueller and
D. Schiff, 1998b, ``Medium-induced radiative energy loss; equivalence
between the BDMPS and Zakharov formalisms,'' Nucl. Phys. {\bf B531},
403.

\bibitem{Baier} Baier, R., Yu. L. Dokshitzer, A.H. Mueller, S. Peigne
and D. Schiff, 1997, ``Radiative energy loss of high energy quarks
and gluons in a finite volume quark gluon plasma,''  Nucl. Phys.
{\bf B483}, 297.

\bibitem{baier0}
Baier, R, Yu. L. Dokshitzer, S. Peigne and D. Schiff, 1995,
``Induced gluon radiation in a QCD medium,'' Phys. Lett.
{\bf B345}, 277.

\bibitem{bairmag} 
Baier, V. N. and V. M. Katkov, 1972, ``Production of bremsstrahlung
during collisions of high-energy particles in a magnetic field,''
Dokl. Akad. Nauk. SSSR {\bf 207}, 68 [Sov. Phys. Dokl., {\bf 17},
1068 (1973)].

\bibitem{baierlpm} Baier, V. N. and V. M. Katkov, 1997a, ``The theory
of the Landau, Pomeranchuk, Migdal effect,'' hep-ph/9709214;
Phys. Rev. {\bf D57}, 3146 (1998).

\bibitem{baierlpm2} Baier, V. N. and V. M. Katkov, 1997b, ``The
Landau-Pomeranchuk-Migdal effect in a thin target,'' preprint
hep-ph/9712524.

\bibitem{baierlpm3} Baier, V. N. and V. M. Katkov, 1999,
``Multi-photon effects in energy losses spectra,'' 
Phys. Rev. {\bf D59}, 056003.

\bibitem{bairmag2} 
Baier, V. N., V. M. Katkov and V. M. Strakhovenko, 1988, ``Radiation
at collision of relativistic particles in media in the presence of
external field,'' Zh. Eksp. Teor. Fiz. {\bf 94}, 125 [Sov. Phys. JETP
{\bf 67}, 70 (1988)].

\bibitem{bairrvw1}
Baier, V. N., V. M. Katkov and V. M. Strakhovenko, 1989, ``Interaction
of high-energy electrons and photons with crystals,''
Usp. Fiz. Nauk. {\bf 159}, 455 [Sov. Phys. Usp. {\bf 32}, 972 (1989)].


\bibitem{baiertalk}
Baier, V. N., 1998, private communication.

\bibitem{bak} Bak, J. F., J. B. B. Petersen, E. Uggerh\o j, K. 
\O stergaard, S. P. M\o ller and A. H. S\o rensen, 1986, ``Influence of
transition radiation and density effect on atomic K-Shell
excitation,'' Phys. Scripta {\bf 33}, 147.

\bibitem{bak}
Bak J. F., {\it et al.}, 1988, ``Channeling radiation from 2 to 20 GeV/c
electrons and positrons (II).,'' Nucl. Phys., {\bf B302}, 525.

\bibitem{PDG} Barnett, R.M. \etal\ (Particle Data Group), 1996,
``Review of particle properties,'' Phys. Rev. {\bf D54, 1}.

\bibitem{Bell}
Bell, J. S., 1958, ``Bremsstrahlung from multiple scattering,''
Nucl. Phys. {\bf 8}, 613.

\bibitem{BH} Bethe, H. A. and W. Heitler, 1934, ``On the stopping of
fast particles and the creation of positive electrons,'' Proc. Royal
Soc. {\bf A146}, 83.

\bibitem{bird}
Bird, D. J., {\it etal.}, 1994,
``The cosmic-ray energy spectrum observed by the Flys Eye,''
Astrophys. J. {\bf 424}, 491.


\bibitem{blankdrell}
Blankenbecler, R. and S. D. Drell, 1996, ``The Landau-Pomeranchuk-Migdal
effect for finite targets,'' Phys. Rev. {\bf D53}, 6265.

\bibitem{blanken1}
Blankenbecler, R., 1997a, ``Structured targets and the 
Landau-Pomeranchuk-Migdal effect,'' Phys. Rev. {\bf D55}, 190.

\bibitem{blanken2}
Blankenbecler, R., 1997b , ``Multiple scattering and functional integrals,''
Phys. Rev. {\bf D55}, 2441.

\bibitem{blankx}
Blankenbecler, R., 1997c, private communication.

\bibitem{brodsky}
Brodsky, S. J. and P. Hoyer, 1993, ``A bound on the energy loss of partons in
nuclei,'' Phys. Lett. {\bf B298}, 165.

\bibitem{Capelle}
Capelle, K. S., J. W. Cronin, G, Parente and E. Zas, 1998,
``On the detection of ultrahigh-energy neutrinos with the Auger
Observatory,'' Astropart. Phys. {\bf 8}, 321.

\bibitem{capdev}
Capdevielle, J. N. and R. Attallah, 1992, ``Limits of classical
detection for GAS,'' Nucl. Phys. B (Proc. Suppl.), {\bf 28B},
90.

\bibitem{proposal}
Cavalli-Sforza, M. {\it et al.}, 1992, ``A proposal for an experiment
to study the interference between multiple scattering and 
bremsstrahlung,'' SLAC-Proposal-E-146, June, 1992.

\bibitem{beam}
Cavalli-Sforza M., \etal, 1994, ``A method of obtaining parasitic
$e^+$ or $e^-$ beams during SLAC linear collider operation,'' IEEE
Trans. Nucl. Sci. {\bf 41}, 1374.

\bibitem{chenklein}
Chen, P. and S. Klein, 1992, ``The Landau-Pomeranchuk-Migdal effect
and suppression of beamstrahlung and bremsstrahlung in linear
colliders,'' in {\it Proc. of the Advanced Accelerator Conference
Workshop}, ed. J. S. Wurtele (AIP, New York), pg. 921.

\bibitem{Cherry}
Cherry, M. L., 1978, ``Measurements of the spectrum and energy
dependence of X-ray transition radiation,'' Phys. Rev. {\bf D17}, 2245.

\bibitem{fb2} Feinberg, E. L., 1994, ``Effect confirmed 40 years
later,'' Priroda, No 1. pg. 30.

\bibitem{feinberg}
Feinberg, E. L. and I. Pomeranchuk, 1956, ``High energy inelastic
diffraction phenomena,'' Supplemento Al Volume III,
Series X, Del Nuovo Cimento, 652.

\bibitem{fomin}
Fomin, P. I., 1958, ``Radiative corrections to bremsstrahlung,''
Zh. Eksp. Teor. Fiz. {\bf 35,} 707
[Sov. Phys. JETP {\bf 35}, 491 (1959)].

\bibitem{fomin2}
Fomin, S. P. and N. F. Shul'ga, 1986, ``On the space-time evolution of
the process of ultrarelativistic electron radiation in a thin
layer of substance,'' Phys. Lett. {\bf 114A}, 148.

\bibitem{fowler}
Fowler, P. H., D. H. Perkins and K. Pinkau, 1959, ``Observation of the
suppression effect on bremsstrahlung,'' Phil. Mag. {\bf 4}, 1030.

\bibitem{galitsky}
Galitsky, V. M. and I.~I.~Gurevich, 1964, ``Coherence effects in
ultra-relativistic electron bremsstrahlung,'' Il Nuovo Cimento {\bf
32}, 396.

\bibitem{garibyan}
Garibyan, G. M., 1960, ``Radiation of a particle moving across the
interface of two media with account of multiple scattering,''
Zh. Eksp. Teor. Fiz. {\bf 39}, 332
[Sov. Phys. JETP {\bf 12}, 237 (1961)].

\bibitem{goldman} 
Gol'dman, I. I., 1960, ``Bremsstrahlung at the boundary of a medium
with account of multiple scattering,''
JETP {\bf 38}, 1866 
[Sov. Phys. JETP {\bf 11}, 1341 (1960)].

\bibitem{Heckler}
Heckler, A. F., 1995, ``On the formation of a Hawking-radiation
photosphere around microscopic black holes,'' Phys. Rev. {\bf D55},
480.

\bibitem{jackson}
Jackson, J. D., 1975, {\it Classical Electrodynamics,} (John
Wiley and Sons, New York).

\bibitem{jones} Jones, L. W. {\it et al.}, 1993, ``An experimental
test of the Landau-Pomeranchuk-Migdal effect,'' Fermilab Proposal
P-813 (unpublished).

\bibitem{Kalmykov}
Kalmykov, N. N., S. S. Ostapchenko and A. I. Pavlov, 1995, ``Influence
of the Landau-Pomeranchuk-Migdal Effect on the features of extensive
air showers,'' Yad. Fiz. {\bf 58}, 1829 [Phys. Atomic Nuclei {\bf 58},
1728 (1995)].

\bibitem{Kasahara} Kasahara, K., 1985, ``Experimental examination of
the Landau-Pomeranchuk-Migdal effect by high-energy electromagnetic
cascades in lead'', Phys. Rev. {\bf D31}, 2377.

\bibitem{Kasahara2} Kasahara, K., 1996, ``The LPM and geomagnetic
effects on the development of air showers in the GZK cutoff region,''
in {\it Proc. of Int. Sympo. on Extremely High Energy Cosmic Rays:
Astrophysics and Future Observations}, ed.  M. Nagano, Institute for
Cosmic Ray Research, Tokyo.

\bibitem{katkov} 
Katkov, V. N. and V. M. Strakhovenko, 1977,
``Bremsstrahlung in collisions of electrons in a magnetic field,''
Yad. Fiz. {\bf 25}, 1245
[Sov. J. Nucl. Phys. {\bf 25}, 660 (1977)].

\bibitem{klein} Klein, S. R. {\it et al.}, 1993, ``A measurement of the
LPM effect'' in {\it Proc. XVI Int. Symp. Lepton and Photon
Interactions at High Energies}, edited by P.~Drell and D.~Rubin, 
(AIP, New York), p. 172.

\bibitem{klein2} Klein, S. R., 1998, ``Bremsstrahlung and pair
creation: suppression mechanisms and how they affect air showers,'' in
{\it Workshop on Observing Giant Cosmic Ray Air Showers From $\ge
10^{20}$ eV Particles from Space}, AIP Conference Proceedings 443,
edited by J. F. Krizmanic, J. F. Ormes and R. E. Streitmatter
(AIP, New York), p. 132.

\bibitem{knoll1} Knoll, J. and D. N. Voskresensky, 1995,
``Non-equilibrium description of bremsstrahlung in dense matter
(Landau-Pomeranchuk-Migdal effect),'' Phys. Lett. {\bf B351}, 43.

\bibitem{knoll} Knoll, J. and D. N. Voskresensky, 1996, ``Classical
and quantum many-body description of bremsstrahlung in dense matter,''
Ann. Phys. {\bf 249}, 532.

\bibitem{konishi}
Konishi, E., A. Adachi, N. Takahasi and A. Misaki, 1991,
``On the characteristics of individual cascade showers with the
LPM effect at extremely high energies'', J. Phys. G {\bf 17}, 719. 

\bibitem{krizmanic} Krizmanic, J. F., J. F. Ormes and
R. E. Streitmatter, 1998, {\it Workshop on Observing Giant Cosmic Ray
Air Showers From $\ge 10^{20}$ eV Particles from Space}, AIP
Conference Proceedings No. 433 (AIP, New York).

\bibitem{landau1}
Landau, L. D and I.~J.~Pomeranchuk, 1953a, ``The limits of applicability
of the theory of bremsstrahlung by electrons and of the
creation of pairs at large energies,'' Dokl. Akad. Nauk. SSSR {\bf 92}
535. This paper is available in English in (Landau, 1965).

\bibitem{landau2} Landau, L. D. and I.~J.~Pomeranchuk, 1953b,
``Electron-cascade processes at ultra-high energies,''
Dokl. Akad. Nauk. SSSR {\bf 92}, 735.  This paper is available in
English in (Landau, 1965).

\bibitem{landau3}
Landau, L. D., 1965,  {\it The Collected Papers of L.~D.~Landau}, Pergamon
Press.

\bibitem{Laskin} Laskin, N. V., A. S. Mazmanishvili and N. F. Shul'ga,
1984, ``The path-integral approach to inclusion of the influence of
multiple scattering on the radiation by high-energy particles in
crystals and amorphous media,'' Dokl. Akad. Nauk. SSSR {\bf 227}, 850
[Sov. Phys. Dokl. {\bf 29}, 638 (1984)].

\bibitem{Laskin2} Laskin, N. V., A. S. Mazmanishvili, N. N. Nasonov
and N. F. Shul'ga, 1985, ``Theory of emission by relativistic
particles in amorphous and crystalline media,'' Zh. Eksp.
Teor. Fiz. {\bf 89}, 763 [Sov. Phys. JETP {\bf 62}, 438 (1985)].

\bibitem{Learned}
Learned J. G. and S. Pakvasa, 1995, ``Detecting tau-neutrino oscillations
at PeV energies,'' Astropart. Phys. {\bf 3}, 267.

\bibitem{Lohrmann}
Lohrmann, E., 1961,  ``Investigation of bremsstrahlung and pair production at
energies $>10^{11}$ eV,'' Phys. Rev. {\bf 122} 1908.

\bibitem{mies}
Miesowicz, M., O. Stanisz and W. Wolther, 1957,
``Investigation of an electromagnetic cascade of very high energy in
the first stage of its development,'' Nuovo Cim.  {\bf 5}, 513.

\bibitem{migdal}
Migdal, A. B., 1956, ``Bremsstrahlung and pair production in condensed media
at high energies,'' Phys. Rev. {\bf 103}, 1811.

\bibitem{mgidal2}
Migdal, A. B., 1957, ``Bremsstrahlung and pair production at high energies
in condensed media,'' Zh. Eksp. Teor. Fiz. {\bf 32}, 633
[Sov. Phys. JETP {\bf 5}, 527 (1957)]. 

\bibitem{misaikiwater}
Misaki, A., 1990, ``A study of electromagnetic cascade showers with the
LPM effect in water for the detection of extremely high energy
neutrinos,'' Forschr. Phys. {\bf 38}, 413.

\bibitem{misaki2}
Misaki, A., 1993, ``The Landau - Pomeranchuk - Migdal (LPM) effect
and its influence on electromagnetic cascade showers at extremely
high energies,'' Nucl. Phys. B (Proc. Suppl.) {\bf 33A,B}, 192.

%
\bibitem{pafomov2}
Pafomov, V. E., 1964, ``Effect of multiple scattering on transition
radiation,'' 
Zh. Eksp. Teor. Fiz.  {\bf 47}, 530
[Sov. Phys. JETP, {\bf 20}, 353 (1965)].

\bibitem{pafomov}
Pafomov, V. E., 1965, ``Concerning bremsstrahlung,'' 
Zh. Eksp. Teor. Fiz. {\bf 49}, 1222
[Sov. Phys. JETP {\bf 22}, 848 (1966)].

\bibitem{pafomovoptical}
Pafomov, V. E., 1967, ``Optical bremsstrahlung in an absorbing medium,''
Zh. Eksp. Teor. Fiz.  {\bf 52}, 208
[Sov. Phys. JETP, {\bf 25}, 135 (1967)].

\bibitem{parnell}
Parnell, T. A. \etal, 1989, ``Spectra, composition and interactions of
nuclei with magnet interaction chambers,'' in
{\it Particle Astrophysics: The NASA Cosmic Ray Program for the
1990's and Beyond}, ed. W. V. Jones, F. J. Kerr and J. F. Ormes,
AIP, New York.

\bibitem{palazzi}
Palazzi, G. D., 1968, ``High-energy bremsstrahlung and electron pair
production in thin crystals,'' Rev. Mod. Phys. {\bf 40}, 611.

\bibitem{perl}
Perl, M. L., 1994, ``Notes on the Landau, Pomeranchuk, Migdal effect:
experiment and theory,'' in {\it Proc. 1994 Les Rencontres de Physique
de la Vallee D'Aoste}, (Editions Frontieres, Gif-sur-Yvette,
France,). Ed. M.~Grego, p.~567.

\bibitem{Pomanskii} Pomanskii, A. A., 1970, ``The mean free path of
ultra-high energy electrons in ground. (Cascade theory in the
ultra-relativistic region),'' Izv. Akad Nauk. SSR Ser. Fiz. {\bf 34},
2008.

\bibitem{raffelt}
Raffelt, G. and D. Seckel, 1991, ``Multiple-scattering suppression of
bremsstrahlung emission of neutrinos and axions in supernovae,''
Phys. Rev. Lett.  {\bf 67}, 2605.

\bibitem{Rossi}
Rossi, B., 1952, {\it High Energy Particles}, Prentice Hall, Inc.

\bibitem{schiff} 
Schiff, L. I., 1968, ``Quantum Mechanics'', 3rd edition, (McGraw Hill,
New York).

\bibitem{scott} 
Scott, W. T., 1963, ``The theory of small-angle multiple
scattering of fast charged particles,'' Rev. Mod. Phys.  {\bf 35},
231.

\bibitem{shulga2}
 Shul'ga, N. F. and S. P. Komin, 1978, ``Suppression of radiation in an
amorphous medium and in a crystal,'' 
Pis'ma Zh. Eksp. Teor. Fiz. {\bf 27}, 126
[JETP Lett.  {\bf 27}, 117 (1978)].

\bibitem{shulga} 
Shul'ga, N. F. and S. P. Fomin, 1996, ``On the experimental verification
of the Landau-Pomeranchuk-Migdal effect,'' 
Pis'ma Zh. Eksp. Teor. Fiz. {\bf 63}, 837
[JETP Lett., {\bf 63}, 873 (1996)].

\bibitem{shulga} Shul'ga, N. F. and S. P. Fomin, 1998, ``Effect of
multiple scattering on the emission of ultrarelativistic electrons in
a thin layer of matter'', Zh. Eksp Teor. Fiz. {\bf 113}, 58 [JETP {\bf
86}, 32 (1998)].

\bibitem{ehe} Sokolsky, P., P. Sommers and B. R. Dawson, 1992,
``Extremely High-Energy Cosmic Rays,'' Phys. Rep. {\bf 217}, 225.

\bibitem{soren2}
S\o rensen, A. H., 1992, ``On the suppression of the gluon radiation
for quark jets penetrating a dense quark gas,''
Z. Phys. C {\bf 53}, 595.

\bibitem{sorensen} S\o rensen, A. H., 1996, ``Channeling,
bremsstrahlung and pair creation in single crystals,''
Nucl. Instrum. \& Meth., {\bf B119}, 1.

\bibitem{stanev}
Stanev T., Ch. Vankov, R. E. Streitmatter, R. W. Ellsworth and
T. Bowen, 1982, ``Development of ultrahigh-energy electromagnetic
cascades in water and lead, including the Landau-Pomeranchuk-Migdal
effect,'' Phys. Rev. {\bf D25}, 1291.

\bibitem{stanev2}
Stanev, T. and H. P. Vankov, 1997, ``Nature of the highest energy
cosmic rays,'' Phys. Rev. {\bf D55}, 1365.

\bibitem{strausz}
Strausz, S. C. {\it et al.}, 1991, ``A measurement of the Landau Pomeranchuk
Migdal effect in electromagnetic showers,'' in {\it Proc. of the 22nd
Intl. Cosmic Ray Conf., Dublin, Ireland},
Vol. 4, pg. 233.

\bibitem{AGASA} Takeda M. {\it et al.}, 1998, ``Extension of the
cosmic ray energy spectrum beyond the predicted
Greisen-Zatsepin-Kuz'min cutoff,'' Phys. Rev. Lett. {\bf 81}, 1163.

\bibitem{tm1}
Ter-Mikaelian, M. L., 1953a, ``Scatter of high energy electrons
in crystals,'' Zh. Eksp. Teor. Fiz. {\bf 25}, 289.

\bibitem{tm2} Ter-Mikaelian, M. L., 1953b, ``The interference emission
of high-energy electrons,'' Zh. Eksp. Teor. Fiz. {\bf 25}, 296.

%
\bibitem{termikaelianbook} Ter-Mikaelian, M. L., 1972, {\it High
Energy Electromagnetic Processes in Condensed Media,} (John Wiley \&
Sons, New York).

\bibitem{ternovskii}
Ternovskii, F. F., 1960, ``On the theory of radiative processes in
piecewise homogeneous media,'' 
Zh. Eksp. Teor. Fiz. {\bf 39}, 171
[Sov. Phys. JETP {\bf 12}, 123 (1961)].

\bibitem{toptygin}
Toptygin, I. N., 1963, ``On the theory of bremsstrahlung and pair
production in a medium,'' 
Zh. Eksp. Teor. Fiz. {\bf 46}, 851
[Sov. Phys. JETP {\bf 19}, 583 (1964)].

\bibitem{tsai}
Tsai, Y. S., 1974, ``Pair production and bremsstrahlung of
charged leptons,'' Rev. Mod. Phys. {\bf 46}, 815.

%
\bibitem{crx} Varfolomeev, R. R.~I.~Gerasimova, I.~I.~Gurevich,
L.~A.~Makar'ina, A.~S.~Romantseva and S.~A.~Chueva, 1960, ``Influence
of the medium density on bremsstrahlung in electron-photon showers in
the energy range $10^{11}$ - $10^{13}$ eV,'' Zh. Eksp. Teor. Fiz. {\bf
38}, 33 [Sov. Phys. JETP {\bf 11}, 23 (1960)].

\bibitem{protvino}
Varfolomeev A. {\it et al.}, 1975, ``Effect of the medium on the
bremsstrahlung spectrum of 40-GeV electrons,''
Zh. Eksp. Teor. Fiz. {\bf 69}, 429 [Sov. Phys. JETP {\bf 42}, 218 (1976)].

\bibitem{wang2}
Wang, X. N., M. Gyulassy, and M. Pl\"umer, 1995, 
``Landau-Pomeranchuk-Migdal effect in QCD and radiative energy
loss in a quark-gluon plasma,'' Phys. Rev. {\bf D51}, 4346.

\bibitem{wang}
Wang, X. N., Z. Huang and I. Sarcevic, 1996, ``Jet quenching in the
direction opposite to a tagged photon in high-energy heavy-ion
collisions,'' Phys. Rev. Lett. {\bf 77}, 231.

\bibitem{williams}
Williams, E. J., 1935, ``Correlation of certain collision problems
with radiation theory,'' Kgl. Danske Videnskab. Selskab,
Mat. Phys. Medd. {\bf 13}, Number. 4, Pg. 1.

\bibitem{zakharov}
Zakharov B. G., 1996a, ``Fully quantum treatment of the 
Landau-Pomeranchuk-Migdal effect in QED and QCD,'' 
Pis'ma v. ZhETF {\bf 63}, 906
[JETP Lett. {\bf 63} 952 (1996)].
 
\bibitem{zakharov2} 
Zakharov, B. G., 1996b, ``Landau-Pomeranchuk-Migdal effect for
finite-size targets,'' 
Pis'ma v. ZhETF {\bf 64}, 781
[JETP Lett. {\bf 64}, 781 (1996)].

\bibitem{zakharov3}
Zakharov, B. G., 1997, ``Radiative energy loss of high energy quarks in
finite-size nuclear matter and quark-gluon plasma,'' 
Pis'ma v. ZhETF {\bf 65}, 685
[JETP Lett. {\bf 65}, 615 (1997)].

\bibitem{zakharov4} Zakharov, B. G., 1998a, ``Light-cone path integral
approach to the Landau-Pomeranchuk-Migdal effect,'' Yad. Fiz. {\bf
61}, 924 [Phys. Atom. Nucl. {\bf 61}, 838 (1998)].

\bibitem{zakharov5}
Zakharov, B. G., 1998b, ``Light-cone path integral approach to the
Landau-Pomeranchuk-Migdal effect and the SLAC data on bremsstrahlung
from high energy electrons,'' preprint hep-ph/9805271.

\bibitem{zas}
Zas, E., and J. Alvarez-Muniz, ``\v Cerenkov radio pulses from EeV
neutrino interactions: the LPM effect'', Phys. Lett. {\bf B411}, 218.

\end{references}
\end{document}